\documentclass[12pt]{elsarticle}
\usepackage{latexsym, amssymb, amsmath, mathtools, hyperref, braket, accents}

\DeclareMathOperator{\tr}{tr}

\newcommand{\dif}{\text{d}}
\newcommand{\fdif}{\text{D}}

\newcommand{\C}{\mathbb{C}}

\newcommand{\fA}{\mathcal{A}}
\newcommand{\fC}{\mathcal{C}}
\newcommand{\fD}{\mathcal{D}}

\newcommand{\fH}{\mathcal{H}}
\newcommand{\fK}{\mathcal{K}}
\newcommand{\fL}{\mathcal{L}}

\newcommand{\fO}{\mathcal{O}}
\newcommand{\fU}{\mathcal{U}}

\newcommand{\cl}{\mathrm{c}}
\newcommand{\q}{\mathrm{q}}
\newcommand{\ta}{\mathrm{a}}
\newcommand{\ts}{\mathrm{s}}
\newcommand{\tR}{\mathrm{R}}
\newcommand{\tA}{\mathrm{A}}
\newcommand{\tK}{\mathrm{K}}

\newcommand{\tB}{\mathrm{B}}

\newcommand{\tT}{\mathrm{T}}

\newlength{\dhatheight}
\newcommand{\doublehat}[1]{
    \settoheight{\dhatheight}{\ensuremath{\hat{#1}}}
    \addtolength{\dhatheight}{-0.35ex}
    \hat{\vphantom{\rule{1pt}{\dhatheight}}
    \smash{\hat{#1}}}}
\newcommand{\cev}[1]{\reflectbox{\ensuremath{\vec{\reflectbox{\ensuremath{#1}}}}}}

\begin{document}
\begin{frontmatter}

\title{Field Theory of Many-Body Lindbladian Dynamics}

\author{Foster Thompson1}
\author{Alex Kamenev1,2\\1 School of Physics and Astronomy, University of Minnesota, Minneapolis, 55414, MN, USA\\2 William I. Fine Theoretical Physics Institute, University of Minnesota, Minneapolis, 55414, MN, USA}


\begin{abstract}
We review and further develop the Keldysh functional integral technique  for the study of Lindbladian evolution of  many-body driven-dissipative quantum systems. A systematic and pedagogical account of the dynamics of generic bosonic and fermionic Lindbladians is presented.  
Our particular emphasis is on unique properties of the stationary distribution function, determined by the Lyapunov equation.   
  This framework is applied to study examples of Lindbladian dynamics in the context of band theory, disorder, collisionless collective modes, and mean-field theory. 
\end{abstract}

\end{frontmatter}

\tableofcontents

\section{Introduction}\label{1}
Non equilibrium dynamics of open quantum systems has attracted a lot of attention in recent years \cite{QComp1,QComp2,DiehlKeldyshLindblad,Trajectories,DrivenOpenColdAtom}.  The interest is mostly stimulated by a rapid progress in design and manufacturing of prototypical quantum computers with dozens and hundreds of qubits \cite{TrappedIon1,TrappedIon2,TrappedIon3,TrappedIon4,SCqubit1,SCqubit2,SCqubit3,SCqubit4,ManyBodyLight,AMO,IsingMachine}. The very essence of qubits as controllable quantum systems dictates both their non-equilibrium nature as well as their coupling to extensive number of external degrees of freedom. 

An economic and, in many cases, justified way of treating such driven dissipative quantum systems  is to employ the Markovian (i.e. time-local) approximation. Under such assumption dynamics of a {\em reduced} density matrix is given by a Lindblad equation \cite{Lindblad,GKS}. Historically, the study of Lindbladian dynamics was primarily restricted to systems with a few degrees of freedom, with most of the focus coming from the quantum optics literature \cite{OpenQM,QuantumNoise,QuantumOptics,QuantumOptics2}. For example,
Lindbladian evolution of a single two-level system is fully equivalent to the set Bloch equations.
Other examples include the dynamics of parametrically driven oscillators \cite{Dykman1,Dykman2,Dykman3}, cavity QED and coupled cold atom-cavity systems \cite{ColdAtomCavity,TrappedIonColdAtom}, etc. Their considerations lead to a number of insightful physical results and powerful theoretical approaches.

The modern quantum computation platforms, such as Josephson or ionic traps, fall squarely into the realm of {\em many-body} systems. In addition to these, we also mention driven-dissipative quantum fluids and Bose condensates \cite{QuantumLightFluid,BEC,BEC2,BEC3,BECMB1,BECMB2,BECMB3,BECMB4,BECMB5}, dynamics of large networks of coupled parametric oscillators and optical cavities \cite{PDOMB1,PDOMB2,PDOMB3,PDOMB4,PDOMB5,PDOMB6,PDOMB7}, and monitored dynamics of spatially extended systems \cite{Monitored1,Monitored2,Monitored3,Monitored4}, all of which implicitly involve extensively large numbers of degrees of freedom.
Indeed, already $N=50$ connected qubit devices gives rise to the Hilbert space dimension ${\cal N}=2^{50}$, which is well beyond traditional single-particle matrix manipulation techniques.  This calls for the developing of {\em many-body} field theoretical techniques geared towards description of the Lindbladian (as opposed to von Neumann) evolution.

An important step in this direction is consideration of many-body bosonic or fermionic systems, 
traditionally described in the occupation number basis via the algebra of creation/annihilation operators. In this approach both many-body effective Hamiltonian and a set of many-body 
quantum {\em jump operators} are all expressed as polynomials of such creation/annihilation 
operators. It is important to remember, though, that despite of a deceptively simple appearance 
all these operators act in the exponentially large (in the number of the degrees of freedom) Hilbert space.  
Therefore a brute force numerical solution of the corresponding Lindblad equation requires diagonalization of ${\cal N}^2\times {\cal N}^2$ matrix (for, e.g., fermionic case). Clearly this is not a productive direction. 

Various techniques have been developed for studying dynamics of {\em quadratic} Lindbladians, i.e. those with the Hamiltonian  given by a quadratic form, while all quantum jump operators by  linear forms of the creation/annihilation operators. One such approach is provided by the so-called ``third quantization" technique \cite{3rdQuantBose,3rdQuantFerm,3rdQuantSpectralThm,3rdQuantSols,3rdQuantCorrelationFunctions}, based on the use of algebras of bosonic or fermionic {\em superoperators}.  This approach shows that ${\cal N}^2$ eigenvalues of a quadratic many-body Lindbladian may be constructed from $N$ complex eigenvalues of a certain $N\times N$ non-Hermitian matrix, using conventional bosonic or fermionic occupation numbers.

Let us also mention the notable topological classification of Lindbladian fermions \cite{Top1,Top2,Top3} and various results pertaining to both bosonic and fermionic Gaussian states \cite{GaussianBose1,GaussianBose2,GaussianBose3,GaussianBose4-Heff,GaussianFerm1,GaussianFerm2,GaussianFerm3,GaussianFerm4}.
These techniques have been variously applied to the study numerous problems, including the Bogoliubov spectrum of driven-dissipative condensates \cite{BEC}, exact solutions of nonlinear integrable systems \cite{ExactQuarticOscillator}, and the study topological properties of various low-dimensional dissipative systems \cite{KitaevFerm1,KitaevFerm2,TopLowD,SSH,KitaevBose}.

The purpose of this manuscript is to review and further develop an alternative apparatus, based on coherent state functional integral field-theoretical treatment of the Lindbladian dynamics, pionered by Sieberer, Buchhold, and  Diehl \cite{DiehlKeldyshLindblad}.  This technique originates from the Keldysh theory \cite{Kamenev} of the underlying Von Neumann dynamics of the interacting system-bath pair. Upon integrating out the bath degrees of freedom and adopting Markovian approximation for the bath-induced self-energy, one ends up with a time-local effective action. The latter is fully equivalent to the many-body operator Lindblad equation \cite{DiehlKeldyshLindblad}.  It constitutes, however, a more convenient starting point for calculation of various observables, correlation functions, linear response characteristics, collective modes, etc. 
It is also indispensable for  generalizations beyond the quadratic theory.  

In the quadratic approximation this approach naturally reproduces the results of the third quantization for the spectra of the Lindbladian superoperators. Its advantage is in making unmistakably clear that this information is only part of the whole picture. While in equilibrium systems, 
both statistical weights and the dynamics are determined by the same set of energies, this is {\em not} the case for driven-dissipative Lindbladian dynamics. In this case the complex spectrum of the dynamical relaxation is not directly related to (real) statistical weights of the stationary (but non-equilibrium) density matrix. 
The latter is determined by the stationary distribution function $\check F_\mathrm{st}$, which naturally emerges as one of the main building blocks of the Keldysh treatment. 

One of the main goals of this text is to 
draw distinctions between the transient relaxation spectra (derived from the eigenvalues of a certain non-Hermitian $N\times N$ matrix $\check H$) and a Hermitian $N\times N$ stationary distribution $\check F_\mathrm{st}$.  Already for quadratic Lindbladians finding $\check F_\mathrm{st}$ requires solving a linear kinetic equation.  The latter  takes the form of the so-called continuous-time {\em Lyapunov matrix equation}, well-known in the dynamical systems literature in the context of stability and control of linear systems \cite{Godunov}.  We show that, on the one hand,  $\check F_\mathrm{st}$ determines a host of observables and correlation function of physics interest. On the other hand, its properties are often qualitatively  different from those of $\check H$. 
While the latter are frequently emphasized in the literature, the former are undeservedly overlooked.   For example, certain non-analyticities (associated with the {\em exceptional} points) in the $\check H$ spectra, do not 
show up in the $\check F_\mathrm{st}$ spectra. Similarly, while the band structure of $\check H$ often exhibits interesting topological characteristics, the band structure of the corresponding $\check F_\mathrm{st}$ may be topologically trivial.

The structure of the manuscript is rather straightforward. In section \ref{2} we develop major aspects of the field-theoretical treatment of the Lindbladian dynamics for bosonic and fermionic many-body systems. Here we emphasize the role of the stationary distribution and explain the origin of the Lyapunov equation. We also derive generic expressions for observables and linear response and consider exceptional points.   Section \ref{3} is devoted to a number of pedagogic examples illustrating various aspects of many-body Lindbladian dynamics. Some technicalities are delegated to appendices.

\section{Formalism}\label{2}
This section discusses the general formalism of quadratic theories of both bosons and fermions.  It begins with an introduction to the Keldysh formalism as it relates to Lindbladians.  With this established, one can obtain the many-body Lindbladian spectrum and stationary density matrix using the Keldysh Green's function formalism.

\subsection{Lindblad and Keldysh}\label{2.1}

In the operator formalism approach to open quantum systems, one studies the {\em reduced} density matrix, $\rho$, of the system of interest, resulting from tracing out the environment Hilbert space.  The tracing out of environmental degrees of freedom coupled to the system generates additional terms in the evolution equation for $\rho$ alongside the standard von-Neumann part, resulting in an effective non-equilibrium dynamics.  In situations where memory effects may be neglected, time evolution of $\rho$ is described by the Lindblad master equation \cite{OpenQM},
\begin{subequations}\label{LindbladEq}
\begin{equation}
\partial_t\rho=\doublehat\fL\rho,
\end{equation}
\begin{equation}
\doublehat\fL=-i[\hat\fH,\cdot]+\sum_v\Big(\hat\fL_v\cdot\hat\fL_v^\dagger-\frac{1}{2}\{\hat\fL_v^\dagger\hat\fL_v,\cdot\}\Big),
\end{equation}
\end{subequations}
where the latter equation defines the Lindbladian superoperator.  The Hermitian operator $\hat\fH$ is an effective (possibly renormalized by the environment)  Hamiltonian of the system.  The jump operators $\hat\fL_v$ (in general non-Hermitian) specify channels through which the system is coupled to its environment.

The Lindbladian plays a role analogous to the Hamiltonian in closed quantum systems in that determining its eigenvectors and eigenvalues provides complete knowledge of the system's dynamics.  One thus seeks to solve the superoperator eigenvalue problem $\doublehat\fL\rho_\Lambda=\Lambda\rho_\Lambda$.  A given density matrix $\rho$ will be a superposition of the Lindbladian eigenvectors $\rho_\Lambda$.  The corresponding Lindbladian eigenvalues will generically be complex, coming in complex-conjugate pairs, with
the real and imaginary parts corresponding to the rates of decay and coherent (phase) rotation of the $\rho_\Lambda$ component of $\rho$.

Dynamical stability at long times requires all $\Lambda$ to have  non-positive real parts. This requirement is comparable to a Hamiltonian spectrum being bounded from below in dynamics of a closed quantum system.  The $\rho_\Lambda$ with purely imaginary $\Lambda$ play a special role: they do not dissipate.  There is always at least one zero eigenvalue, corresponding to a stationary state $\rho_\mathrm{st}$.  In general there may be multiple stationary states spanning a multidimensional operator subspace.  The structure and dimension of the space of stationary states is determined by the symmetries of the system \cite{ProsenSym,NonAbelianSym,AlbertGeom,NonEqSSKeldysh,AlbertSym,Santos2020}.  For a generic Lindbladian without additional symmetry however, the stationary state is unique.

The focus of this manuscript is on {\em many-body} Lindbladians, in which the Hilbert space of states is either a bosonic or fermionic Fock space.  It is thus convenient to use the creation/annihilation operator basis $\hat a_j$ and $\hat a_j^\dagger$, where $j\leq N$ is a generic internal index encompassing e.g. spin, flavor, orbital number, space or momentum, etc.  In this basis, $\hat\fH=\fH(\hat a_j^\dagger,\hat a_j)$ and $\hat\fL_v=\fL_v(\hat a_j^\dagger,\hat a_j)$ where $\fH$ and $\fL_v$ without hats are polynomial functions.  The focus below is on $\fH$ quadratic and $\fL_v$ linear.  In such theories, the Lindbladian dynamics is akin to non-interacting Hamiltonian dynamics:  it is possible to solve exactly the full many-body problem from studying single-particle quantities.  In particular, a generic many-body eigenvalue takes the form:
\begin{equation}\label{LindbladEw}
\Lambda_{n_1...n_{2N}}=-i\sum_sn_s\epsilon_s,
\end{equation}
where the $2N$ quantum numbers $n_s$ are integer-valued occupation numbers and $\epsilon_s$ are the solutions to a single-particle non-Hermitian eigenvalue problem.  The stationary state density matrix can be obtained from stationary solution to the single-particle quantum kinetic equation.

The formal machinery used to extract this information is the Keldysh path integral and the corresponding theory of non-equilibrium Green's functions \cite{Kamenev}.  Using the formalism of \cite{DiehlKeldyshLindblad}, the dynamics encoded in the Lindblad equation can be mapped to a coherent state path integral.  In broad strokes, this is achieved by expressing the density matrix at time $t$ in terms its value at an initial time $t_0$ via a time evolution superoperator $\rho(t)=\doublehat\fU_t\rho(t_0)$, defined by $\exp(t\doublehat\fL)$.  One may then introduce the Keldysh partition function as the superoperator analogue of the propagator,
\begin{equation}\label{Z}
Z=\tr\Big(\doublehat\fU_t\rho(t_0)\Big).
\end{equation}
which is always identically equal to 1 due to the density matrix normalization.  $Z$ can be brought into the form a path integral by cutting the time interval into infinitesimal slices via the Trotter formula, so that one may write $\exp(\delta_t\doublehat\fL)\simeq1+\delta_t\doublehat\fL$ on each time slice where $\delta_t$ is the duration of a slice.  By inserting factors of a coherent state resolution of identity in-between each slice on both sides of the density matrix, operators are converted into fields.  For a single bosonic mode, one uses the bosonic coherent states defined by $\hat a\ket{\phi}=\phi\ket{\phi}$ where $\phi$ is a complex number (generalization to $N$ bosons is automatic, for details on fermionic Keldysh integrals, see Section 2.6).  Each coherent state component $\ket{\phi^+_1}\bra{\phi^-_1}$ of the density matrix at time $t$, being acted upon by the superoperator $\doublehat\fL$, leads to the following matrix elements:  
\begin{equation}\label{Overlap}
\bra{\phi^+_2}\doublehat\fL(\ket{\phi^+_1}\bra{\phi^-_1})\ket{\phi^-_2}=-ie^{\bar\phi^+_2\phi^+_1+\bar\phi^-_1\phi^-_2}\fK(\bar\phi^+_2,\phi^+_1,\bar\phi^-_1,\phi^-_2),
\end{equation}
where the Keldysh ``Hamiltonian" $\fK$ has the same form as the Lindbladian upon replacing creation operators $\hat a$, that multiply the density matrix on the left/right,  with $\phi^\pm$ respectively.  Provided the functional form of the Hamiltonian $\fH$ and jump operators $\fL_v$ are normal ordered, one may write:
\begin{subequations}\label{KeldHam&Diss}
\begin{equation}\label{KeldHam}
\fK(\bar\phi^+,\phi^+,\bar\phi^-,\phi^-)=\fH^+-\fH^-+i\fD(\bar\phi^+,\phi^+,\bar\phi^-,\phi^-),
\end{equation}
\begin{equation}\label{DissipativeKernel}
\fD(\bar\phi^+,\phi^+,\bar\phi^-,\phi^-)=\sum_v\Big(\bar\fL_v^-\fL_v^+-\frac{1}{2}\bar\fL_v^+\fL_v^+-\frac{1}{2}\bar\fL_v^-\fL_v^-\Big),
\end{equation}
\end{subequations}
where $\fH^\pm=\fH(\bar\phi^\pm,\phi^\pm)$ and $\fL_v^\pm=\fL_v(\bar\phi^\pm,\phi^\pm)$.  Note that the second equality holds for quadratic theories considered here because the normal ordering of the product $\hat\fL_v^\dagger\hat\fL_v$ is equivalent to the product of the normal ordering up to a trivial constant.

Upon re-exponentiation in the limit $\delta_t\to0$, this retrieves a functional integral in terms of two sets of fields $\phi^\pm(t),\bar\phi^\pm(t)$ corresponding to multiplication of the density matrix by the creation/annihilation operators $\hat a,\hat a^\dagger$ on the left or right side in the Lindbladian.  It is conventional to present this functional integral using the Keldysh rotated basis of ``classical" and ``quantum" fields, $\phi^{\cl,\q}=(\phi^+\pm\phi^-)/\sqrt2$.  All together, one has for the partition function \cite{DiehlKeldyshLindblad},
\begin{equation}\label{KeldZBose}
Z=\int\fdif\bar\phi^\alpha\fdif\phi^\alpha e^{iS[\bar\phi^\alpha,\phi^\alpha]},
\end{equation}
with $\alpha=\cl,\q$.  The Keldysh action $S$ is the given by the time integral of a Lagrangian defined by the Keldysh Hamiltonian $\fK$ defined as a function of $\phi^\pm$ by eq.~(\ref{KeldHam}),
\begin{equation}\label{KeldActionBose}
S=\int\dif t\Big(\bar\phi^\q i\partial_t\phi^\cl+\bar\phi^\cl i\partial_t\phi^\q-\fK(\bar\phi^\alpha,\phi^\alpha)\Big).
\end{equation}
The value of density matrix $\rho(t_0)$ at the initial time is contained in the boundary conditions of the path integral.

With the Keldysh path integral in hand, $n$-point correlation functions can be calculated via operator insertion at different times on either side of the density matrix.  By extension, the expectation of a quadratic observable $\hat\fO=\hat\fA^\dagger\check O\hat\fA$, where $\hat\fA=[\hat a\ \hat a^\dagger]$ is the Nambu space spinor and $\check O$ is a matrix, at a finite time can be computed as the expectation inside the function integral,
\begin{equation}
\braket{\hat\fO(t)}=\int\fdif\bar\phi^\alpha\fdif\phi^\alpha e^{iS}\fO\big(\bar\phi^-(t),\phi^+(t)\big),
\end{equation}
where $\fO(\bar\phi,\phi)=\bar\Phi\check O\Phi$ is the classical function of operators replaced with fields, with $\Phi=[\phi\ \bar\phi]$ the Nambu space vector.  The integral can performed by expressing $\fO$ as a function of the Keldysh fields.  As an example which will be relevant below, one may consider the single-particle bosonic covariance matrix $\langle\{\hat\fA,\hat\fA^\dagger\}\rangle$.  This Nambu space matrix-valued expectation reduces to the equal time two-point expectation of classical fields $\langle\{\hat\fA,\hat\fA^\dagger\}\rangle(t)=\langle\Phi^\cl(t)\bar\Phi^\cl(t)\rangle$.

\begin{figure}
\begin{center}
\scalebox{.15}{\includegraphics{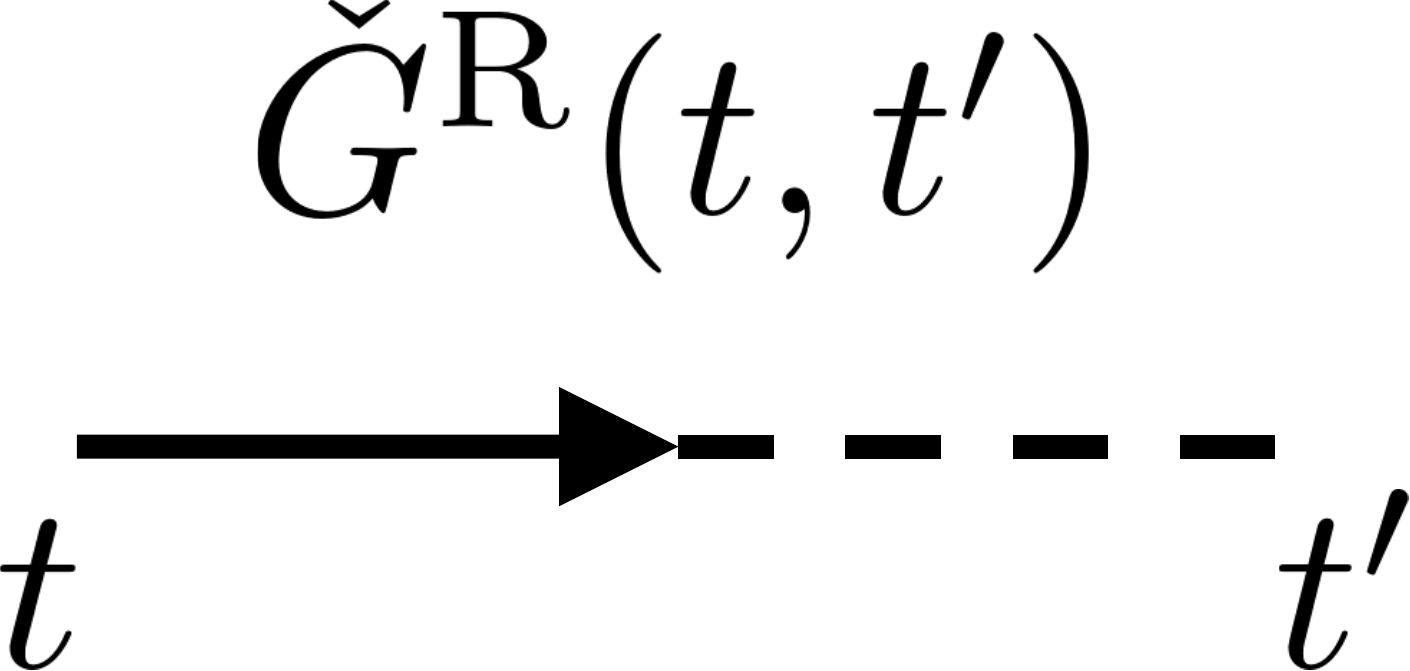}}\qquad\qquad
\scalebox{.15}{\includegraphics{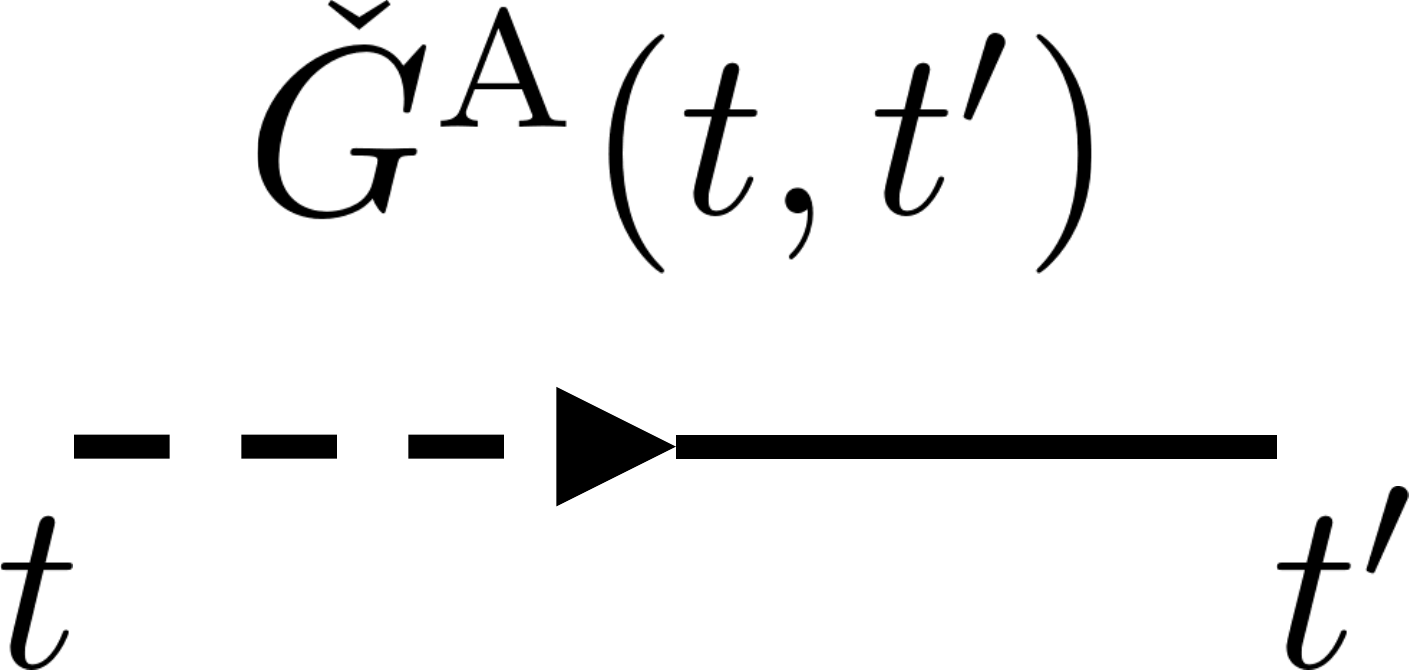}}\qquad\qquad
\scalebox{.15}{\includegraphics{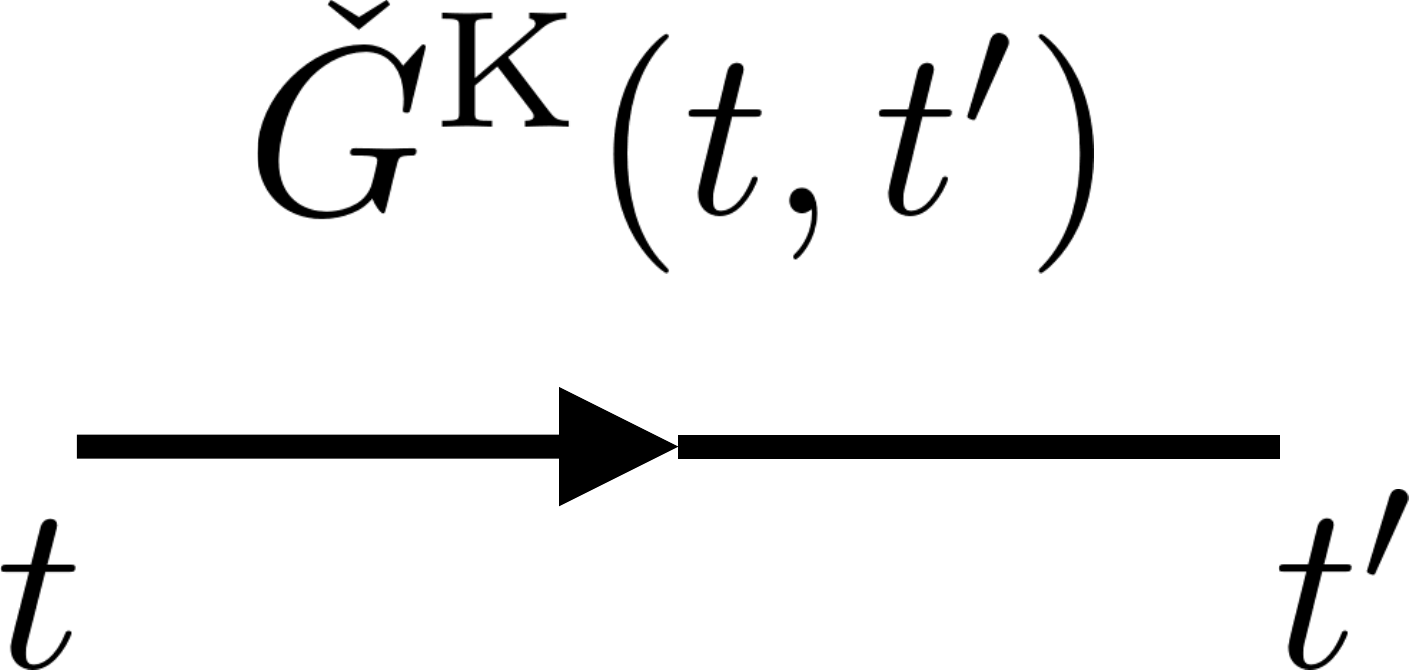}}
\caption{Diagrammatic representations of the three Green's functions from the Keldysh theory.  The solid lines denotes the classical field $\phi^\cl$ and the dashed line, the quantum field $\phi^\q$.}
\label{Lines}
\end{center}
\end{figure}

One of the main advantages of the Keldysh formalism is that this procedure is straight-forwardly generalized to expectations of fields with different time arguments.  For a quadratic theory, the two-point functions are the most important as they can be used to compute all higher-order $n$-point functions via Wick's theorem.  Combining all four fields together into a single Keldysh-Nambu vector $\Phi=[\phi^\cl\ \bar\phi^\cl\ \phi^\q\ \bar\phi^\q]$, the two-point functions define the spectral and Keldysh Green's functions,
\begin{equation}
\begin{bmatrix}i\check G^\tK(t,t')&i\check G^\tR(t,t')\\i\check G^\tA(t,t')&0\end{bmatrix}=\braket{\Phi(t)\bar\Phi(t')}.
\end{equation}
Note that Green's functions defined using this convention are matrices on the Nambu space so as to account for the possibility of non-zero anomalous expectations e.g. $\langle\phi^\alpha(t)\phi^\beta(t')\rangle$.  The Keldysh Green's function $\check G^\tK$ contains information about the distribution function of the system; at equal times one can see that it is just the covariance matrix from the example above, $\check G^\tK(t,t)=\langle\{\hat\fA,\hat\fA^\dagger\}(t)\rangle$.  It is conventional to represent the Green's functions diagrammatically, as shown in fig.~\ref{Lines}.  This relation generalizes to $N$ particles and, as discussed below, can be used to compute the stationary state density matrix.  The spectral Green's functions $\check G^{\tR,\tA}$ contain purely dynamical information and are independent of the state of the system.  As discussed in the following section, they are related to the single-particle eigenvalue spectrum.

\subsection{Bosons}\label{2.2}
A generic quadratic Lindbladian for a system of $N$ Bosons is defined by a quadratic Hamiltonian and a set of linear Jump operators,
\begin{subequations}\label{Ham&JumpBose}
\begin{equation}\label{HamBose}
\hat\fH=\sum_{i,j}^N\big(\Delta_{ij}\hat a^\dagger_i\hat a_{j}+\lambda_{ij}\hat a_i\hat a_{j}+\lambda^*_{ij}\hat a^\dagger_i\hat a^\dagger_{j}\big),
\end{equation}
\begin{equation}
\hat\fL_v=\sum_j^N(\mu_{vj}\hat a_j+\nu_{vj}\hat a^\dagger_j),
\end{equation}
\end{subequations}
where the hat $\hat a_j$'s are bosonic creation operators, $[\hat a_j,\hat a^\dagger_i]=\delta_{ij}$.  One can in principle allow the parameter matrices to vary as functions time, but for simplicity here only time-independent models are considered.  The corresponding Keldysh action of eq.~(\ref{KeldActionBose}) can be arranged into a quadratic form:
\begin{equation}\label{ActionBoson}
S=\frac{1}{2}\int\dif t\ \bar\Phi\begin{bmatrix}0&\check\tau^3i\partial_t-\check H_0-i\check Q\\\check\tau^3i\partial_t-\check H_0+i\check Q&i\check D\end{bmatrix}\Phi,
\end{equation}
where $\check\tau^i$ are the Pauli matrices acting in Nambu space and the fields $\phi^\alpha$ are understood as vectors with $N$ entries $\phi^\alpha_j$ so that the Keldysh-Nambu spinor $\Phi$ has a total of $4N$ entries.

The operators $\check H_0$, $\check Q$, and $\check D$ are Hermitian $2N\times2N$ Nambu space matrices:
\begin{subequations}\label{ParamMatBose}
\begin{equation}\label{BlochMatBoson}
\check H_0=\begin{bmatrix}\Delta&2\lambda^\dagger\\2\lambda&\Delta^\tT\end{bmatrix},
\end{equation}
\begin{equation}
\check Q=\frac{1}{2}\begin{bmatrix}\gamma-\tilde\gamma&\eta_\mathrm{a}^\dagger\\\eta_\mathrm{a}&\tilde\gamma^\tT-\gamma^\tT\end{bmatrix},
\end{equation}
\begin{equation}
\check D=\begin{bmatrix}\gamma+\tilde\gamma&\eta_\mathrm{s}^\dagger\\\eta_\mathrm{s}&\gamma^\tT+\tilde\gamma^\tT\end{bmatrix},
\end{equation}
\end{subequations}
where 
\begin{equation}\label{GammaEta}
\gamma_{ij}=\sum_v\mu^*_{vi}\mu_{vj},\quad \tilde\gamma_{ij}=\sum_v\nu_{vi}\nu^*_{vj},\quad \eta_{ij}=\sum_v\nu^*_{vi}\mu_{vj},
\end{equation}
and $\eta_{\ts,\ta}=\eta\pm\eta^\tT$.  The $N\times N$ parameter matrices have index symmetries:
\begin{equation}
\Delta=\Delta^\dagger,\quad\lambda=\lambda^\tT,\quad \gamma=\gamma^\dagger,\quad\tilde\gamma=\tilde\gamma^\dagger,\quad\eta_{\ts,\ta}=\pm\eta_{\ts,\ta}^\tT.
\end{equation}
Note that the matrix $\check H_0$ is nothing more than the single-particle Hamiltonian, $\hat\fH=\frac{1}{2}\hat\fA^\dagger\check H_0\hat\fA$.  The matrices $\check Q$ and $\check D$ are determined entirely by the coupling to the environment through the jump operators.

\begin{figure}
\begin{center}
\scalebox{.7}{\includegraphics{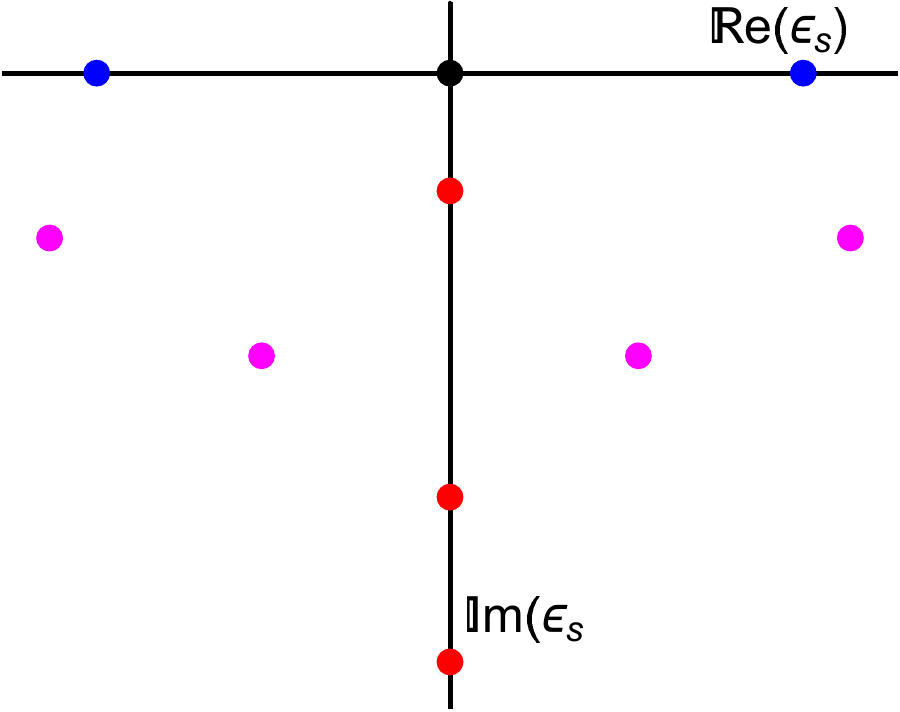}}
\caption{Example spectra of $\check H$ plotted in the complex plane of the eigenvalues $\epsilon_s$.  The eigenvalues labeled in red correspond to eigenvectors that evolve in a purely dissipative way.  The purple eigenvalues correspond to modes with simultaneous dissipation and coherent rotation.  The black eigenvalue at the origin denotes the stationary state(s), which can be compared to energy eigenstates in a closed quantum system.  The blue eigenvalues aligned along the real axis correspond to limiting cycles which do not decay at long times; these are analogous to coherent superpositions of different energy eigenstates.}
\label{SpectrumExample}
\end{center}
\end{figure}

The dynamic matrix $\check H=\check\tau^3(\check H_0-i\check Q)$ is a non-Hermitian matrix that replaces the single-particle Hamiltonian in coherent many-body systems.  Its $2N$ eigenvalues are the obtained by solving the non-Hermitian eigenvalue problem,
\begin{equation}
\check H\ket{s}=\epsilon_s\ket{s},
\end{equation}
where $\epsilon_s$ are the eigenvalues of $\check H$ obtained through diagonalization by a generically non-Unitary matrix $\check U$,
\begin{equation}
\big(\check U\check H\check U^{-1}\big)_{ss'}=\epsilon_s\delta_{ss'}.
\end{equation}
Note that the complex eigenvalues of $\check H$ come in complex-conjugate pairs on due to the Nambu space particle-hole symmetry.  The $\epsilon_s$ act as eigenvalues of the single-particle sector of the Lindbladian.  In the standard quantum theory, the absence of interactions implies the energies of the full many-body system to be determined by filling the single particle states with various numbers of particles.  This is essentially true in the Lindbladian theory as well, except that the single-particle states $\ket{s}$ have a finite lifetime on account of the complex single-particle ``energies" $\epsilon_s$ having migrated into the bottom half of the complex plane, see Fig.~(\ref{SpectrumExample}).  The expression in eq.~(\ref{LindbladEw}) for a many-body eigenvalue is just the assigning of bosonic occupation numbers to this single-particle sector.  Demonstrating this fact explicitly can be achieved either by semiclassical quantization of the Keldysh action (see Appendix \ref{AppA}) or through the third quantization formalism \cite{3rdQuantBose} (see Appendix \ref{AppB} for connections between the Keldysh and third quantization formalisms).

To gain further intuition about the nature of the dynamics, consider the the classical mechanics of the Keldysh action eq.~(\ref{ActionBoson}).  The equation of motion of the classical field $\Phi^\cl=[\phi^\cl\ \bar\phi^\cl]$ with the quantum field set to zero $\Phi^\q=[\phi^\q\ \bar\phi^\q]=0$ is
\begin{equation}\label{BosonClassicalEOM}
i\partial_t\Phi^\cl=\check H\Phi^\cl.
\end{equation}
This is a non-Hermitian Schr\"odinger equation where the classical field acts as the single-particle wave function.  This can equivalently be conceptualized in first-quantized language as an equation of motion for the coordinate on the $N$-particle phase space.  Due to the non-Hermiticity of the dynamic matrix, the classical mechanics this equation encodes is dissipative: the phase portrait will consist of spiraling paths centered at the origin.  The dynamics are only stable when all of the eigenvalues of $\check H$ have non-positive imaginary part, so that all phase space trajectories fall into the origin rather than running to infinity.  This behaviour is not ensured for generic choices of the parameter matrices and can fail if the magnitudes of $\lambda$ or $\tilde\gamma$ are large compared to other parameters.  These situations are unphysical, being associated with either an unstable Hamiltonian in which the potential of one or more coordinates in the phase space is inverted or with a situation where the rate of particle gain is greater than loss, resulting in an uncontrolled pumping of quanta into the system.  At the threshold of such an instability the eigenvalues of $\check H$ can be purely imaginary, resulting in a coherent orbiting around the origin.  This corresponds to a closing of the dissipative gap in the Lindbladian spectrum and stable long-time dynamics beyond a single stationary state.

The spectral Green's functions $\check G^{\tR,\tA}(t,t')$ contain the same dynamical information in their pole structure.  They can be read off as the off-diagonal blocks of the inverse of the quadratic form in eq.~(\ref{ActionBoson}).  This is equivalent to inverting the differential operator in eq.~(\ref{BosonClassicalEOM}).  The spectral Green's functions are independent of the distribution of the system and are thus always functions of the difference of their time arguments,
\begin{equation}\label{GreenBose}
\check G^\tR(t,t')=-i\theta(t-t')e^{-i(t-t')\check H}\check\tau^3,\quad\quad \check G^\tA(t,t')=i\theta(t'-t)\check\tau^3e^{-i(t-t')\check H^\dagger}.
\end{equation}
Upon Fourier transform with respect to the difference of the time arguments $t-t'$, they are adopt a simple form of the resolvent of the dynamic matrix,
\begin{equation}\label{GreenSpectralBose}
\check G^\tR(\epsilon)=\frac{1}{\epsilon-\check H}\check\tau^3,\quad\quad\check G^\tA(\epsilon)=\check\tau^3\frac{1}{\epsilon-\check H^\dagger}.
\end{equation}
The poles of the Green's functions are located at the eigenvalues $\epsilon_s$.  This can be compared to the association between the energies of single-particle states and poles in standard quantum theory.

\subsection{Lyapunov Equation}\label{2.3}
In this section, the stationary state of the Lindbladian is discussed.  Contrasting to the spectral Green's functions discussed above, the Keldysh Green's function $\check G^\tK$ depends on the distribution of the system.  Conventionally, one parameterizes the Keldysh Green's function in terms of the spectral Green's functions and a Hermitian matrix $\check F(t,t')$,
\begin{equation}\label{BoseFDT}
\check G^\tK=\check G^\tR\circ\check\tau^3\check F-\check F\check\tau^3\circ\check G^\tA,
\end{equation}
where the composition $\circ$ denotes matrix composition both in the time argument and the Nambu space.  Note the additional factor of $\check\tau^3$ here compared to the standard convention in \cite{Kamenev} is a consequence of the symplectic structure of the bosonic Nambu space.  The matrix $\check F$ acts as a single-particle distribution matrix.  Acting on this equation on the left by $\check\tau^3\check G^{\tR-1}$ and on the right by $\check G^{\tA-1}\check\tau^3$ retrieves the quantum kinetic equation for $\check F$,
\begin{equation}\label{KinEqBose}
[\partial_t\overset{\circ}{,}\check F]=-i\big(\check H\check F-\check F\check H^\dagger\big)+\check\tau^3\check D\check\tau^3,
\end{equation}
where the $\partial_t$ and $\check D$ are understood to be diagonal functions of their two time arguments, i.e. coming with factors of $\delta(t-t')$.  Note the use of the relation $i\check D=-\check G^{\tR-1}\circ\check G^\tK\circ\check G^{\tA-1}$ obtained by inverting the quadratic form in the action eq.~(\ref{ActionBoson}) to derive this equation.  This is valid because the path integral in eq.~(\ref{KeldZBose}) is Gaussian on the bulk of the time contour, except possibly at the initial time $t_0$ in the case of a non-Gaussian initial density.  The initial density lives on the boundary of the time contour and determines the boundary condition of the quantum kinetic equation.

In a stationary state, $\check G^\tK$ and by extension $\check F$ are independent of the the central time $t+t'$.  This nullifies the left-hand side of eq.~(\ref{KinEqBose}), meaning the right-hand side must be independently nullified by a stationary solution.  This is achieved by a time-diagonal ansatz, $\check F(t-t')=\check F_\mathrm{st}\delta(t-t')$, where $\check F_\mathrm{st}$ is a time-independent matrix in the orbital and Nambu spaces obeying the relation:
\begin{equation}\label{KinEqBoseStationary}
0=-i\big(\check H\check F_\mathrm{st}-\check F_\mathrm{st}\check H^\dagger\big)+\check\tau^3\check D\check\tau^3.
\end{equation}
This is a complex Lyapunov equation.  There is a unique solution provided that all of the eigenvalues of $\check H$ have finite imaginary parts \cite{Godunov}.  This implies that the Lindblad equation possesses a unique stationary state that arbitrary initial conditions converge towards at long times.  The existence of additional stationary states for bosonic quadratic Lindbladians thus occurs only at the brink of dynamical instability, when a parameter is tuned so that the dissipative gap closes.

Provided the stationary state is unique, one can solve eq.~(\ref{KinEqBoseStationary}) in the eigenbasis of $\check H$,
\begin{equation}\label{Fsscomponent}
\big(\check U\check F_\mathrm{st}\check U^\dagger\big)_{ss'}=\frac{-i}{\epsilon_s-\epsilon^*_{s'}}\big(\check U\check \tau^3\check D\check \tau^3\check U^\dagger\big)_{ss'}.
\end{equation}
Note that $\check D$ is generically not diagonal in this basis, meaning that off-diagonal elements of $\check F_\mathrm{st}$ are finite.  This can be compared to the equilibrium theory, in which the preferred stationary $\check F$ is the thermal distribution, which is diagonal in Nambu space in the eigenbasis of the single-particle Hamiltonian.  The relation in eq.~(\ref{BoseFDT}) is equivalent to the Fluctuation-Dissipation theorem for each particle species.  In the Lindbladian setting, $\check F_\mathrm{st}$ is $\epsilon$-independent and generically develops off-diagonal elements in the eigenbasis of $\check H$ and so is more naturally thought of as a matrix.  An alternative but equivalent interpretation of $\check F$ is given by integrating eq.~(\ref{BoseFDT}), giving the relation $i\check G^\tK(t,t)=\check F_\mathrm{st}$.  That is, $\check F_\mathrm{st}$ is equivalent to the covariance matrix $\braket{\{\hat\fA,\hat\fA^\dagger\}}$ discussed in section 2.1.

As an alternative to eq.~(\ref{Fsscomponent}), one can instead express $\check F_\mathrm{st}$ in its eigenbasis. Letting $\check U_F$ be a diagonalizing transformation of $\check\tau^3\check F_\mathrm{st}$, one has \cite{CovEws}:
\begin{equation}\label{FewsBose}
\check U_F\check\tau^3\check F_\mathrm{st}\check U_F^\dagger=\mathrm{diag}\big(\coth(\beta_1/2),...,-\coth(\beta_1/2),...\big),
\end{equation}
where the $N$ numbers $\beta_j$ parametrize the eigenvalues of $\check F_\mathrm{st}$ and act as effective inverse temperatures for the $j$th eigenvector.  For a dynamically stable theory, $0<\beta_j\leq\infty$.  As mentioned above, this is in general {\em not} the same basis in which the dynamic matrix $\check H$ is diagonal, $\check U\neq\check U_F$.  This is in stark contrast to the equilibrium theory of quadratic Hamiltonians, in which the thermal state is the Gaussian state with $\hat\fH_\mathrm{st}=\hat\fH$.  In equilibrium, the bases in which the dynamics and the distribution are diagonal are the same.  Out of equilibrium, as is the case for the Lindbladian theory, this is generically untrue.

The form of the stationary density matrix $\rho_\mathrm{st}$ can be obtained using the identity of $\check F_\mathrm{st}$ as the covariance matrix.  The covariance matrix is a central object in the theory of Gaussian states and is known to be equivalent to full knowledge of such a state \cite{GaussianBose1,GaussianBose2,GaussianBose3}.  A Gaussian state is a state with a density matrix given by the exponentiation of some quadratic operator of the form of eq.~(\ref{HamBose}).  For a quadratic Lindbladian with a unique stationary state, the state will be Gaussian.  As such, one can write the stationary density matrix $\rho_\mathrm{st}$ in terms of an effective {\em Hermitian} Hamiltonian,
\begin{subequations}
\begin{equation}\label{rhoss}
\rho_\mathrm{st}\propto\exp(-\hat\fH_\mathrm{st}),
\end{equation}
\begin{equation}
\hat\fH_\mathrm{st}=\frac{1}{2}\hat\fA^\dagger\check H_\mathrm{st}\hat\fA,
\end{equation}
\end{subequations}
where the proportionality is determined by the normalization $\tr(\rho)=1$ and $\check H_\mathrm{st}$ is generically not the same as $\check H_0$.  The effective Hamiltonian can be found from the stationary distribution matrix through the relation \cite{GaussianBose4-Heff}:
\begin{equation}
\check F_\mathrm{st}\check\tau^3=\coth(\check\tau^3\check H_\mathrm{st}/2).
\end{equation}
In the eigenbasis of $\check F_\mathrm{st}$, it adopts a particularly simple form,
\begin{equation}
\hat\fH_\mathrm{st}=\sum_j^N\beta_j\hat b^\dagger_j\hat b_j,
\end{equation}
where the diagonal basis bosons are defined by $[\hat b\ \hat b^\dagger]=\check U_F[\hat a\ \hat a^\dagger]$.  The $\beta_j$ determine the average populations of the $\hat b$ bosons in the stationary state.  Note that in situations where there is not a unique stationary state, some stationary states may be non-Gaussian and the value of the Keldysh Green's function at long times depends on the initial conditions.

\subsection{Observables and Response}\label{2.4}
With the stationary distribution in hand, one can compute the stationary expectation of observables.  As an example, consider a quadratic observable $\hat\fO=\hat\fA^\dagger\check O\hat\fA$.  The expectation at arbitrary finite time after reaching the stationary state is:
\begin{equation}\label{QuadraticObsBose}
\braket{\hat\fO}=\frac{1}{2}\tr\Big(\check O\big(\check F_\mathrm{st}-\check\tau^3\big)\Big).
\end{equation}
The quantity $\frac{1}{2}(\check F_\mathrm{st}-\check\tau^3)$ thus acts like am effective {\em single-particle} density matrix.  Alternatively, naming the classical and quantum parts of the observable $\fO^{\cl,\q}=\frac{1}{2}(\bar\Phi^+\check O\Phi^+\pm\bar\Phi^-\check O\Phi^-)$, one has
\begin{equation}\label{QuadraticObsBoseC}
\braket{\fO^\cl}=\frac{1}{2}\tr\big(\check O\check F_\mathrm{st}\big)
\end{equation}
Thus, single-particle traces with the distribution matrix by itself generates moments of the classical (Weyl-ordered) parts of observables.  Correlations of different observables at different times and observables containing products of more than two field operators can be obtained using Wick's theorem.

The response of the system in its stationary state can be studied by introducing perturbations.  It is assumed that the system has been prepared then allowed to relax to its stationary state.  Formally, this amounts to pushing the initial time into the infinite past $t_0\to\infty$ so that the system retains no memory of its initial condition.  Then, at a later finite time $t_i$, some potentially time-dependent perturbations are switched on, $\doublehat\fL\to\doublehat\fL+\delta\doublehat\fL(t)\theta(t-t_i)$.  One may of course consider perturbing the system directly by modifying the Hamiltonian $\hat\fH\to\hat\fH+\delta\hat\fH(t)$.  For a quadratic perturbation, this is equivalent to changing the single-particle Hamiltonian by the inclusion of a matrix-valued classical source $\check H_0\to\check H_0+\delta\check H_0(t)$.  In a Lindbladian problem, one may additionally consider variations to the dissipative part of the evolution, either by introducing a new jump operator or by varying an existing jump operator.  In both cases, this leads to a modification of the other two parameter matrices $\check Q\to\check Q+\delta\check Q(t)$ and $\check D\to\check D+\delta\check D(t)$.  In the prior case, $\delta \check Q$ and $\delta \check D$ are of the same form as in eq.~(\ref{ParamMatBose}).
In the latter, one may take perturbations to the jump operators by modifying $\mu\to\mu+\delta\mu(t)$ and $\nu\to\nu+\delta\nu(t)$ in eq.~(\ref{GammaEta}) and keeping only whatever order in $\delta\mu$ and $\delta\nu$ is required.

In the path integral formalism, this is equivalent to introducing a perturbation to the action $S\to S+\delta S$.  This translates to a perturbation of the Keldysh Hamiltonian in eq.~(\ref{KeldHam}), $\fK\to\fK+\delta\fK(t)$.  The perturbations from varying $\check H_0$, $\check Q$, and $\check D$ respectively are given by:
\begin{subequations}\label{deltaK}
\begin{equation}
\delta_{H_0}\fK(t)=\frac{1}{2}\check\sigma^1_{\alpha\beta}\bar\Phi^\alpha\delta\check H_0(t)\Phi^\beta
\end{equation}
\begin{equation}
\delta_Q\fK(t)=-\frac{1}{2}\check\sigma^2_{\alpha\beta}\bar\Phi^\alpha\delta\check Q(t)\Phi^\beta
\end{equation}
\begin{equation}
\delta_D\fK(t)=-\frac{i}{2}\bar\Phi^\q\delta\check D(t)\Phi^\q
\end{equation}
\end{subequations}
where $\check\sigma^i$ denotes the Pauli matrices in the space of Keldysh indices.  For weak perturbations, it suffices to keep only the first order correction to the measure.  This gives the linear response, for which one obtains for the expectation of an observable $\hat\fO$ at any finite time,
\begin{equation}\label{Kubo}
\braket{\hat\fO}(t)\simeq\braket{\hat\fO}_\mathrm{st}-i\int_{t_i}^t\dif t'\braket{\fO^\cl(t)\delta\fK(t')}_\mathrm{st}.
\end{equation}
This is nothing but the Kubo formula generalized to the Lindbladian context.  The first term in this expression is given by eq.~(\ref{QuadraticObsBose}).  The latter term includes only the classical part $\fO^\cl$ because only $\check F$ in eq.~(\ref{QuadraticObsBose}) receives perturbative corrections.  It can be computed using Wick's theorem, which leads to bubble diagram contributions depicted in fig.~\ref{Bubbles}.  Note that one must be careful to only keep terms corresponding to fully connected diagrams, see Appendix \ref{AppC}.

\begin{figure}
\begin{center}
\scalebox{.15}{\includegraphics{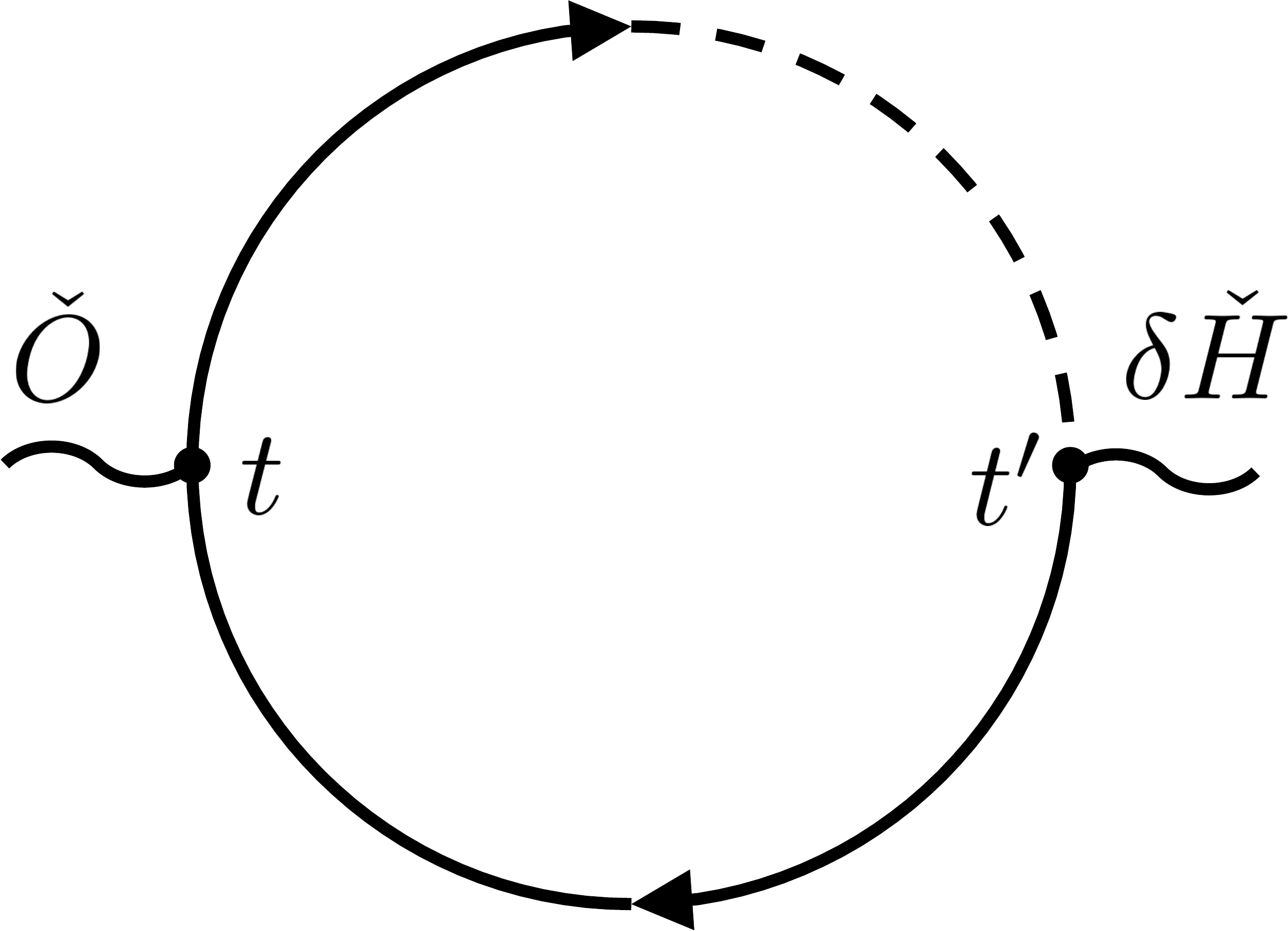}}\qquad
\scalebox{.15}{\includegraphics{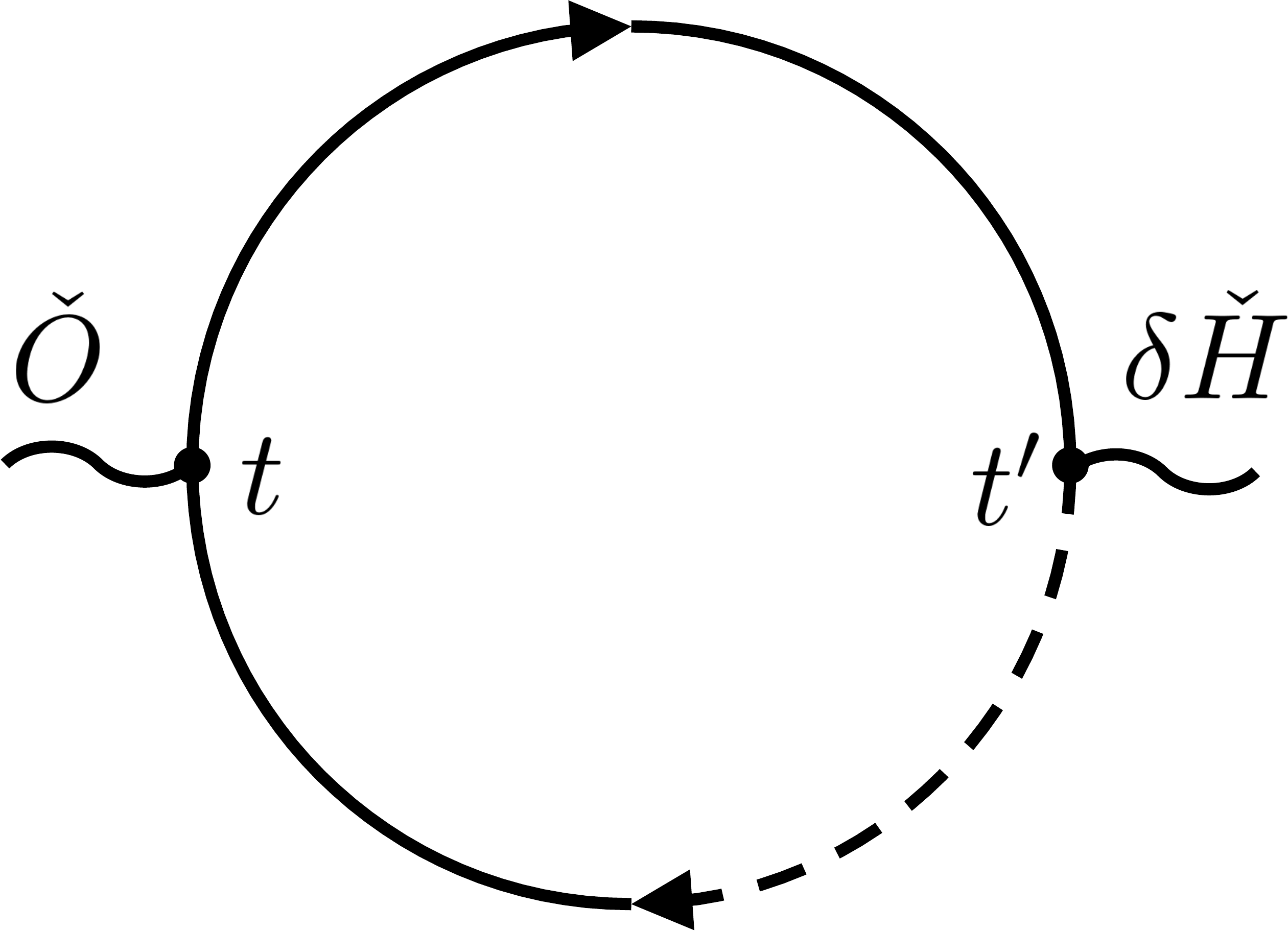}}\qquad
\scalebox{.15}{\includegraphics{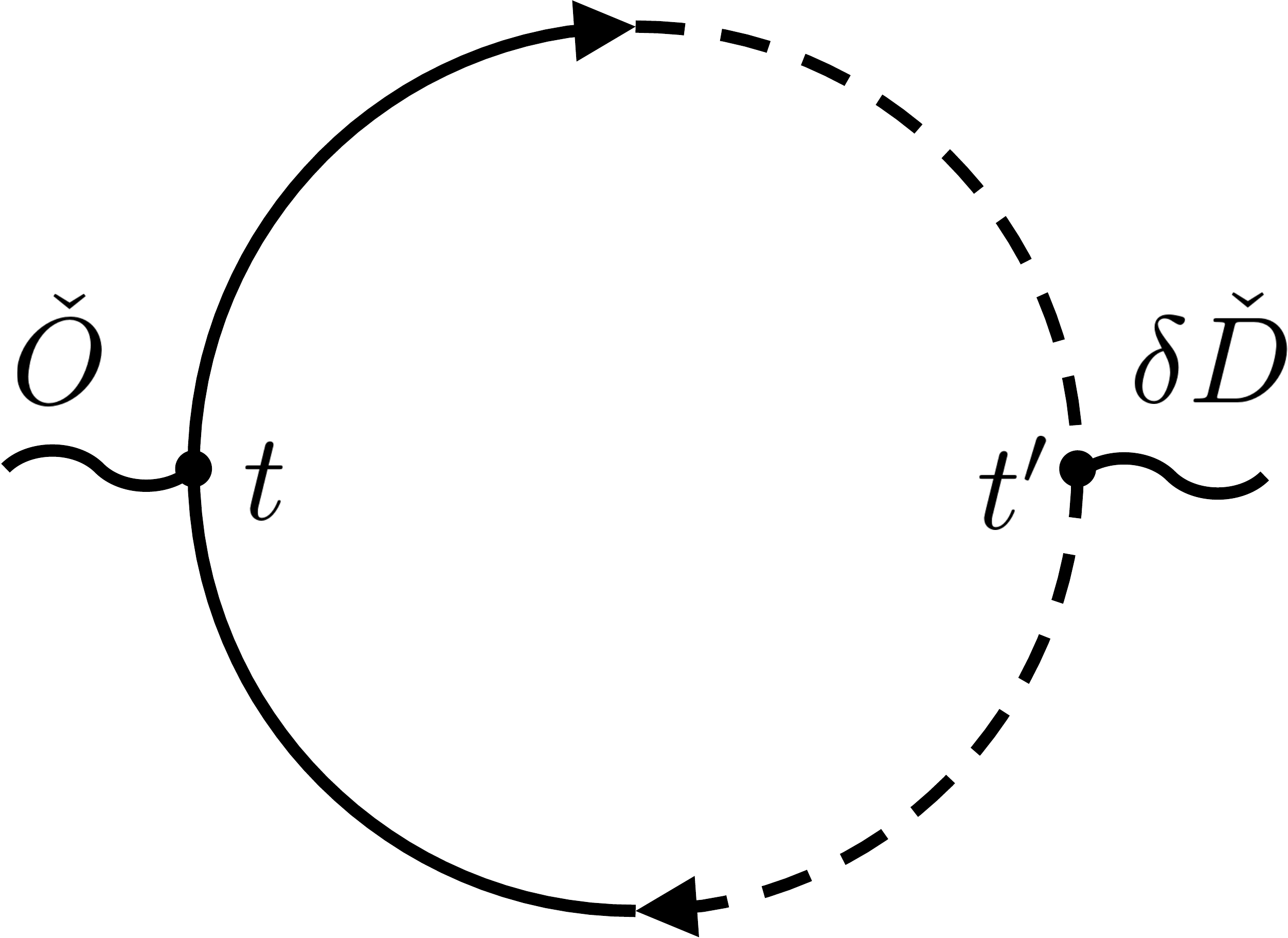}}
\caption{Bubble diagrams contributing to the linear response of a Lindbladian perturbation.  A Hamiltonian perturbation as in eq.~(\ref{LinRespH}) corresponds to the difference of the retarded and advanced polarization bubbles, as depicted by the leftmost two diagrams.  Dissipative perturbations modifying $\check Q$ as in eq.~(\ref{LinRespQ}) are comparably given by the of these two bubbles.  The rightmost diagram appears in perturbations modifying $\check D$ as in eq.~(\ref{LinRespD}).  This diagram in contrast plays no role in the coherent linear response theory.}
\label{Bubbles}
\end{center}
\end{figure}

For a purely Hamiltonian perturbation, one finds a correction in the standard form of the retarded response function,
\begin{align}\begin{split}\label{LinRespH}
\braket{\fO^\cl(t)\delta_{H_0}\fK(t')}_\mathrm{st}&=\braket{\fO^\cl(t)\delta\fH^\q(t')}_\mathrm{st}
\\&=-\frac{1}{2}\tr\Big(\check G^\tR(t,t')\big(\check\tau^3\check F_\mathrm{st}\delta\check H_0(t')-\delta\check H_0(t')\check F_\mathrm{st}\check\tau^3\big)\check G^\tA(t',t)\check O\Big).
\end{split}\end{align}
Perturbations to the dissipative couplings cannot be expressed as expectations of products of the quantum and classical parts of observables.  They do however admit simple expressions in terms of the distribution matrix,
\begin{equation}\label{LinRespQ}
\braket{\fO^\cl(t)\delta\fK_Q(t')}_\mathrm{st}=-\frac{i}{2}\tr\Big(\check G^\tR(t,t')\big(\check\tau^3\check F_\mathrm{st}\delta\check Q(t')+\delta\check Q(t')\check F_\mathrm{st}\check\tau^3\big)\check G^\tA(t',t)\check O\Big),
\end{equation}
\begin{equation}\label{LinRespD}
\braket{\fO^\cl(t)\delta\fK_D(t')}_\mathrm{st}=\frac{i}{2}\tr\Big(\check G^\tR(t,t')\delta\check D(t')\check G^\tA(t',t)\check O\Big).
\end{equation}
Note that both expressions appropriately have a retarded causality, despite not being expectations of classical and quantum observables like eq.~(\ref{LinRespH}).  Also note that the linear response theory for Lindbladians was studied in \cite{AlbertResponse} using the superoperator formalism.  The above formulas are comparable to the results presented there specialized to many-body bosons.

To go beyond the linear response, rather than keeping higher orders in the perturbation theory one can instead solve the quantum kinetic equation eq.~(\ref{KinEqBose}) to determine the full the non-stationary $\check F$.  The assumption that the system reached its stationary state before the perturbation is encoded in the boundary condition $\check F(t,t')=\check F_\mathrm{st}$ for all $t,t'<t_i$.  Because the perturbation to the Lindbladian is assumed to be local in time, one can always seek a solution that is time-diagonal, with $\check F(t,t')=\delta(t-t')\check F(t)$.  With this ansatz, the kinetic equation adopts the local form,
\begin{equation}\label{KinEqBoseRed}
\partial_t\check F(t)=-i\big(\check H(t)\check F(t)-\check F(t)\check H^\dagger(t)\big)+\check\tau^3\check D(t)\check\tau^3.
\end{equation}
Assuming one can solve this equation, the expectations of quadratic observables at finite times can be computed using the appropriate generalizations of eq.s~(\ref{QuadraticObsBose}) and (\ref{QuadraticObsBose}),
\begin{subequations}\label{QuadraticObsBoseTime}
\begin{equation}
\braket{\hat\fO}(t)=\frac{1}{2}\tr\Big(\check O\big(\check F(t)-\check\tau^3\big)\Big),
\end{equation}
\begin{equation}
\braket{\fO^\cl(t)}=\frac{1}{2}\tr\big(\check O\check F(t)\big).
\end{equation}
\end{subequations}
The classical parts of observables containing products more than two field operators can again be obtained using Wick's theorem.  This is valid even with the non-stationary distribution because the path integral is still a Gaussian functional integral with the time dependent perturbations; the density matrix remains a Gaussian state as it evolves in time.

As a check, one can compare the two approaches by combining and rearranging eq.s~(\ref{LinRespH}), (\ref{LinRespQ}), and (\ref{LinRespD}).  The correction to $\braket{\hat\fO}$ in eq.~(\ref{Kubo}) is given by the trace of $\check O$ multiplied by the object:
\begin{equation}
\int_{t_i}^t\dif t'\check G^\tR(t,t')\check\tau^3\Big(-i\big(\delta\check H(t')\check F_\mathrm{st}-\check F_\mathrm{st}\delta\check H^\dagger(t')\big)+\check\tau^3\delta\check D(t')\check\tau^3\Big)\check\tau^3\check G^\tA(t',t),
\end{equation}
which is just the leading-order perturbative solution to eq.~(\ref{KinEqBoseRed}).

\subsection{Exceptional Points}\label{2.5}
This section addresses subtleties that emerge due non-Hermiticity that have thus far been ignored.  The dynamic matrix $\check H$, and by extension the Lindbladian itself, may be non-diagonalizable.  This occurs at so-called exceptional points of the parameter space, at which two or more eigenvalues merge.  This occurs in a fundamentally different way than in standard Hermitian quantum mechanics, in which the crossing of energy levels is generally avoided and degeneracies are traditionally understood to be a consequence of some underlying symmetry.  The coalescing of eigenvalues at an exceptional point should instead marks a bifurcation in the dynamics and is unrelated to dynamical symmetry.

\begin{figure}
\begin{center}
\scalebox{.45}{\includegraphics{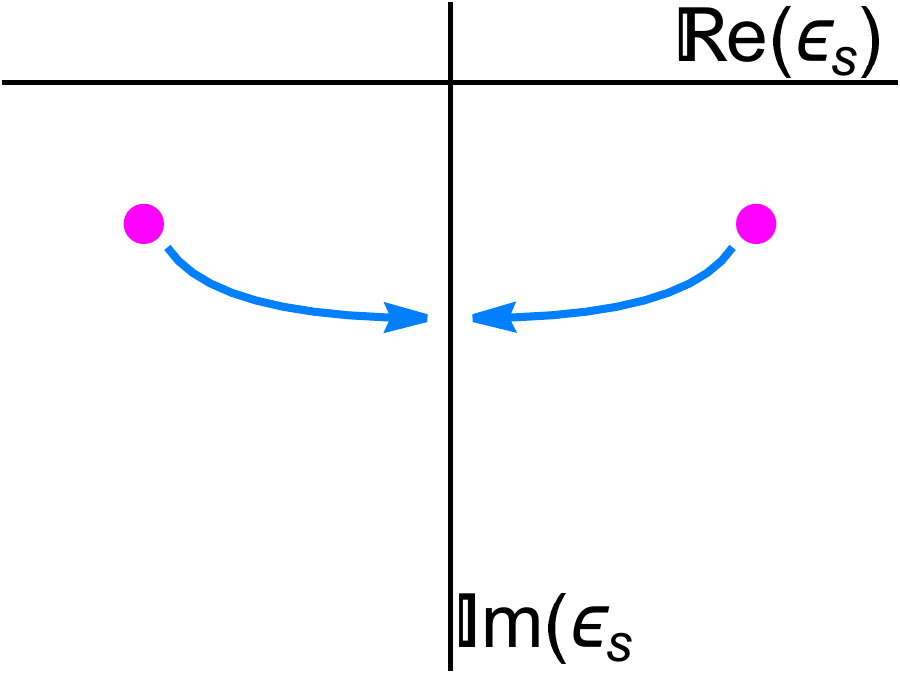}}
\scalebox{.45}{\includegraphics{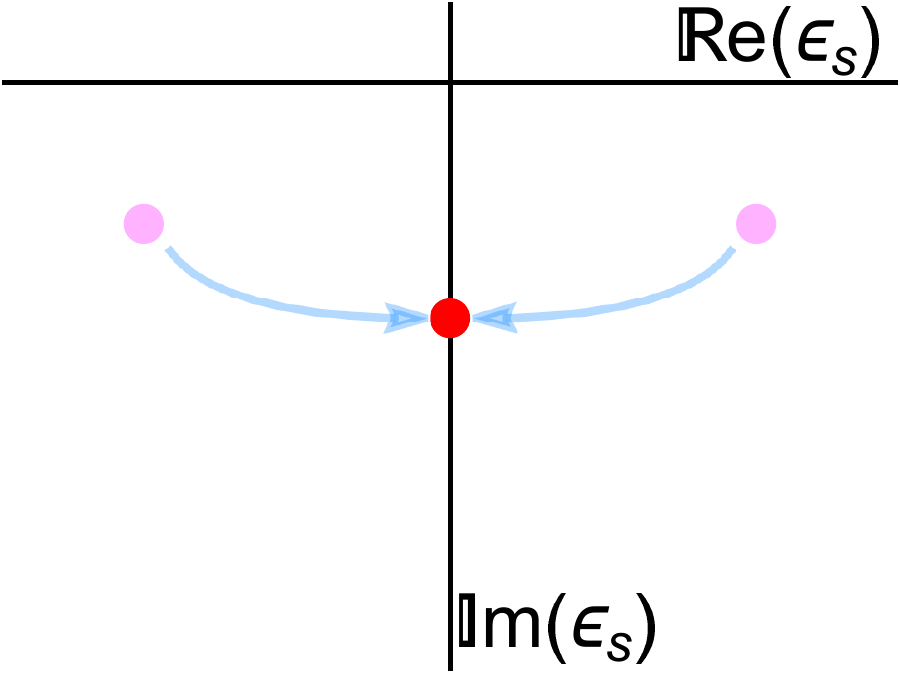}}
\scalebox{.45}{\includegraphics{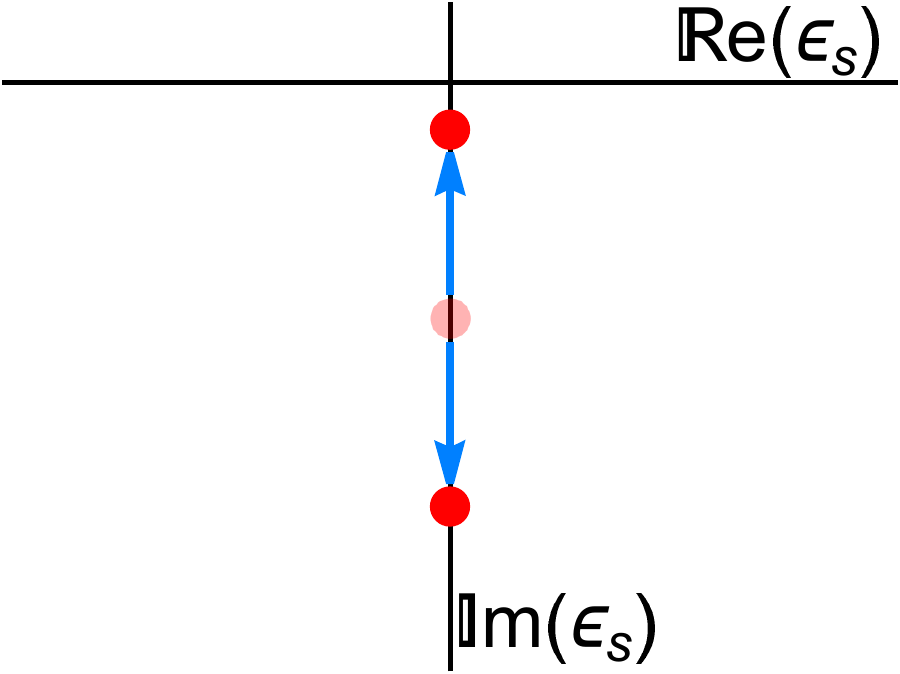}}
\caption{Example of tuning a parameter through an exceptional point separating an under-damped to over-damped dynamical bifurcation.  As a parameter is tuned, a pair of eigenvalues initially located as a conjugate pair in the left half of the complex plane begin moving toward the real axis toward one another.  After colliding, they will remain stuck on the real axis but split and move in opposite directions.}
\label{EP}
\end{center}
\end{figure}

The prototypical example for how this occurs is the collision of two eigenvalues on the real axis, see Fig.~(\ref{EP}).  The eigenvalues of the matrix $-i\check H$, and by extension eigenvalues of the Lindbladian, come in complex-conjugate pairs.  An exceptional point on the real axis thus corresponds to the spontaneous breaking of this `particle-hole symmetry.'  This signals an under-damped to over-damped bifurcation, in which the corresponding eigenmodes undergoing damped coherent rotation before the collision experience pure dissipation after.  There is a resonant damping at the exceptional point, resulting in the transient algebraic gain of one eigenmode.  This transient gain is the generic signature of exceptional points.

To see how this works, suppose the dynamic matrix $\check H$ has an exceptional point where $M$ eigenvalues have collided at the value $\epsilon_s$.  The dynamic matrix is non-diagonalizable but it can be brought to Jordan canonical form by a non-unitary similarity transformation $\check U$,
\begin{equation}\label{diagH}
\check U\check H\check U^{-1}=\begin{bmatrix}
\ddots&\phantom{0}&\phantom{0}&\phantom{0}&\phantom{0}&\phantom{0}\\
\phantom{0}&\epsilon_s&1&\phantom{0}&\phantom{0}&\phantom{0}\\
\phantom{0}&\phantom{0}&\epsilon_s&\ddots&\phantom{0}&\phantom{0}\\
\phantom{0}&\phantom{0}&\phantom{0}&\ddots&1&\phantom{0}\\
\phantom{0}&\phantom{0}&\phantom{0}&\phantom{0}&\epsilon_s&\phantom{0}\\
\phantom{0}&\phantom{0}&\phantom{0}&\phantom{0}&\phantom{0}&\ddots
\end{bmatrix}.
\end{equation}
In this basis, $\check H$ is almost diagonal except on the $M\times M$ block for the eigenvalue $\epsilon_s$, for which there are factors of $1$ above the upper diagonal.  As a consequence, there are fewer than $2N$ total eigenvectors of $\check H$.  As a technical replacement for the missing eigenvectors, it is convenient to introduce additional basis vectors of the Jordan block.  Letting $\ket{s,1}$ denote an eigenvector of $\check H$ for the eigenvalue $\epsilon_s$, one may introduce $\ket{s,n}$ with $n\leq M$ defined through the relation,
\begin{equation}
\check H\ket{s,n}=\epsilon_s\ket{s,n}+\ket{s,n-1}.
\end{equation}
The vectors $\ket{s,n}$ comprise a complete basis spanning the single-particle Hilbert space.

In the presence of such a Jordan block the appearance the spectral Green's functions develop higher-order poles.  In particular, for $s\leq s_1<s_2\leq s+M$ in the same Jordan block, there appears the factor:
\begin{equation}
\Big(\check U\frac{1}{\epsilon-\check H}\check U^{-1}\Big)_{s_1s_2}=\Big(\frac{1}{\epsilon-\epsilon_s}\Big)^{1+s_2-s_1}.
\end{equation}
Written as functions of the time $t$, these off-diagonal components possess polynomial coefficients in front of the exponent $|t|^{s_2-s_1}\exp(-i\epsilon_st)$, resulting in transient algebraic gain of certain initial correlations.
This behavior does not survive away from the exceptional point: generic perturbations restore the diagonalizability of $\check H$.  There is an extreme sensitivity to perturbations at an exceptional point.  This manifests in the analytic structure of the eigenvalues, which develop fractional power law non-analyticities \cite{Kato}.  This is anomalous compared to conventional, fully analytic Hermitian perturbation theory.  Consequences of these non-analyticities are explored in some of the examples in section \ref{3}.

It is natural to wonder if there is some analytic signature of an exceptional point present in the stationary density.  Examining the Lyapunov equation eq.~(\ref{KinEqBoseStationary}), one can see the answer to be negative.  Both the matrices $\check H$ and $\check D$ are analytic functions in neighborhoods of exceptional points in parameter space \cite{Kato}.  By expanding all terms in series and matching powers, one can see that only integer powers are permitted for $\check F_\mathrm{st}$.  As a consequence, $\check F_\mathrm{st}$ and by extension $\check H_\mathrm{st}$ are analytic functions on the parameter space even at exceptional points.  Thus, there will generically be no residual signature of the anomalous nature of the dynamics left over at long times.

\subsection{Fermions}\label{2.6}
In this section, the above formalism is adapted to study fermionic Lindbladians.  To begin, one needs a fermionic version of the Keldysh path integral.  This is obtained in essentially the same way as its bosonic counterpart, though some additional care must be taken with respect to the ordering of the anti-commuting fields.  The fundamental building block for the path integral is the fermionic coherent state defined by $\hat c\ket{\psi}=\psi\ket{\psi}$, where $\psi$ is a complex Grassmann number.  The relevant overlap formula for the Lindbladian action on two sets of Grassmann coherent states is the mirror of eq.~(\ref{Overlap}),
\begin{equation}\label{OverlapFerm}
\bra{\psi^+_2}\doublehat\fL(\ket{\psi^+_1}\bra{\psi^-_1})\ket{\psi^-_2}=-ie^{\bar\psi^+_2\psi^+_1+\bar\psi^-_1\psi^-_2}\fK(\bar\psi^+_2,\psi^+_1,\bar\psi^-_1,\psi^-_2),
\end{equation}
where the Keldysh Hamiltonian is given by eq.~(\ref{KeldHam&Diss}) with Grassmann fields $\psi^\pm$ in place of the bosonic fields.  Note that in eq.~(\ref{DissipativeKernel}) the ordering of fields in the first term is non-trivial, chosen so that backwards fields $(\bar\psi^-,\psi^-)$ always appear before the forwards fields $(\bar\psi^+,\psi^+)$ in the dissipative term $\fD$.  With this, one can massage the partition function eq.~(\ref{Z}) into the form of a fermionic functional integral.  It is standard to use the Larkin-Ovchinnikov convention, in which the Keldysh-rotated fields defined $\psi^{1,2}=(\psi^+\pm\psi^-)/\sqrt2$ and $\bar\psi^{1,2}=(\bar\psi^+\mp\bar\psi^-)/\sqrt2$.  The resulting partition function is:
\begin{subequations}
\begin{equation}
Z=\int\fdif\bar\psi^a\fdif\psi^ae^{iS[\bar\psi^a,\psi^a]},
\end{equation}
\begin{equation}
S=\int\dif t\Big(\bar\psi^1i\partial_t\psi^1+\bar\psi^2i\partial_t\psi^2-\fK(\bar\psi^a,\psi^a)\Big).
\end{equation}
\end{subequations}
Grouping all four fields together into the Keldysh-Nambu vector $\Psi=[\Psi^1\ \Psi^2]$, where  $\Psi^1=[\psi^1\ \bar\psi^2]$ and $\Psi^2=[\psi^2\ \bar\psi^1]$, the matrix of two-point functions defines the fermion Green's functions:
\begin{equation}
\begin{bmatrix}
i\check G^\tR(t,t')& i\check G^\tK(t,t')\\0&i\check G^\tA(t,t')\end{bmatrix}=\braket{\Psi(t)\bar\Psi(t')}.
\end{equation}
These play the same role as in the bosonic theory.

A quadratic Lindbladian system of $N$ fermions is defined by the Hamiltonian and jump operators:
\begin{subequations}
\begin{equation}\label{HamFerm}
\hat\fH=\sum_{i,j}^N\big(\varepsilon_{ij}\hat c^\dagger_i\hat c_j+\Delta_{ij}\hat c_i\hat c_j+\Delta_{ij}^*\hat c^\dagger_j\hat c^\dagger_i\big),
\end{equation}
\begin{equation}
\hat\fL_v=\sum_j^N(\mu_{vj}\hat c_j+\nu_{vj}\hat c^\dagger_j),
\end{equation}
\end{subequations}
where the $\hat c_j$'s are fermion creation operators, $\{\hat c_i,\hat c_j^\dagger\}=\delta_{ij}$.  The corresponding Keldysh action is:
\begin{equation}\label{ActionFerm}
S=\frac{1}{2}\int\dif t\bar\Psi\begin{bmatrix}i\partial_t-\check H_0+i\check Q&i\check D\\0&i\partial_t-\check H_0-i\check Q\end{bmatrix}\Psi,
\end{equation}
The operators $\check H_0$, $\check Q$, and $\check D$ are Hermitian $2N\times 2N$ matrices in Nambu space:
\begin{subequations}\label{ParamMatFerm}
\begin{equation}
\check H_0=\begin{bmatrix}\varepsilon&2\Delta^\dagger\\2\Delta&-\varepsilon^\tT\end{bmatrix},
\end{equation}
\begin{equation}
\check Q=\frac{1}{2}\begin{bmatrix}\gamma+\tilde\gamma&\eta_\mathrm{s}^\dagger\\\eta_\mathrm{s}&\gamma^\tT+\tilde\gamma^\tT\end{bmatrix},
\end{equation}
\begin{equation}
\check D=\begin{bmatrix}\gamma-\tilde\gamma&\eta_\mathrm{a}^\dagger\\\eta_\mathrm{a}&\tilde\gamma^\tT-\gamma^\tT\end{bmatrix},
\end{equation}
\end{subequations}
where $\gamma$, $\tilde\gamma$, and $\eta$ are defined the same as in the bosonic theory as per eq.~(\ref{GammaEta}).  The $N\times N$ parameter matrices have the index symmetries:
\begin{equation}
\varepsilon=\varepsilon^\dagger,\quad\Delta=-\Delta^\tT,\quad\gamma=\gamma^\dagger,\quad\tilde\gamma=\tilde\gamma^\dagger,\quad\eta_\mathrm{s,a}=\pm\eta_\mathrm{s,a}^\tT.
\end{equation}
The matrix $\check H_0$ is the single-particle Hamiltonian, $\hat\fH=\frac{1}{2}\hat\fC^\dagger\check H_0\hat\fC$, where $\hat\fC=[\hat c\ \hat c^\dagger]$ is the Nambu space vector of fermion creation operators.

Note that the definitions of the matrix $\check Q$ and $\check D$ are reversed compared to the bosonic theory.  The consequence is that dynamical stability is guaranteed for all choices of the parameters.  To see why, define the fermion single-particle dynamic matrix $\check H=\check H_0-i\check Q$.  Analogous to the bosonic theory, dynamical stability of the theory requires all eigenvalues of this matrix to have non-positive imaginary parts.  To see why this is guaranteed, first introduce the Nambu space vectors $\ket{v_{1v}}=[\mu_v^*\ \nu^*_v]$ and $\ket{v_{2v}}=[\nu_v\ \mu_v]$.  Then $\check Q=\frac{1}{2}\sum_{a,v}\ket{v_{av}}\bra{v_{av}}$ is just a sum of one-dimensional projection operators.  As a consequence, all of its eigenvalues are strictly non-negative.  Now consider an eigenvector $\ket{\epsilon}$ of $\check H$.  Then the imaginary part of the corresponding eigenvector is $\mathrm{Im}(\epsilon)=-\braket{\epsilon|\check Q|\epsilon}\leq0$.  The physical reason behind this is Pauli exclusion: there is a limit to the number of fermions each state can hold, preventing an uncontrolled number of particles from entering the system even when the rate of particle pumping is greater than the rate of loss.

As for bosons, the many-body Lindbladian eigenvalues are given by assigning occupation numbers to each eigenvalue of $\check H$ via eq.~(\ref{LindbladEw}).  For fermions, this must be done with the understanding that the $n_s$ are fermionic occupation numbers, equal only to either 0 or 1.  

The fermionic spectral Green's functions are given by:
\begin{equation}
\check G^\tR(t-t')=-i\theta(t-t')e^{-i(t-t')\check H},\quad\quad\check G^\tA(t-t')=i\theta(t'-t)e^{-i(t-t')\check H^\dagger}.
\end{equation}
In frequency space, this becomes the resolvent of $\check H$:
\begin{equation}
\check G^\tR(\epsilon)=\frac{1}{\epsilon-\check H},\quad\quad\check G^\tA(\epsilon)=\frac{1}{\epsilon-\check H^\dagger}.
\end{equation}
The poles of the Green's functions are located at the complex eigenvalues of $\check H$.

The Keldysh Green's function is parametrized through the distribution matrix $\check F(t,t')$ through:
\begin{equation}
\check G^\tK=\check G^\tR\circ\check F-\check F\circ\check G^\tA.
\end{equation}
The distribution matrix obeys the quantum kinetic equation:
\begin{equation}\label{KinEqFerm}
[\partial_t\overset{\circ}{,}\check F]=-i\big(\check H\check F-\check F\check H^\dagger\big)+\check D,
\end{equation}
where $\check D$ and $\partial_t$ come with factors of $\delta(t-t')$.  In the stationary limit, one has $\check F(t,t')=\check F_\mathrm{st}\delta(t-t')$ where $\check F_\mathrm{st}$ obeys the Lyapunov equation,
\begin{equation}\label{KinEqFermStationary}
0=-i\big(\check H\check F_\mathrm{st}-\check F_\mathrm{st}\check H^\dagger\big)+\check D.
\end{equation}
Paralleling eq.~(\ref{Fsscomponent}), in the eigenbasis of $\check H$, one may solve for the components of $\check F_ss$ as:
\begin{equation}
\big(\check U\check F_\mathrm{st}\check U^\dagger\big)_{ss'}=\frac{-i}{\epsilon_s-\epsilon^*_{s'}}\big(\check U\check D\check U^\dagger\big)_{ss'}.
\end{equation}
The stationary distribution matrix $\check F_\mathrm{st}$ is equivalent to the stationary equal time Keldysh Green's function, which itself is equal to the fermion covariance matrix, $\check F_\mathrm{st}=i\check G^\tK(t,t)=\braket{[\hat\fC,\hat\fC^\dagger]}$.

When the stationary state is unique, the stationary density matrix is a fermionic Gaussian state of the form of eq.~(\ref{rhoss}), with
\begin{equation}
\hat\fH_\mathrm{st}=\frac{1}{2}\hat\fC^\dagger\check H_\mathrm{st}\hat\fC.
\end{equation}
The effective Hamiltonian is related to the distribution matrix through \cite{GaussianFerm1}:
\begin{equation}\label{StationaryEwsFerm}
\check F_\mathrm{st}=\tanh(\check H_\mathrm{st}/2).
\end{equation}
Diagonalizing $\check F_\mathrm{st}$, one finds $2N$ eigenvalues of $\check H_\mathrm{st}$ that determine the population numbers of $\rho_\mathrm{st}$ in its diagonal basis \cite{GaussianFerm3}:
\begin{subequations}
\begin{equation}
\check U_F\check F\check U^{-1}_F=\mathrm{diag}\big(\tanh(\beta_1/2),...,-\tanh(\beta_1/2),...\big).
\end{equation}
\begin{equation}
\hat\fH_\mathrm{st}=\sum_j^N\beta_j\hat d_j^\dagger\hat d_j,
\end{equation}
\end{subequations}
with $[\hat d\ \hat d^\dagger]=\check U[\hat c\ \hat c^\dagger]$.  In contrast to the bosonic theory, there are no bounds on the inverse effective temperatures $\beta_j$; they may be negative.

The expectations of observables, in the presence of sources, can be evaluated by solving the fermionic analogue eq.~(\ref{KinEqBoseRed}) and using that of eq.~(\ref{QuadraticObsBoseTime}),
\begin{subequations}
\begin{equation}
\partial_t\check F(t)=-i\big(\check H(t)\check F(t)-\check F(t)\check H^\dagger(t)\big)+\check D(t).
\end{equation}
\begin{equation}
\braket{\hat\fO}(t)=\frac{1}{2}\tr\Big(\check O\big(1-\check F(t)\big)\Big),
\end{equation}
\begin{equation}
\braket{\fO^\cl(t)}=-\frac{1}{2}\tr\big(\check O\check F(t)\big).
\end{equation}
\end{subequations}
For weak perturbations from the stationary state, the linear response formulas which replace eq.s~(\ref{LinRespH}), (\ref{LinRespQ}), and (\ref{LinRespD}) are:
\begin{subequations}
\begin{equation}
\braket{\fO^\cl(t)\delta_{H_0}\fK(t')}_\mathrm{st}=\frac{1}{2}\tr\Big(\check G^\tR(t,t')\big[\check F_\mathrm{st},\delta\check H_0(t')\big]\check G^\tA(t',t)\check O\Big),
\end{equation}
\begin{equation}
\braket{\fO^\cl(t)\delta\fK_Q(t')}_\mathrm{st}=\frac{i}{2}\tr\Big(\check G^\tR(t,t')\big\{\check F_\mathrm{st},\delta\check Q(t')\big\}\check G^\tA(t',t)\check O\Big),
\end{equation}
\begin{equation}
\braket{\fO^\cl(t)\delta\fK_D(t')}_\mathrm{st}=-\frac{i}{2}\tr\Big(\check G^\tR(t,t')\delta\check D(t')\check G^\tA(t',t)\check O\Big).
\end{equation}
\end{subequations}

\section{Examples}\label{3}
Below, the formalism presented above is developed to study examples of Lindbladian band theory, semiclassical kinetics, and mean-field theory.  While by no means an exhaustive list, these examples demonstrate how both quadratic and nonlinear Lindbladians may be studied using the Keldysh language and serve to illustrate important differences compared to the equilibrium theory.

\subsection{Parametrically Driven Oscillator}\label{3.1}
As a warm up, consider first a model with one degree of freedom: a single linear bosonic oscillator in contact with a thermal bath and subjected to a parametric drive.  This simple model is prototypical of much of the phenomena unique to Lindbladian dynamics described above.  The Hamiltonian and jump operators for the parametrically driven oscillator in the rotating frame of a drive are given by:
\begin{subequations}
\begin{equation}
\hat\fH=\Delta\hat a^\dagger\hat a+\lambda(\hat a^{\dagger2}+\hat a^2),
\end{equation}
\begin{equation}
\hat\fL_1=\hat a,\qquad\hat\fL_2=\hat a^\dagger.
\end{equation}
\end{subequations}
The jump operators describe the loss and gain of quanta to and from the environment at the corresponding rates $\gamma$ and $\tilde\gamma$.  The rates can be related to the strength of the coupling between the system and the bath $\kappa$, the bath temperature $T$, and the natural frequency of the system $\omega_0$ by:
\begin{equation}\label{PDOTemp}
2\kappa=\gamma-\tilde\gamma,\qquad\coth(\omega_0/2T)=\frac{\gamma+\tilde\gamma}{\gamma-\tilde\gamma}.
\end{equation}
Note the appearance of the Bose function $2n_\tB+1=\coth(\omega_0/2T)$ at the bath temperature and system frequency.

The dynamic matrix $\check H$ is given by:
\begin{equation}
\check H=\begin{bmatrix}\Delta-i\kappa&2\lambda\\-2\lambda&-\Delta-i\kappa\end{bmatrix}.
\end{equation}
The spectral Green's functions are given by eq.~(\ref{GreenSpectralBose}),
\begin{equation}\label{GRPDO}
\check G^{\tR,\tA}(\epsilon)=\frac{1}{(\epsilon\pm i\kappa)^2-\Omega^2}\begin{bmatrix}\epsilon+\Delta\pm i\kappa&2\lambda\\2\lambda&\epsilon-\Delta\pm i\kappa\end{bmatrix},
\end{equation}
where $\Omega^2=\Delta^2-4\lambda^2$ is the Bogoliubov frequency of the Hamiltonian part of $\check H$.  The eigenvalues of $\check H$ match the poles of the spectral Green's functions from the numerator in the above expression,
\begin{equation}\label{PDOews}
\epsilon_{1,2}=-i\kappa\mp\Omega.
\end{equation}
From this, one can see that the system is stable so long as the coupling $\kappa$ is positive, which occurs when the rate of loss of quanta is greater than the rate of gain, $\gamma>\tilde\gamma$.  Additionally, the model is only stable when the drive strength is small enough, $2\lambda\leq\sqrt{\Delta^2+\kappa^2}$.  When the bound is saturated, one of the two eigenvalues is tuned to zero and the theory is at the brink of instability.  The matrix $\check U$ performing the diagonalization is equivalent to the coherent Bogoliubov rotation matrix in the absence of the coupling to the bath:
\begin{equation}\label{PDOU}
\check U=\frac{1}{\sqrt2}\begin{bmatrix}\sqrt{\frac{\Delta}{\Omega}+1}&\sqrt{\frac{\Delta}{\Omega}-1}\\\sqrt{\frac{\Delta}{\Omega}-1}&\sqrt{\frac{\Delta}{\Omega}+1}\end{bmatrix}
\end{equation}
The matrix $\check D$ is proportional to the identity matrix by the factor $2\kappa\coth(\omega_0/2T)$.
The Keldysh Green's function is given by:
\begin{multline}\label{GKPDO}
\check G^\tK(\epsilon)=\frac{-2i\kappa\coth(\omega_0/2T)}{\big((\epsilon+i\kappa)^2-\Omega^2\big)\big((\epsilon-i\kappa)^2-\Omega^2\big)}\\\times\begin{bmatrix}(\epsilon+\Delta)^2+\kappa^2+4\lambda^2&-4\lambda(\Delta+i\kappa)\\-4\lambda(\Delta-i\kappa)&(\epsilon-\Delta)^2+\kappa^2+4\lambda^2\end{bmatrix}.
\end{multline}
Solving eq.~(\ref{KinEqBose}) gives the stationary distribution matrix:
\begin{equation}\label{PDOFss}
\check F_\mathrm{st}=\frac{\coth(\omega_0/2T)}{\kappa^2+\Omega^2}\begin{bmatrix}\Delta^2+\kappa^2&-2\lambda(\Delta+i\kappa)\\-2\lambda(\Delta-i\kappa)&\Delta^2+\kappa^2\end{bmatrix}.
\end{equation}
The eigenvalues of $\check F_\mathrm{st}$ are related to the diagonal frequency $\beta$ of the effective Hamiltonian $\hat\fH_\mathrm{st}$ through eq.~(\ref{FewsBose}),
\begin{equation}
\coth(\beta/2)=\coth(\omega_0/2T)\sqrt\frac{\kappa^2+\Delta^2}{\kappa^2+\Omega^2}.
\end{equation}
The diagonalizing transformation $\check U_F$ is given by:
\begin{equation}
\check U_F=\frac{1}{\sqrt2}\begin{bmatrix}e^{-i\theta}\sqrt{\sqrt{\frac{\kappa^2+\Delta^2}{\kappa^2+\Omega^2}}+1}&e^{i\theta}\sqrt{\sqrt{\frac{\kappa^2+\Delta^2}{\kappa^2+\Omega^2}}-1}\\e^{-i\theta}\sqrt{\sqrt{\frac{\kappa^2+\Delta^2}{\kappa^2+\Omega^2}}-1}&e^{i\theta}\sqrt{\sqrt{\frac{\kappa^2+\Delta^2}{\kappa^2+\Omega^2}}+1}\end{bmatrix},
\end{equation}
where $\theta=\arg(\Delta+i\kappa)$.  As discussed above, the bases in which the distribution and dynamic matrices are diagonal are not the same, $\check U\neq\check U_F$.

To gain more intuition for the model, consider the limits of strong versus weak dissipation.  When the dissipation is very weak compared to all other scales in the problem $\kappa\to0$, the two Lindbladian eigenvalues $-i\epsilon_s$ are complex conjugates with a small negative real part.  The dynamics in this limit are weakly under-damped coherent rotation.  Up to corrections of order $\kappa$ the stationary state is diagonal in the basis Bogoliubov quasi-particles of the coherent problem.  That is, $\check U_F=\check U+O(\kappa)$.  With this, the stationary state effective Hamiltonian in its diagonal basis is given by:
\begin{equation}
\hat\fH_\mathrm{st}=\beta\hat b^\dagger\hat b,
\end{equation}
with $[\hat b\ \hat b^\dagger]=\check U_F[\hat a\ \hat a^\dagger]$.  This is equal to the original Hamiltonian $\hat\fH$ up to a proportionality constant $\beta=\Omega/T_\mathrm{eff}$,
\begin{equation}
\coth(\Omega/2T_\mathrm{eff})=\frac{\Delta}{\Omega}\coth(\omega_0/2T).
\end{equation}
The stationary state is thus a thermal state of the coherent dynamics, but with a different effective temperature than the bath temperature.  Note that even in the limit $T\to0$, the effective temperature $T_\mathrm{eff}$ is finite.  This phenomenon is known as quantum heating \cite{Dykman1,Dykman2,Dykman3}.

Alternatively, one may consider the the limit of strong dissipation, $\kappa\to\infty$.  In this limit, the drive strength can stably be larger than the detuning, $2\lambda>\Delta$, past the point of coherent stability.  In this situation, the Lindbladian eigenvalues $-i\epsilon_s$ are purely imaginary and the dynamics are that of over-damped pure dissipation without any coherent rotation.  Up to corrections of the order of the Hamiltonian parameters $\Delta$ and $\lambda$, the dynamic matrix $\check H$ is proportional to the identity with a factor of $\kappa$.  Thus, the diagonal basis for the dynamics is the starting basis of the problem, $\check U=\check1$.  Similarly, examining eq.~(\ref{PDOFss}) in this limit, one sees that $\check F_\mathrm{st}$ is also proportional to the identity, so that $\check U_F=\check1$.  The stationary state effective Hamiltonian is given by:
\begin{equation}
\hat\fH_\mathrm{st}=\frac{\omega_0}{T}\, \hat a^\dagger\hat a.
\end{equation}
The stationary state is a thermal state of the un-driven Hamiltonian $\omega_0\hat a^\dagger\hat a$ with a temperature equal to the bath temperature.

Thus, tuning the dissipation strength between the extreme limits the extreme limits $\kappa\to0$ and $\kappa\to\infty$ changes the diagonal basis of both the dynamics and stationary state between the Bogoliubov basis of the $(\hat b,\hat b^\dagger)$ bosons and the original basis of the $(\hat a,\hat a^\dagger)$ bosons.  In both of these limits, these basis are the same, $\check U=\check U_F$; this will cease the be the case in between these two extremes.  Interstitial between these two regimes, there is an exceptional point at which the two Lindbladian eigenvalues $-i\epsilon_s$ coalesce to the value $-\kappa$ and the dynamics is a resonantly damped dissipation.  This occurs at the threshold of coherent instability $2\lambda=\Delta$.  At this point, the matrix $\check H$ is non-diagonalizable and is brought to Jordan canonical form by the similarity transformation:
\begin{equation}
\check U=\begin{bmatrix}0&-1\\2\lambda&2\lambda\end{bmatrix}
\end{equation}
Note that this expression is not the equal to the limiting form of eq.~(\ref{PDOU}), which itself does not exist.  In contrast, the distribution matrix is a smooth function of the parameters even at this point, as are the diagonal frequencies of the effective Hamiltonian and the diagonal basis of bosons determined by $\check U_F$.  These are given by their limiting forms in terms of the above general expressions.  The exceptional point is reflected in the limiting form of the spectral Green's functions as second-order pole:
\begin{equation}
\check G^{\tR,\tA}(\epsilon)=\frac{1}{(\epsilon\pm i\kappa)^2}\begin{bmatrix}\epsilon+2\lambda\pm i\kappa&2\lambda\\2\lambda&\epsilon-2\lambda\pm i\kappa\end{bmatrix}.
\end{equation}
The non-diagonalizability results in the linear gain of certain non-stationary densities.  

To exemplify this, consider increasing $\Delta$ slightly above or below the critical value of $2\lambda$.  This removes the eigenvalue degeneracy, thus eliminating the polynomial gain of any initial correlations.  The resulting decay at rates are determined by differences of the perturbed eigenvalues $\epsilon_s-\epsilon_{s'}^*$.  Writing $\Delta-2\lambda=\delta$, expanding eq.~(\ref{PDOews}) around the exceptional point $\delta=0$ gives a series in powers of $\delta^{1/2}$:
\begin{equation}
\epsilon_{1,2}=-i\kappa\mp2\sqrt{\lambda\delta}+O(\delta).
\end{equation}
As a consequence, introducing a small detuning causes initial densities with off-diagonal terms in the perturbed basis of $\check H$ eigenvalues to coherently rotate at a rate $4\sqrt{\lambda\delta}$ that is a non-analytic function of the deviation $\lambda$.  This demonstrates a stronger sensitivity to perturbations at the exceptional point than the usual $O(\delta)$ corrections in Hermitian systems.  Further implications of the square root singularity in eigenvalues near exceptional points is explored in the following section.

\subsection{Non-Hermitian Band Theory}\label{3.2}
This section examines two simple Lindbladian tight binding models.  For simplicity the focus is restricted to one-dimensional chains, though the general principles discussed hear can naturally be extended to higher dimensions.  Much like clean coherent models, Lindbladian lattice systems with translational invariance are described in terms of the band theory.  The bands in a Lindbladian system are the momentum-dependent eigenvalues of the dynamic matrix $\check H$, and as such are generically complex.  Beyond having an imaginary part, there are additional complications that emerge due to the non-Hermiticity of $\check H$, specifically relating to the potential existence of exceptional points.  This subtlety is showcased in two simple models below.  Note that the theory of non-Hermitian bands in connection with Lindbladian dynamics is still under construction and the following discussion is far from exhaustive, see \cite{NonHermBand1,NonHermBand2,NonHermBand3,NonHermBand4,NonHermBand5,NonHermBand6,NonHermBand7}.

As a first example, consider a chain of identical parametric oscillators from the preceding section coupled linearly to their nearest neighbors.  The Hamiltonian and jump operators are:
\begin{subequations}
\begin{equation}\label{HamChainBoson}
\hat \fH=\sum_{j=1}^N\Big(\Delta_0\hat a^\dagger_j\hat a_j+\tau(\hat a^\dagger_{j+1}\hat a_j+\hat a^\dagger_j\hat a_{j+1})+\lambda_0(\hat a_j^2+\hat a_j^{\dagger2})\Big),
\end{equation}
\begin{equation}
\hat\fL_{1j}=\hat a_j,\qquad\hat\fL_{2j}=\hat a^\dagger_j.
\end{equation}
\end{subequations}
With periodic boundary conditions, it is convenient to change to the momentum representation,
\begin{equation}
\phi^\alpha(k)=\frac{1}{\sqrt{N}}\sum_j e^{ijk}\phi^\alpha_j,
\end{equation}
where $k\in[0,2\pi)$ is the crystal momentum.  In momentum space, writing $\Phi^\alpha(k)=[\phi^\alpha(k)\ \bar\phi^\alpha(-k)]$ brings the Keldysh action to the standard form of eq.~(\ref{ActionBoson}), where the parameter matrices are diagonal functions of $k$,
\begin{equation}
S=\frac{1}{2}\int\dif t\dif k\ \bar\Phi\begin{bmatrix}0&\check\tau^3i\partial_t-\check H_0(k)-i\check Q(k)\\\check\tau^3i\partial_t-\check H_0(k)+i\check Q(k)&i\check D(k)\end{bmatrix}\Phi.
\end{equation}
The parameter matrices are given by a $k$-dependent version of eq.~(\ref{ParamMatBose}) with $\eta=0$ and the other parameters,
\begin{subequations}
\begin{equation}
\Delta(k)=\Delta_0 +2\tau\cos(k),\qquad\lambda(k)=\lambda_0,
\end{equation}
\begin{equation}
\gamma(k)=2\kappa(n_\tB+1),\qquad\tilde\gamma(k)=2\kappa n_\tB,\qquad\eta(k)=0.
\end{equation}
\end{subequations}
with $n_\tB$ defined by $2n_\tB+1=\coth(\omega_0/2T)$ and $\kappa$, $\omega_0$, and $T$ the dissipative coupling, system natural frequency, and bath temperature defined in eq.~(\ref{PDOTemp}).

\begin{figure}
\begin{center}
\scalebox{.45}{\includegraphics{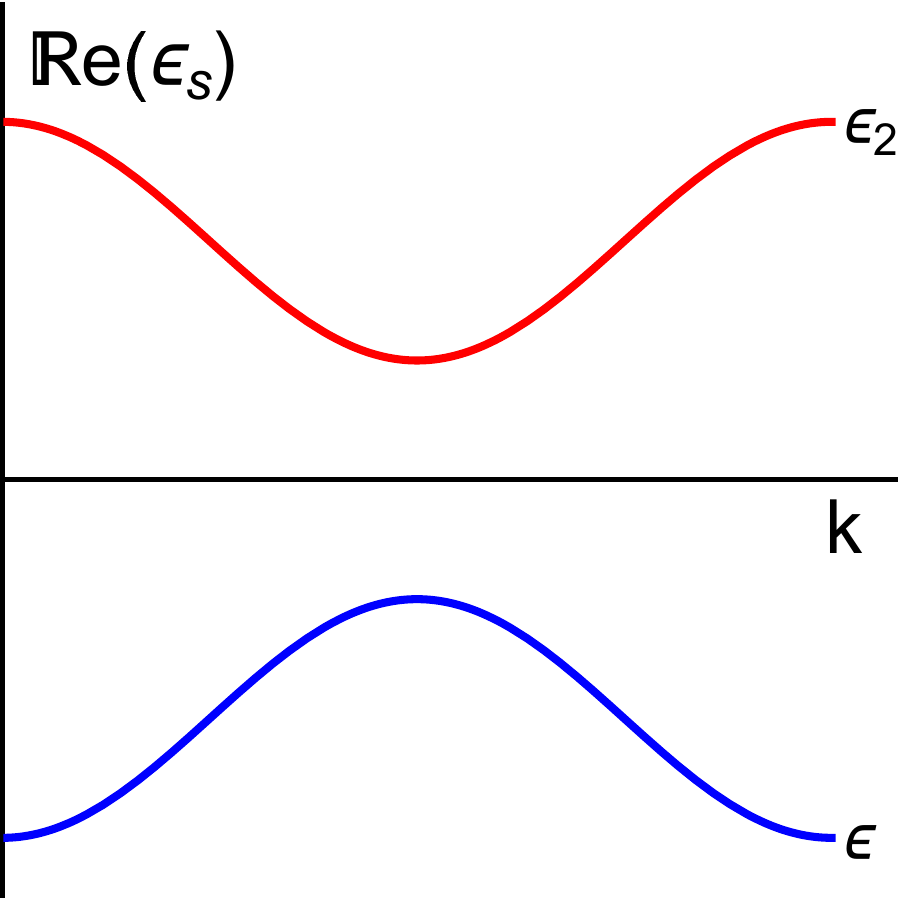}}
\scalebox{.45}{\includegraphics{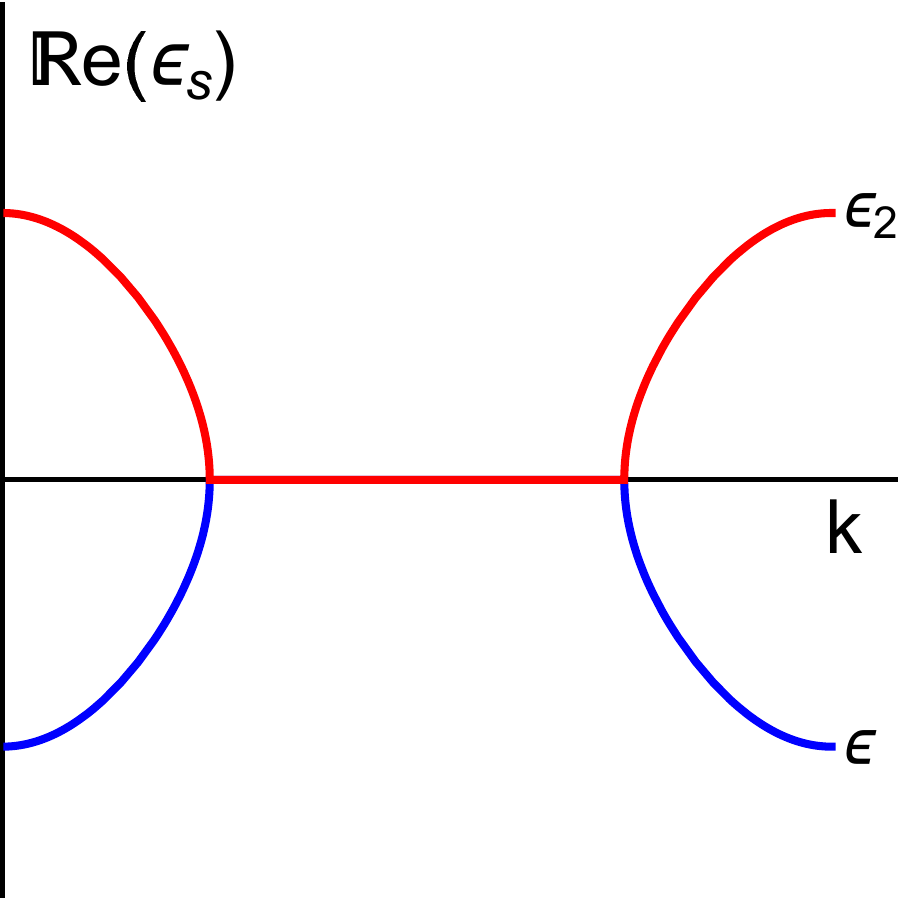}}
\scalebox{.45}{\includegraphics{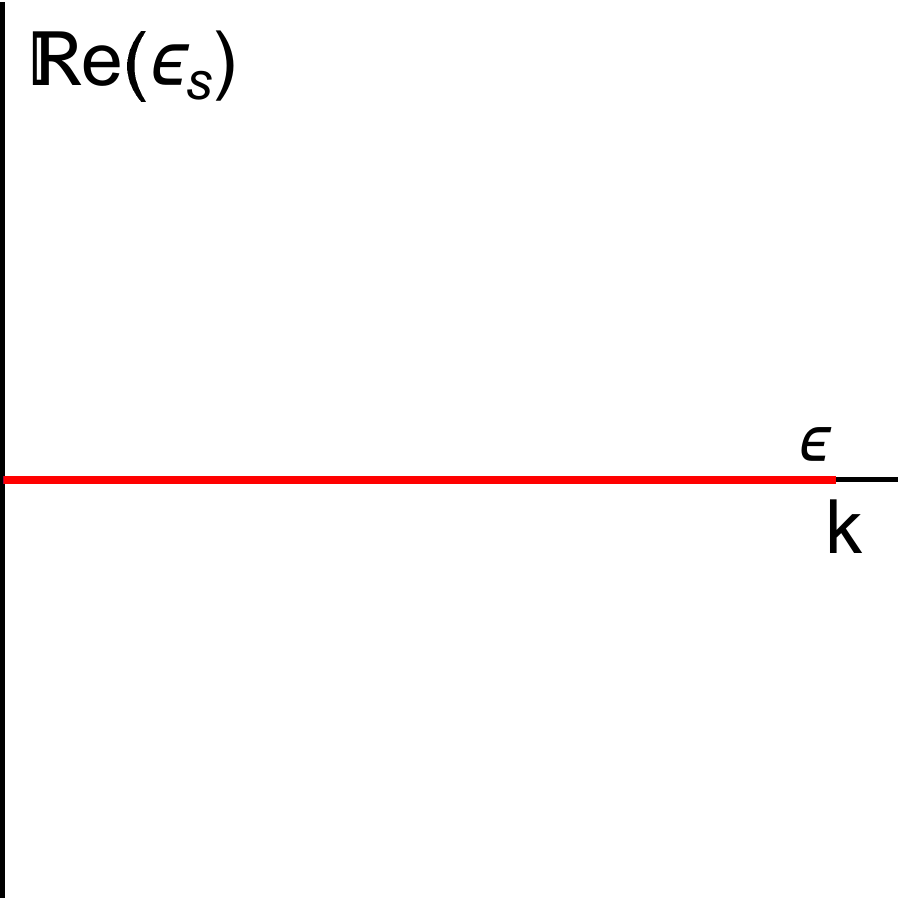}}
\scalebox{.45}{\includegraphics{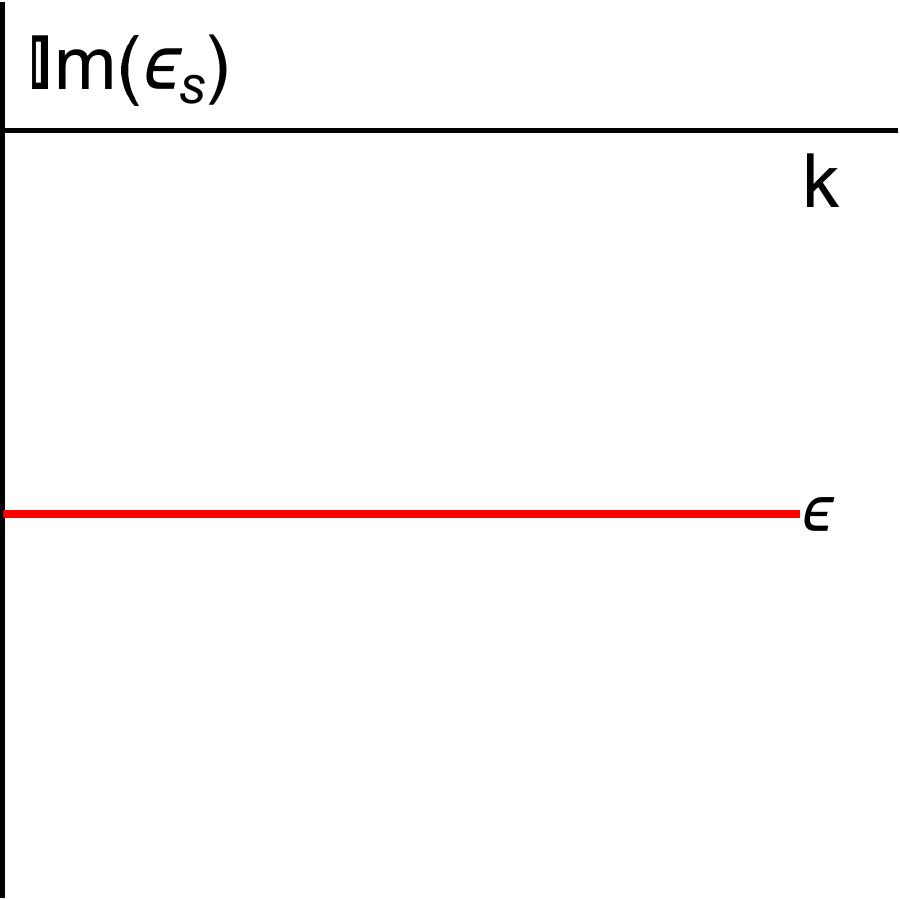}}
\scalebox{.45}{\includegraphics{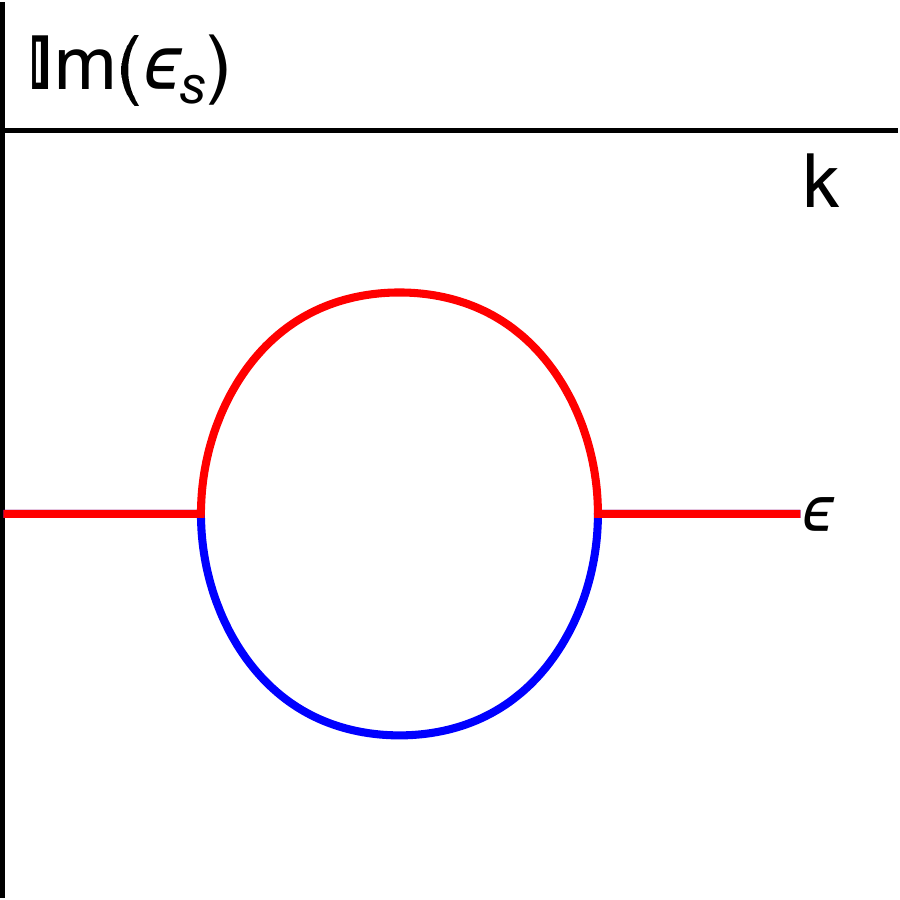}}
\scalebox{.45}{\includegraphics{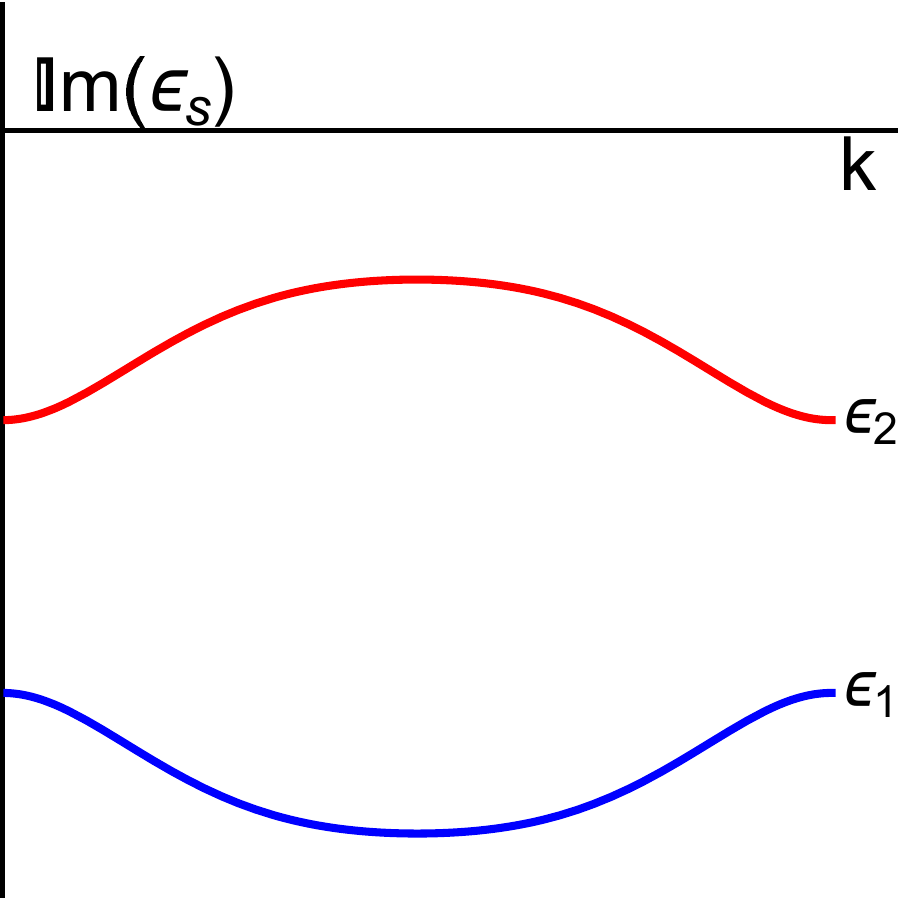}}
\caption{Plots of the real and imaginary parts of the complex bands eq.~(\ref{PDObands}) with $\kappa=5$, $\Delta=2$, $\tau=1/2$.  The three columns show the bands in the completely under-damped, mixed, and completely over-damped regime, corresponding to $\lambda=0,1$, and $1.7$, going from left to right.}
\label{PDObandsFig}
\end{center}
\end{figure}

With this, one can read off the Green's functions from the previous section using eq.s~(\ref{GRPDO}) and (\ref{GKPDO}) and replacing the appropriate quantities with their $k$-dependent analogues.  There are two bands of eigenvalues of $\check H(k)$, given by an $k$-dependent version of eq.~(\ref{PDOews}),
\begin{equation}\label{PDObands}
\epsilon_{1,2}=-i\kappa\mp\sqrt{\big(\Delta_0+2\tau\cos(k)\big)^2-4\lambda_0^2}.
\end{equation}
This defines a pair of complex-valued bands.  As $\lambda_0$ is tuned from being small to large, the two bands go from being complex with a constant imaginary part to purely imaginary.  These two regimes correspond to completely under-damped and over-damped dissipation.  The transition between these two limits occurs via an extended intermediate regime in which the bands touch and both over- and under-damped dissipation occurs in different momentum ranges.  This is depicted in fig.~(\ref{PDObandsFig}).  The points in the Brillouin zone where the bands touch define the exceptional momenta $k_\mathrm{EP}$, here given by:
\begin{equation}
k_\mathrm{EP}=\arccos\bigg(\frac{2\lambda_0-\Delta_0}{2\tau}\bigg),
\end{equation}
which has two solutions when $\Delta_0-2\tau<2\lambda_0<\Delta_0+2\tau$.  Unlike in a Hermitian band touching, the touching of complex bands occurs at exceptional points in the parameter space of $\check H$.  This results in a resonant damping at the exceptional momenta $k_\mathrm{EP}$.

One can solve the kinetic equation for the stationary state by looking for translationally invariant solutions.  This reduces eq.~(\ref{KinEqBoseStationary}) to a $k$-dependent Lyapunov equation,
\begin{equation}
0=-i\big(\check H(k)\check F_\mathrm{st}(k)-\check F_\mathrm{st}(k)\check H^\dagger(k)\big)+\check\tau^3\check D(k)\check\tau^3.
\end{equation}
The solution to this equation can be read off from the solution to the single parametric oscillator in eq.~(\ref{PDOFss}) by appropriately replacing parameters by their $k$-dependent generalization.  Just like the single parametric oscillator, the stationary distribution has no signature of the exceptional points: there are no non-analyticities at the exceptional momenta.  Response to translationally-invariant perturbations can be computed using the formalism from section~\ref{2.4} and replacing matrices with their $k$-dependent versions.  Variations that are not spatially homogeneous but which vary smoothly over large distances can be studied using a semiclassical approach; this is discussed in section \ref{3.4}.

The above example exemplifies that the degeneration of bands in Lindbladian models occurs at exceptional points and thus differs from Hermitian band touching.  This naively suggests that as long as there are no spectral degeneracies, the band theory of Lindbladian systems is similar to conventional Hermitian systems save for the additional imaginary part of each band.  This intuition however is badly wrong.  Due to the non-analytic structure of eigenvalues near exceptional points, non-Hermitian bands can exhibit unusual structures even if they do not cross.  In the simplest situation with a $2\times2$ non-Hermitian matrix, an exceptional point corresponds to a degeneration of two eigenvalues and, as demonstrated in the preceding sections, leads to a square root singularity.  Revolving around the exceptional point in parameter space induces a monodromy in which the two eigenvalues are exchanged upon a single winding rather than returning to where they started.  In a one-dimensional tight binding model with two bands, one can imagine tuning a parameter past the regime of band touching so that the bands remain intertwined and mutually wind around the exceptional point, exhibiting monodromy over one period of the Brillouin zone.

This clearly does not occur in the chain of parametric oscillators discussed above.  One can construct a simple model that demonstrates this phenomenon using a Lindbladian generalization of the SSH model.  Similar models were studied in \cite{NonHermBand2,NonHermBand3}, though not in the context of the Lindbladian dynamics.  The SSH Hamiltonian is defined by assigning two species of fermions, $\hat c_\tA$ and $\hat c_\tB$, to each site on a one-dimensional chain.  Hopping is allowed between species on-site and between nearest-neighbors,
\begin{equation}
\hat \fH=\sum_{j=1}^N\Big(\varepsilon_0(\hat c_{\tA j}^\dagger\hat c_{\tA j}+\hat c_{\tB j}^\dagger\hat c_{\tB j})+\tau_1(\hat c_{\tA j}^\dagger\hat c_{\tB j}+\hat c_{\tB j}^\dagger\hat c_{\tA j})+\tau_2(\hat c^\dagger_{\tA j+1}\hat c_{\tB j}+\hat c^\dagger_{\tB j}\hat c_{\tA j+1})\Big).
\end{equation}
In addition, consider on-site gain and loss of particles via a superposition of the A and B fermions,
\begin{equation}
\fL_{1j}=\sqrt{2\kappa}(\hat c_{\tA j}+e^{i\theta_1}\hat c_{\tB j}),\qquad\fL_{2j}=\sqrt{2\kappa}(\hat c_{\tA j}^\dagger+e^{-i\theta_2}\hat c_{\tB j}^\dagger),
\end{equation}
The restriction to purely loss and gain processes in combination with the lack of pair-creation terms in the Hamiltonian gives the model an overall $U(1)$ symmetry, with the associated charge being the total number of particles.  This symmetry is weak \cite{AlbertSym}, and so does not imply the conservation of particle number in time.  A consequence of this symmetry is that the anomalous off-diagonal blocks of the Nambu space Green's functions vanish.  As such, it is convenient to define by $G$ without a check the upper left block of $\check G$ and similarly for other quantities.

The momentum space Keldysh action is then given by:
\begin{equation}
S=\int\dif t\dif k\ \bar\psi\begin{bmatrix}i\partial_t-H_0(k)+iQ(k)&iD(k)\\0&i\partial_t-H_0(k)-iQ(k)\end{bmatrix}\psi,
\end{equation}
where $\psi=[\psi^1\ \psi^2]$, $\psi^a=[\psi^a_\tA(k),\ \psi^a_\tB(k)]$, and the un-checked parameter matrices are the upper-left blocks of their Nambu-space counterparts in the main text eq.~(\ref{ActionFerm}).  The parameter matrices are:
\begin{subequations}
\begin{equation}
H_0(k)=\begin{bmatrix}\varepsilon_0&\tau_1+\tau_2 e^{ik}\\\tau_1+\tau_2 e^{-ik}&\varepsilon_0\end{bmatrix},
\end{equation}
\begin{equation}
Q(k)=\kappa\begin{bmatrix}2&e^{-i\theta_1}+e^{-i\theta_2}\\e^{i\theta_1}+e^{i\theta_2}&2\end{bmatrix},
\end{equation}
\begin{equation}
D(k)=2\kappa\begin{bmatrix}0&e^{-i\theta_1}-e^{-i\theta_2}\\e^{i\theta_1}-e^{i\theta_2}&0\end{bmatrix}
\end{equation}
\end{subequations}
Due to the symmetry in the problem it suffices to study only the two eigenvalue bands of $H=H_0-iQ$; the other two single-particle eigenvalue bands are given by their complex conjugates.

\begin{figure}
\begin{center}
\scalebox{.49}{\includegraphics{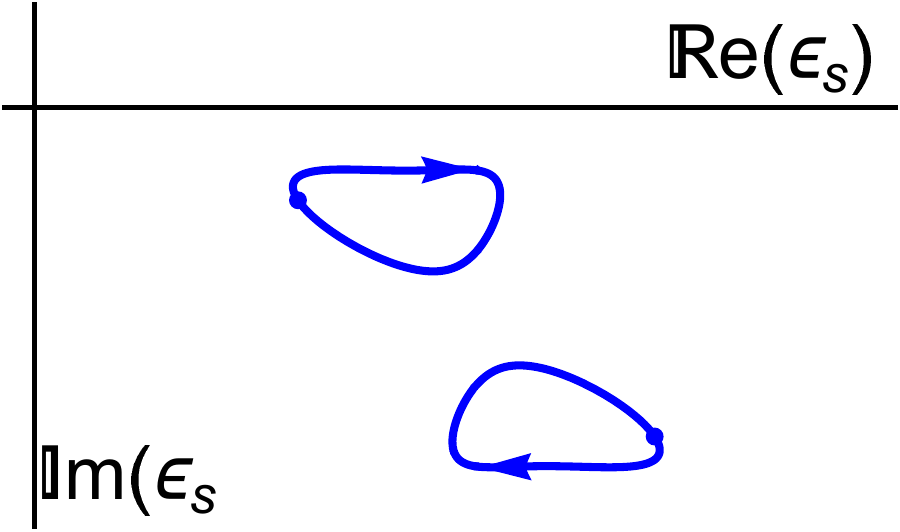}}
\scalebox{.49}{\includegraphics{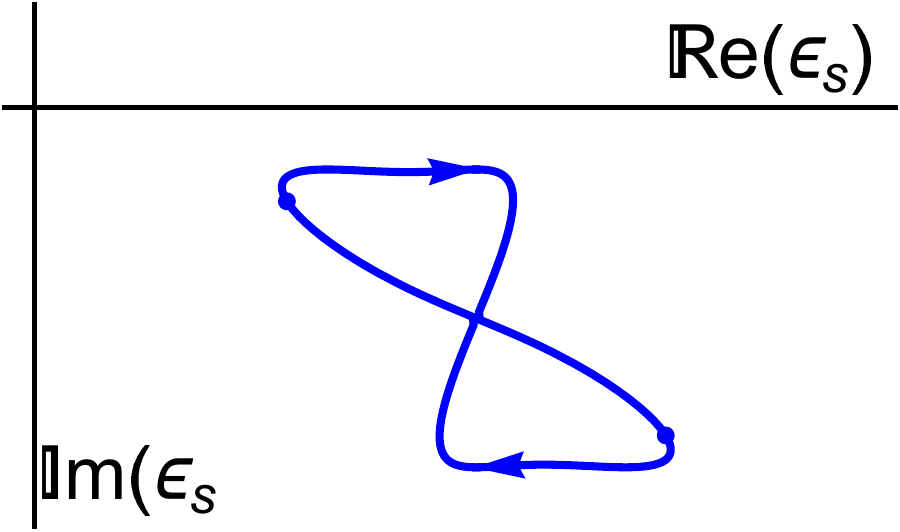}}
\scalebox{.49}{\includegraphics{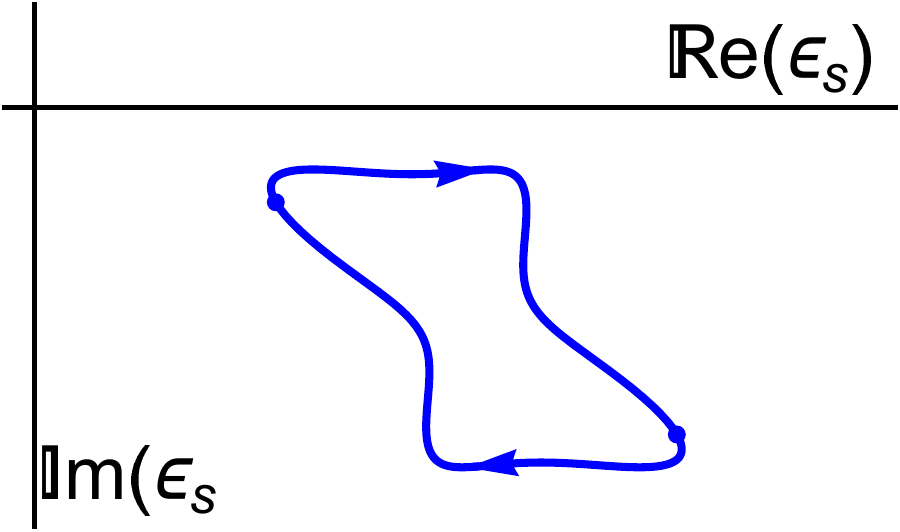}}
\caption{Three regimes of the complex bands of eq.~(\ref{SSHbands}) plotted as paths in the complex plane.  The solid points indicate the values of $\epsilon_j(0)$ and the arrows show the directed path traced out by varying $k$ from $0$ to $2\pi$.  The three regimes correspond to the two disconnected untwisted loops, intersecting loops, and a single untwisted loop for $\tau_2$ less than, equal to, and greater than $\kappa$ respectively.}
\label{EwsBraiding}
\end{center}
\end{figure}

For a simple demonstration of the concept discussed above, one can choose different phases for loss and gain, $\theta_1=\pi/2$ and $\theta_2=0$.  Fixing in addition $\tau_1=\kappa$, the eigenvalues of the dynamic matrix $H(k)$ are given by:
\begin{equation}\label{SSHbands}
\epsilon_{1,2}(k)=\varepsilon_0-2i\kappa\mp\sqrt{\tau_2^2-\kappa^2+2\kappa\tau_2\cos(k)-2i\kappa\Big(\kappa+\tau_2\big(\cos(k)+\sin(k)\big)\Big)}
\end{equation}
From this expression one can see that for $\tau_2=\kappa$, the eigenvalues degenerate to an exceptional point at $k_\mathrm{EP}=3\pi/2$.  For $\tau_2<\kappa$, the eigenvalues are periodic functions of $k$ and the two bands are disconnected.  For $\tau_2>\kappa$ however, the sign of the argument of the square root develops a negative real part for $k$ near $3\pi/2$.  As a consequence, sweeping over the Brillouin zone smoothly traverses from one branch of the square root Riemann sheet to another, introducing the monodromy $\epsilon_{1,2}(k\to2\pi^-)=\epsilon_{2,1}(k\to0^+)$.  These three scenarios are depicted in Fig.~(\ref{EwsBraiding}).  In other words, the square root singularity of the exceptional point effects the analytic structure of the eigenvalue bands even when they do not collide with it.  From encircling of the exceptional point there are no longer two distinct disconnected bands, but rather a single continuous band that double-covers the Brillouin zone.

Recent literature has understood this sort of atypical feature of non-Hermitian band theory in the context of braids and knots \cite{NonHermBand4,NonHermBand5,NonHermBand6,NonHermBand7}.  The eigenvalues of the dynamic matrix $H(k)$ define paths in the complex plain parametrized by $k$.  Due to the periodicity of $k$, the paths must close and so define loops, which can braid together and link up.  This phenomenon does not have a Hermitian analogue, as the eigenvalues of Hermitian matrices live on the real line, a space of too low dimension for the knotting of paths.  In the model considered above, the transition through the exceptional point can be understood as a transition from two disconnected unknots to one.  In the language of braids, this is a transition from two upbraided paths to a pair of paths with a single twist, separated by a point where the paths collide.

In more general models on one-dimensional lattices, eigenvalues may braid multiple times with one another in a more complicated fashion.  This can result in the paths of eigenvalues tracing out collections of linked knots.  In the language of braids, an $N$ band model in a given range of parameters will correspond to an element of the braid group with $N$ generators.  Transitions between different braid group elements/knot configurations occurs via the joining and separating of paths by moving through exceptional points.  Higher-dimensional generalizations of this phenomenon are harder to visualize, as they entail the mapping of higher-dimensional Brillouin zones (tori) into the complex plane.  This problem has received some recent attention (see for example \cite{NonHermBand6}) but in general warrants future study.

As discussed above, the stationary solution to the kinetic theory, $\check F_\mathrm{st}$, displays no signature of exceptional points.  This remains true of the model considered here: the stationary distribution, while non-trivial, varies smoothly across the crossing of the exceptional point and does not differ qualitatively on either side of the transition.  The full expressions for the bands $\beta_j(k)$ of stationary eigenvalues of $\check F_\mathrm{st}(k)$ via eq.~(\ref{StationaryEwsFerm}) are somewhat cumbersome and so are not reproduced here; they are plotted below, at, and above the transition in 
Fig.~(\ref{fig-SSH-Fst}).  There are no apparent signatures these topological features present in the different regimes in the stationary distribution.

\begin{figure}
\begin{center}
\scalebox{.5}{\includegraphics{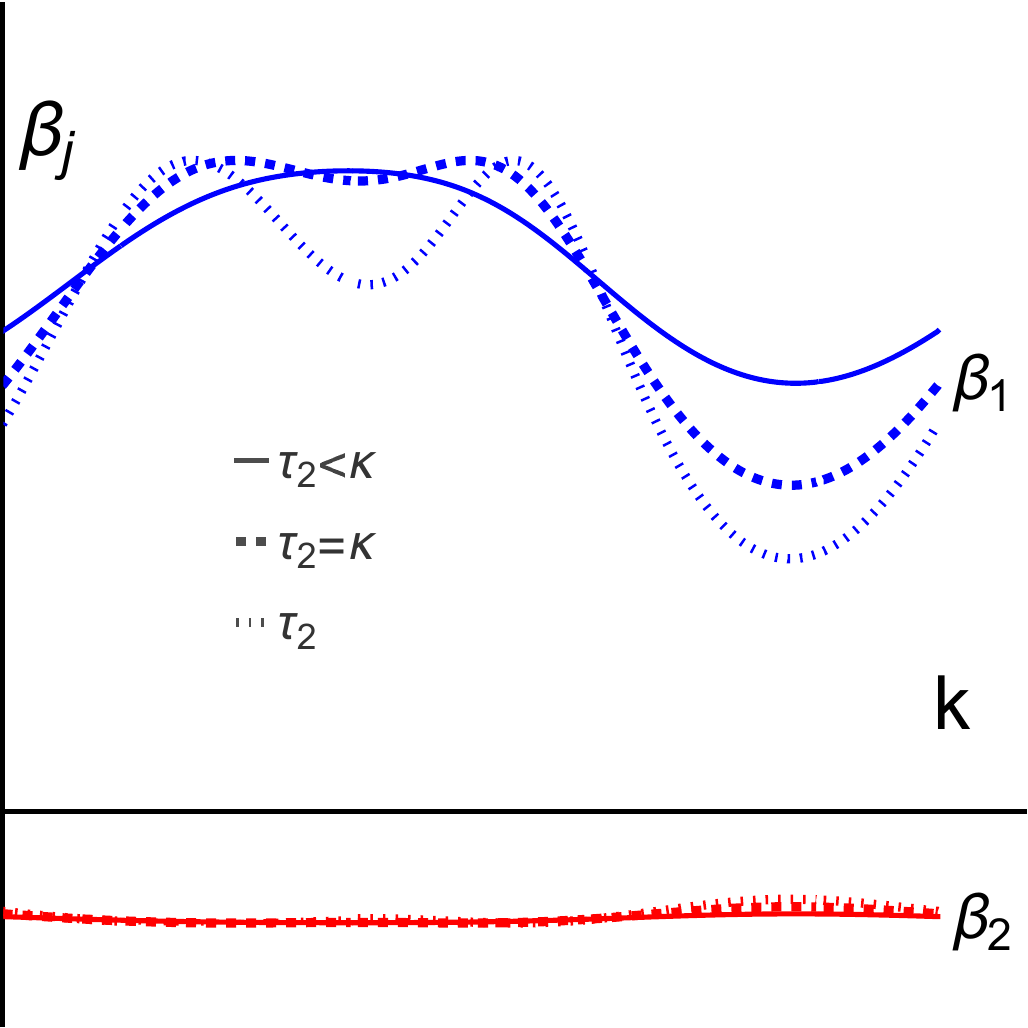}}
\caption{Plot of the two bands of the bands of stationary eigenvalues $\beta_j(k)$ of the Lindbladian SSH chain, showing the three different dynamical regimes.  The non-analytic structure of the dynamic eigenvalues $\epsilon_j(k)$ is not retained by the stationary distribution; at all values of $k$, the curves vary smoothly across the transition.}
\label{fig-SSH-Fst}
\end{center}
\end{figure}

\subsection{Disordered Fermions}\label{3.3}
This section examines a simple model of disordered Lindbladian fermions.  A $U(1)$ symmetric model of $N$ flavors of fermions that are otherwise featureless is studied.  In contrast to the preceding sections discussing `clean' systems, the model considered here gives insight into the behaviour of generic quadratic Lindbladians.  The $U(1)$ symmetry provides a meaningful distinction between loss and gain even in the absence of additional symmetry.  In preparing this manuscript, a paper \cite{RandomQuadratics} appeared up on arXiv which discusses ideas very similar to those presented here.  They focus on Majorana fermions instead of the Dirac fermions considered here and provide a more detailed discussion of the spectrum and level statistics of both dynamic matrix and steady state.

The model examined here can be compared to various studies on Lindbladians in which the Hamiltonian and jump operators are random matrices, which are aimed at understanding generic Lindbladian dynamics in the absence of any additional structure \cite{RandomLindblad1,RandomLindblad2,RandomLindblad3,RandomLindblad4,RandomLindblad5,RandomLindblad6}.  As will be shown below, in certain limits the single-particle quantities of the random quadratic Lindbladian match the pure random matrix results, but generically they different.  This runs counter to the usual intuition from coherent systems.  For equilibrium disordered fermions the many-body spectrum is determined by the single-particle Hamiltonian, which in turn is characterized by Hermitian random matrix theory.  Solving the many-body problem is achieved by solving the pure random matrix quantum mechanics.  In the Lindbladian framework, the single-particle dynamic matrix and stationary distribution are separate quantities and do not define any sort of `single-particle Lindbladian.'  In this sense, the random quadratic problem is not equivalent to the random matrix Lindblad problem.

The Hamiltonian and jump operators defining the model are:
\begin{subequations}
\begin{equation}
\hat\fH=\sum_{i,j}^N(H_0)_{ij}\hat c^\dagger_i\hat c_j,
\end{equation}
\begin{equation}
\hat\fL_{1v}=\sum_j^N\mu_{vj}\hat c_j,\qquad\hat\fL_{2v}=\sum_j^N\nu_{vj}\hat c^\dagger_j,
\end{equation}
\end{subequations}
where there are a total of $M$ jump operators, $v\leq M$.  The parameters $H_0$, $\mu$, and $\nu$ are be Gaussian random variables.  The disorder averaging is defined by:
\begin{equation}
[...]=\int\fdif H_0\fdif\mu\fdif\nu e^{-\frac{N}{2}\tr(H_0^2)-\frac{M}{\gamma}\sum_{j,v}|\mu_{vj}|^2-\frac{M}{\tilde\gamma}\sum_{j,v}|\nu_{vj}|^2}(...).
\end{equation}
Of interest here is the large $N$ limit where the ratio $m=M/N$ is held finite.  In this limit, the model has a total of three parameters: $m$ and the two parameters that control the ratio of loss and gain vs. the strength of the Hamiltonian, $\gamma$ and $\tilde\gamma$.

In the corresponding coherent model, one solves the random matrix theory problem of the single-particle Hamiltonian.  This is exactly solvable in the large $N$ limit, with the density of states $\varrho(\epsilon)\equiv[\tr\delta(\epsilon-H_0)]$ given by the well-known Wigner semicircle distribution.
In the Lindbladian setting, one can determine the spectrum by solving the non-Hermitian random matrix problem for the eigenvalue distribution $\varrho_H(\epsilon)=[\tr\delta(\epsilon-H)]$, which is the probability distribution of the single particle eigenvalues $\epsilon_j$  in the complex $\epsilon$ plain.  This problem turns out to involve more subtle features than its Hermitian counterpart.  In the large $N$ limit, the limit shape of the distribution changes shape as a function of the model parameters, with different phases defined by the number of connected components.
In addition to the spectrum, one can also analyze the stationary state, in which the object of interest is the density of states of the stationary effective Hamiltonian, $\varrho_\mathrm{st}(\beta)=[\tr\delta(\epsilon-H_\mathrm{st})]$, which is the probability distribution of the real-valued $\beta_j$.  For this one must solve a random Lyapunov equation.  Compared to the spectral random matrix problem, this problem is hard to solve exactly, even in the large $N$ limit.  Simple numerical computations show that the transitions in the dynamics are mirrored by transitions in the stationary state.

Consider first the simple case of pure random loss $\gamma\to\infty$, so that the $H_0$ and $\nu$ are dropped.  The single-particle dynamic matrix is a Wishart matrix, given by a sum of one-dimensional projection operators:
\begin{equation}
H_{ij}=-i\sum_v^M\mu^*_{vi}\mu_{vj}.
\end{equation}
In the large $N$ limit, its eigenvalue distribution is given by the well-known Marchenko–Pastur law \cite{MarchenkoPastur}, with the eigenvalue distribution $\varrho_H(\epsilon)$ given by,
\begin{equation}
\varrho_H(\epsilon)=-i\Bigg(\frac{m}{\gamma\epsilon}\sqrt{\frac{\gamma^2}{m}-\frac{1}{4}\bigg(\epsilon-\frac{\gamma}{m}(1+m)\bigg)^2}+\pi(1-m)\theta(1-m)\delta(\epsilon)\Bigg),
\end{equation}
where $\theta$ is the Heaviside function.  When $m>1$, all eigenvalues have finite imaginary part and the stationary state is unique, being given by the Fock vacuum state with zero particles.  When $m<1$, there are always a subset of modes that do not dissipate, reflected in $\varrho_H(\epsilon)$ by the delta measure at the origin.  The stationary state is not unique.  The critical point $m=1$ divides the two regimes.  At this point, the dissipative gap closes and the uniqueness of the stationary state breaks down.  Note that for all values of $m$, the eigenvalues are purely imaginary, so that $\varrho_H(\epsilon)$ is supported only on the imaginary axis.  This is qualitatively different from the results reported for general disordered Lindbladians with random matrix jump operators and no Hamiltonian, in which the distribution of Lindbladian eigenvalues occupies a lemon shaped region in the complex plane with a finite area \cite{RandomLindblad2}.

\begin{figure}
\begin{center}
\scalebox{.36}{\includegraphics{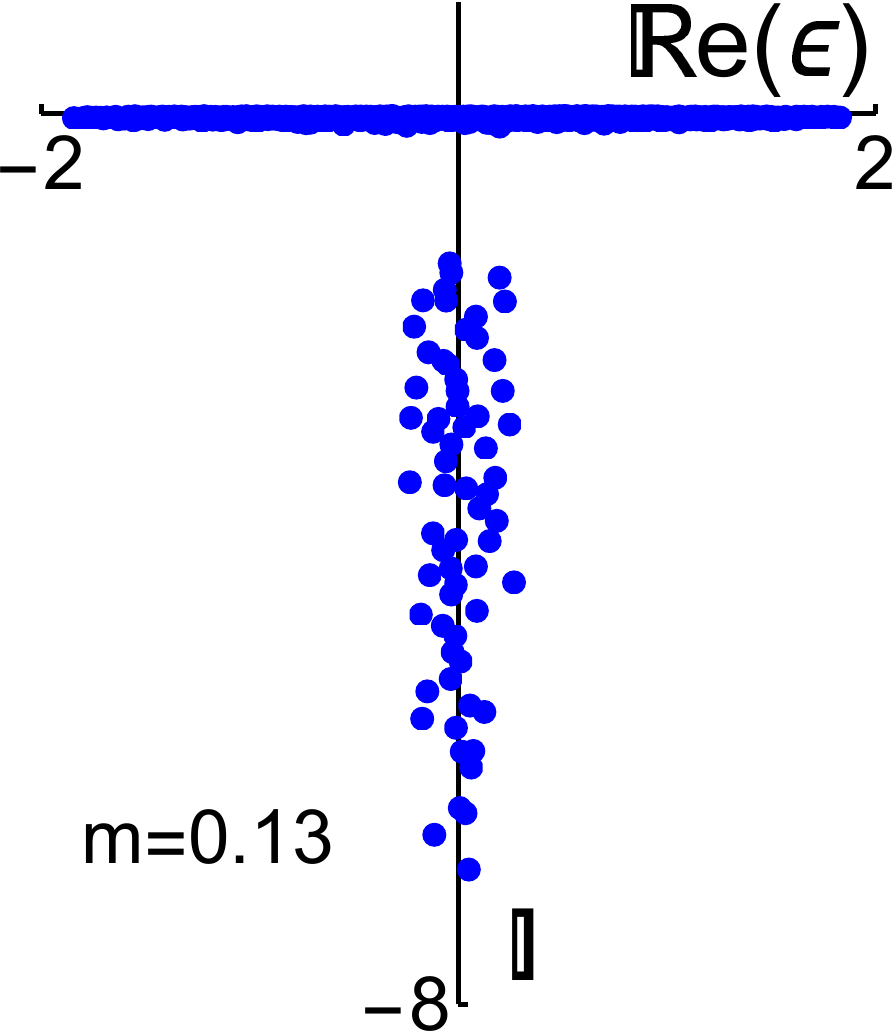}}
\scalebox{.36}{\includegraphics{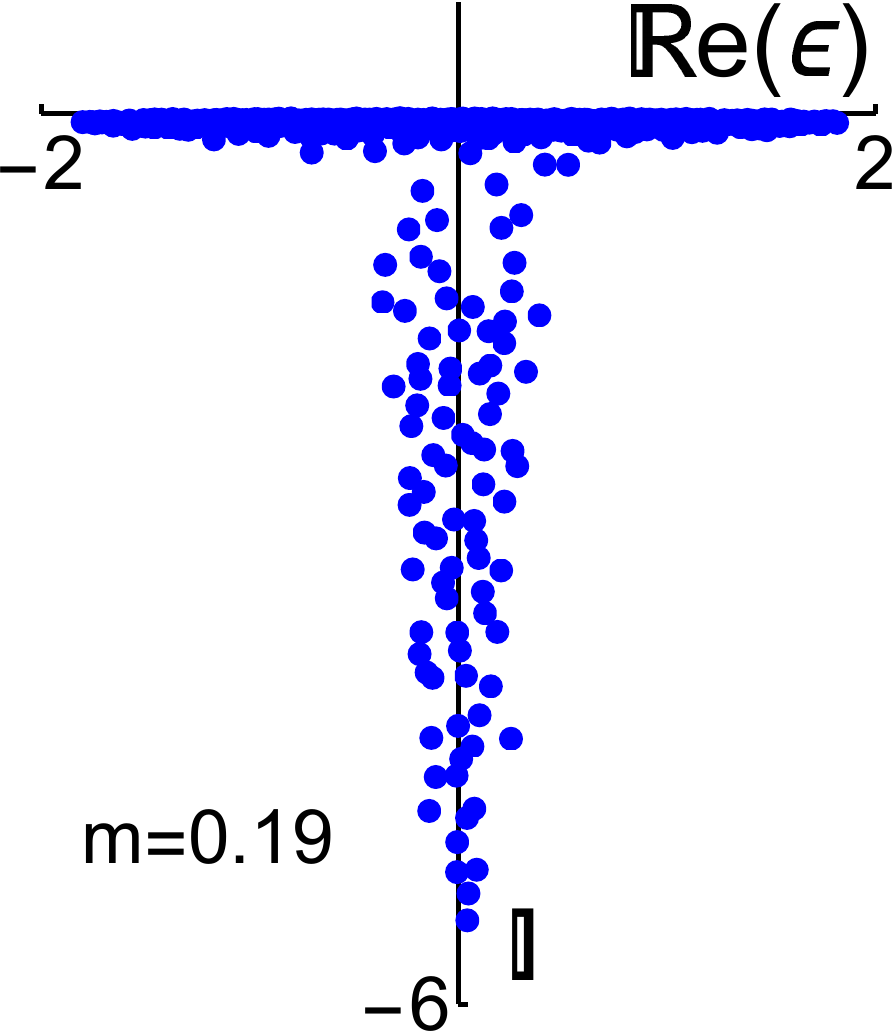}}
\scalebox{.36}{\includegraphics{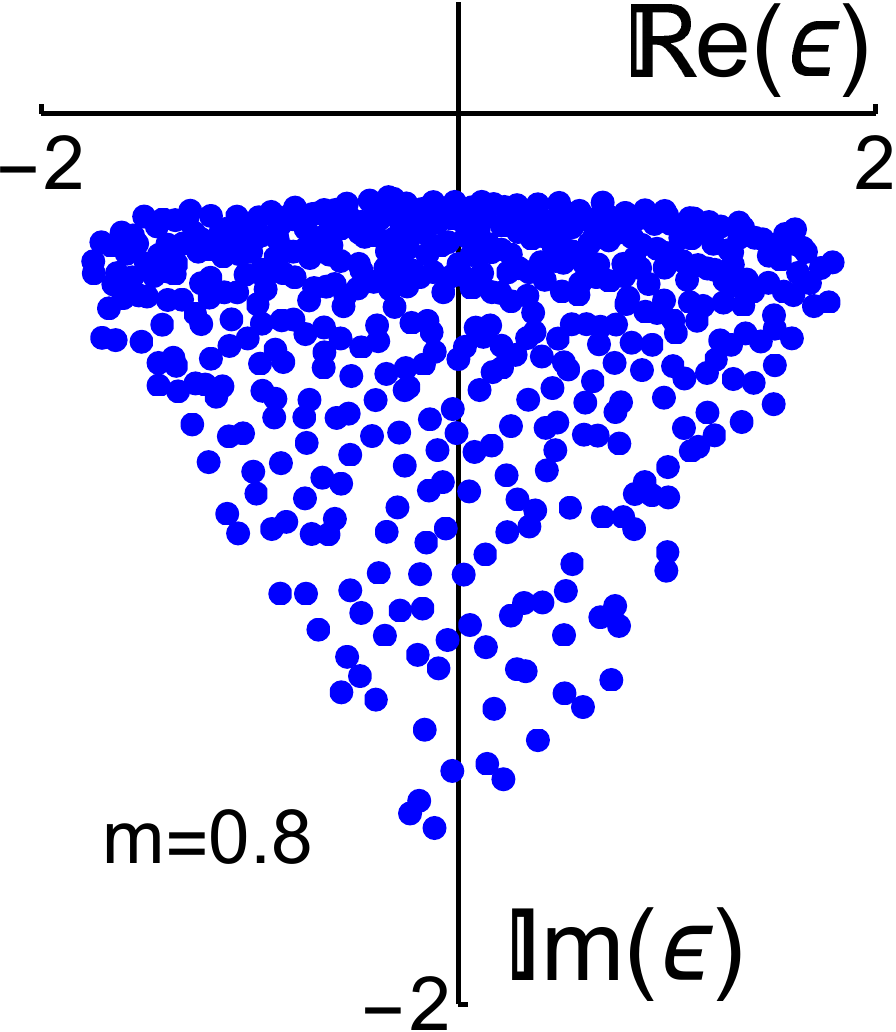}}
\scalebox{.36}{\includegraphics{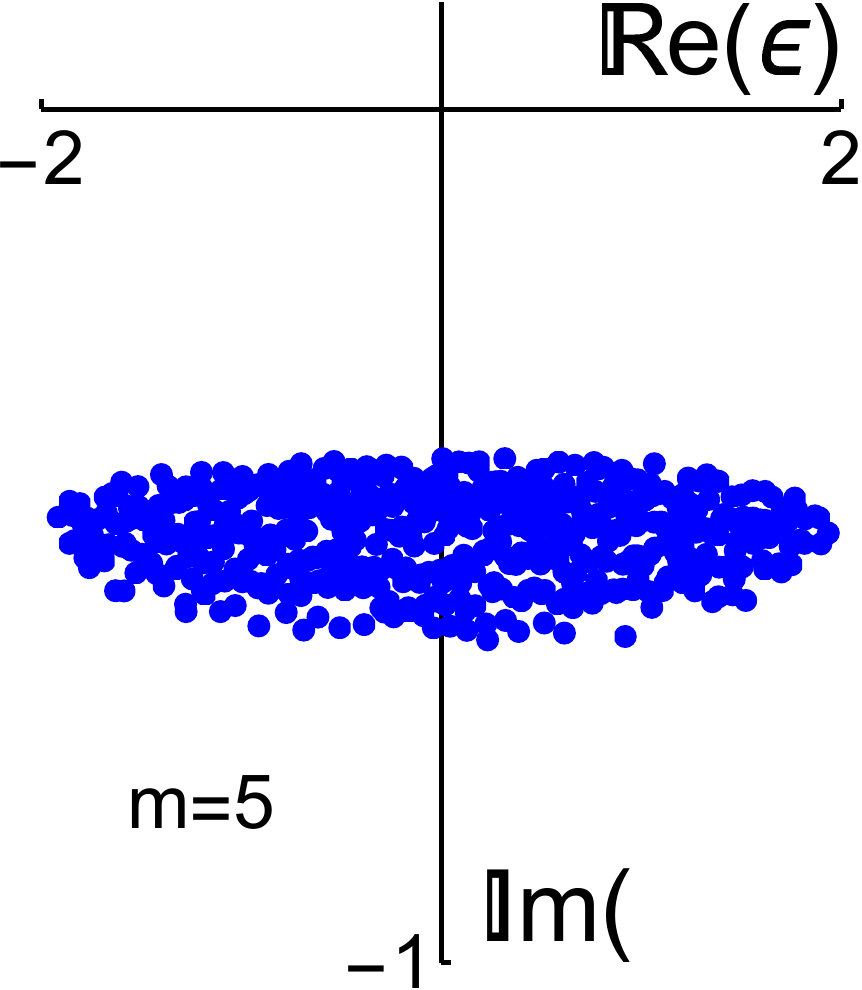}}
\caption{Example spectra of $H$ for different values of $m$, with $N=500$ and $\gamma=1$ and $\tilde\gamma=0$.  The leftmost plot depicts the small $m$ phase in which there are two disconnected connected components of $\varrho_H(\epsilon)$.  Note that in the small $m$ limit, one of the two regions of eigenvalues is very close to real axis, but still possesses a finite imaginary part.  The second to the left is near the critical $m$ (which generically differs from 1 with a Hamiltonian term), in which the two regions are starting to merge.  The two on the right show the large $m$ phase, for which $\varrho_H(\epsilon)$ possesses a single connected component.  The rightmost plot shows the large $m$ limit, in which the limiting shape of the support of $\varrho_H(\epsilon)$ can be seen to approach a Ginibre disk centered around $1/2$ on the imaginary axis.  This limit matches the known behaviour for random matrix Lindbladians with both Hamiltonian and jump operators \cite{RandomLindblad2}.}
\label{SpectrumLoss}
\end{center}
\end{figure}

Adding in the Hamiltonian part but keeping $\nu=0$, the Marchenko-Pastur distribution is deformed into the complex plane.  The distribution $\varrho_H$ is supported on compact subsets of the complex $\epsilon$ plane with a finite area.  For large enough $\gamma$, the large and small $m$ phases deform into phases in which the support of $\varrho_H(\epsilon)$ has one and two connected components respectively. Example spectra in the different phases are shown in Fig.~(\ref{SpectrumLoss}).
The shapes of the eigenvalue distribution can be determined for large $N$ using non-Hermitian random matrix theory.  Similar problems have been studied in the context of chaotic scattering \cite{Scattering1,Scattering2,Scattering3}; for a more detailed discussion of the shape of the distribution for the Lindbladian problem one may refer to \cite{RandomQuadratics}.
With the Hamiltonian part, all eigenvalues now have a finite imaginary part for all $m$.  In contrast to the above situation without a Hamiltonian discussed above, with a Hamiltonian the dissipative gap becomes finite even for small $m$.  As a consequence, the stationary state is always unique.  It is easy to check that the stationary state is still the zero particle Fock vacuum for all values of $m$.

\begin{figure}
\begin{center}
\scalebox{.08}{\includegraphics{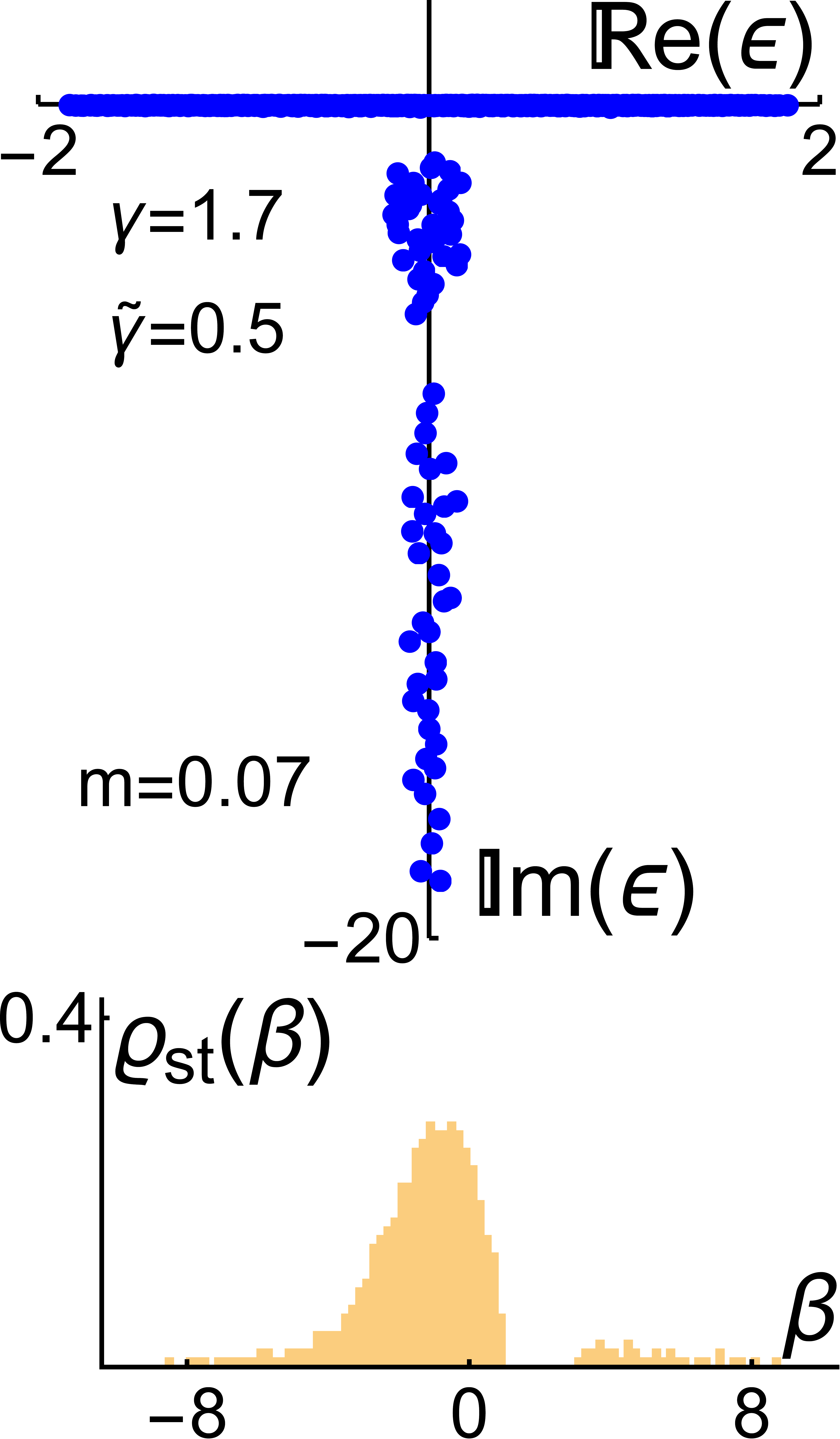}}
\scalebox{.08}{\includegraphics{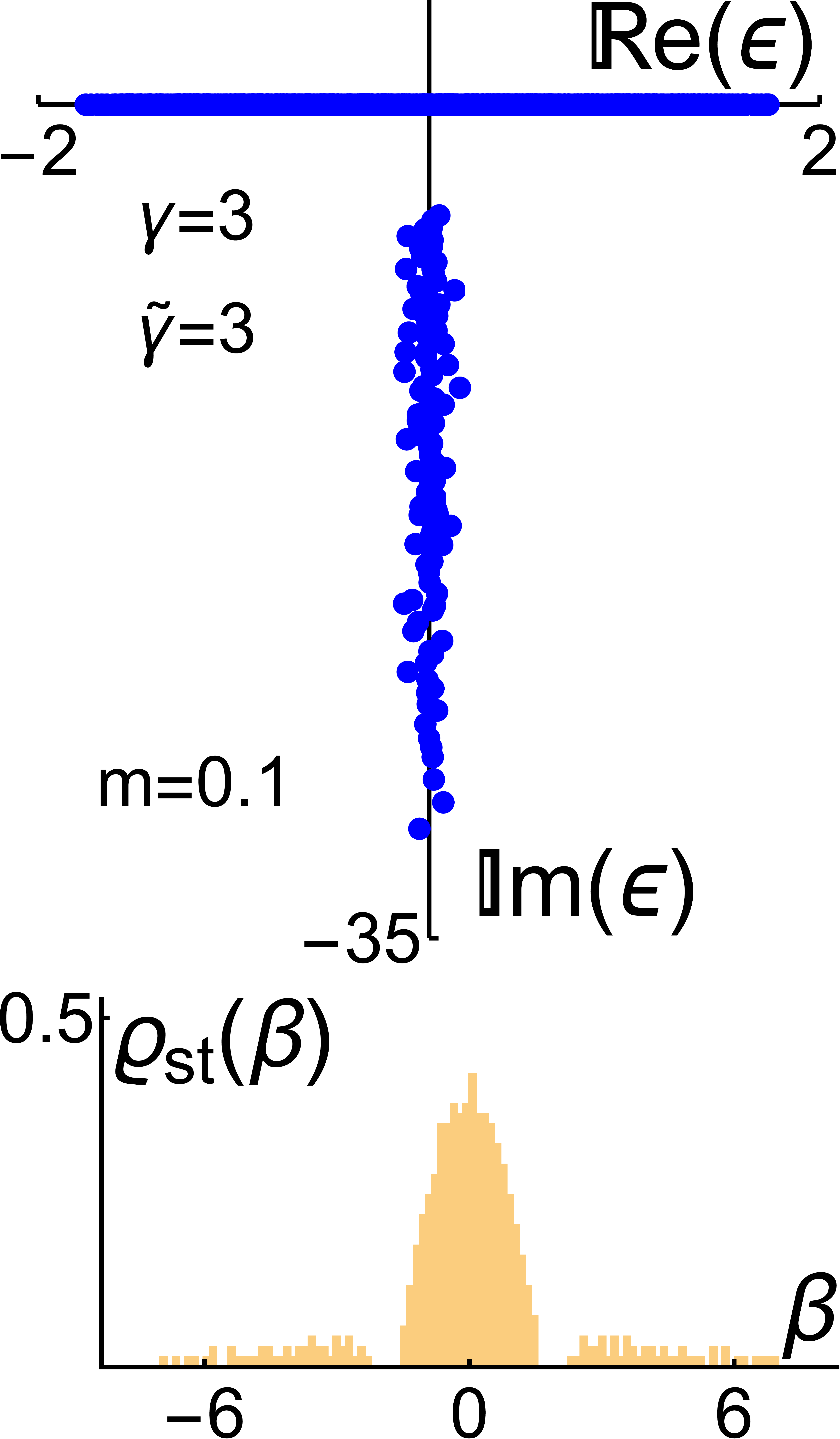}}
\scalebox{.08}{\includegraphics{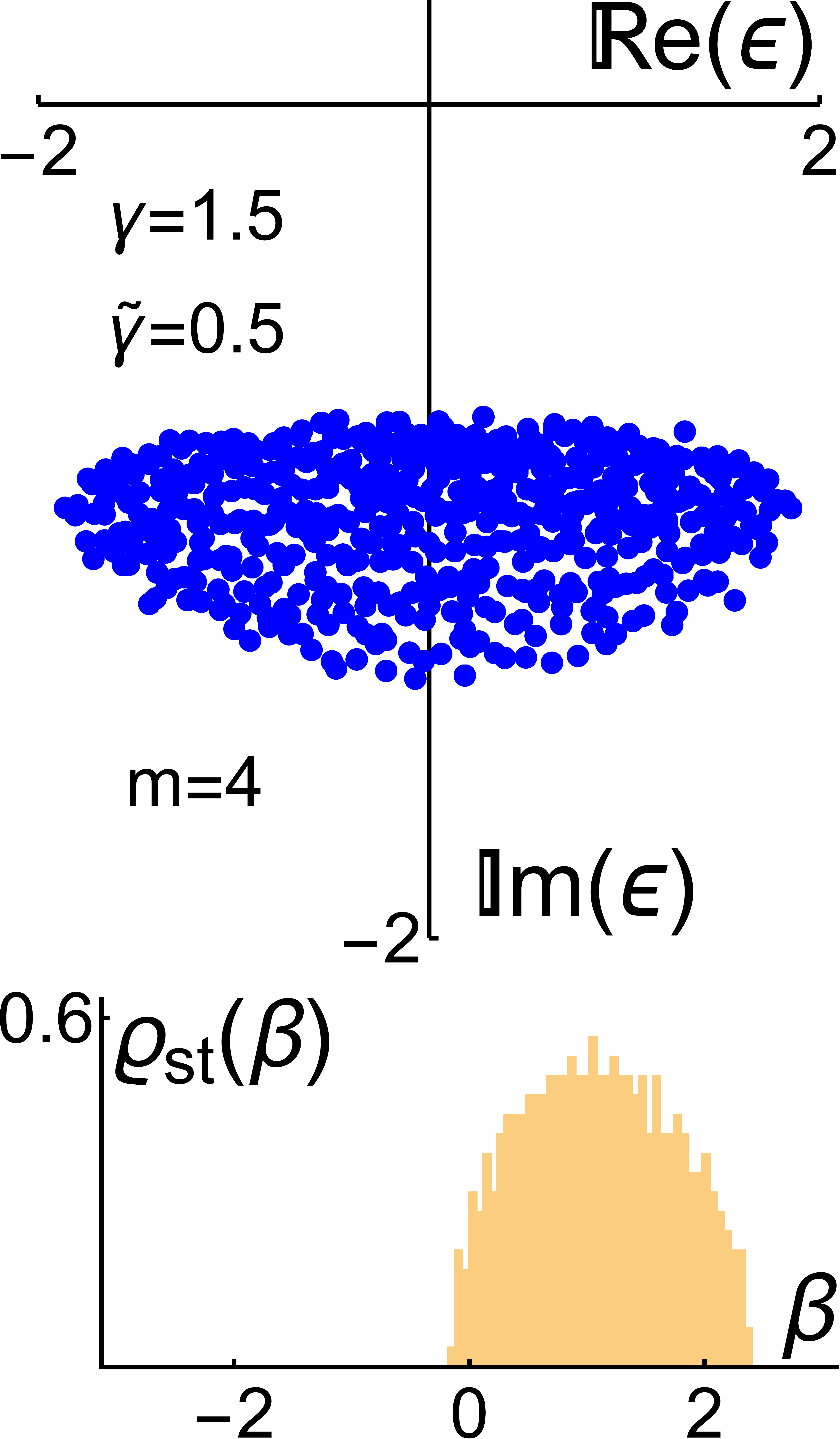}}
\caption{Example spectra of $H$ plotted above the corresponding $\varrho_\mathrm{st}(\beta)$ for finite $\gamma$ and $\tilde\gamma$, with $N=500$.  The leftmost plot depicts a phase with in which the support of $\varrho_H(\epsilon)$ has three connected components, which occurs when $m$ is small and gain is meaningfully smaller or larger than loss.  In this phase, $\varrho_\mathrm{st}(\beta)$ is supported on two disconnected regions, with the density mostly to the left of the origin.  The middle plot shows a two-component phase with large balanced loss and gain.  In this regime, $\varrho_\mathrm{st}(\beta)$ is supported on three disconnected regions, with most measure centered at zero.  In the rightmost plot, $m$ is large and $\varrho_H(\epsilon)$ is approaching the Ginibre disk limit.  The stationary state distribution is nearly a Wigner semicircle with a positive finite mean, corresponding to the fact that $\gamma>\tilde\gamma$.}
\label{SpectrumLossGain}
\end{center}
\end{figure}

Turning finally to the general model with gain and loss, one finds similar geometric transitions depending on the value of $m$.  There are multiple different phases with a maximum of three connected components, which can merge and split in various ways depending on the relative strengths of gain vs. loss vs. Hamiltonian.  The stationary state is generically a non-trivial mixed state.  For large $m$, $\varrho_\mathrm{st}(\beta)$ is a Wigner semicircle, with a width that scales with $1/m$ and is off-centered from zero by an amount determined by $\tilde\gamma/\gamma$.  For smaller $m$, $\varrho_\mathrm{st}(\beta)$ has support on multiple disconnected regions of the real line, each of which deviates from the semi-circle law.  A detailed classification of the various phases is not presented here, but several different examples of $\varrho_M$ and the corresponding $\varrho_\mathrm{st}(\beta)$ are depicted in Fig.~(\ref{SpectrumLossGain}).

\subsection{Lindbladian Gas}\label{3.4}
This section discusses a Fermi gas in $d$ dimensions subject to loss and gain of particles through Markovian exchange with a thermal bath.  This provides a simple example of a theory on a spatial continuum.  Like the preceding section, here a $U(1)$ symmetry is imposed to avoid complexity from the Nambu space.  For simplicity also, no additional matrix structure due to spin, orbitals, flavor, etc. is considered.  Note that even though a fermionic gas is considered here, because of the symmetry, differences between the bosonic and fermionic Nambu spaces do not enter and so the details are essentially the same for the Lindbladian Bose gas.

In terms of the fermionic creation/annihilation operators $\hat c(\mathbf{r})$ and $\hat c^\dagger(\mathbf{r})$, the many-body Hamiltonian can be expressed in terms of the single-particle Hamiltonian $H_0(\mathbf{r},\mathbf{r'})$ as:
\begin{equation}\label{HamGas}
\hat\fH=\int\!\dif\mathbf{r}\,\dif\mathbf{r'}\, \hat c^\dagger(\mathbf{r})H_0(\mathbf{r},\mathbf{r'})\hat c(\mathbf{r'}).
\end{equation}
There are two families of jump operators:
\begin{equation}\label{JumpGas}
\hat\fL_{1v}(\mathbf{r})=\int \dif\mathbf{r'}\, \mu_v(\mathbf{r},\mathbf{r'})\hat c(\mathbf{r'}),\quad\hat\fL_{2v}(\mathbf{r})=\int \dif\mathbf{r'}\, \nu_v(\mathbf{r},\mathbf{r'})\hat c^\dagger(\mathbf{r'}),
\end{equation}
in terms of which the single-particle dissipation matrices are given by:
\begin{subequations}
\begin{equation}
Q(\mathbf{r},\mathbf{r'})=\frac{1}{2}\sum_v\int\dif\mathbf{r_1}\dif\mathbf{r_2}\Big(\mu_v^*(\mathbf{r},\mathbf{r_1})\mu_v(\mathbf{r'},\mathbf{r_2})+\nu_v(\mathbf{r},\mathbf{r_1})\nu_v^*(\mathbf{r'},\mathbf{r_2})\Big),
\end{equation}
\begin{equation}
D(\mathbf{r},\mathbf{r'})=\sum_v\int\dif\mathbf{r_1}\dif\mathbf{r_2}\Big(\mu_v^*(\mathbf{r},\mathbf{r_1})\mu_v(\mathbf{r'},\mathbf{r_2})-\nu_v(\mathbf{r},\mathbf{r_1})\nu_v^*(\mathbf{r'},\mathbf{r_2})\Big).
\end{equation}
\end{subequations}
The action for the Lindbladian Bose gas is given by:
\begin{equation}
S=\int\dif t\,\dif\mathbf{r}\,\dif\mathbf{r'}\begin{bmatrix}\bar\psi^1&\bar\psi^2\end{bmatrix}_\mathbf{r}\begin{bmatrix}i\partial_t-H_0+iQ&iD\\0&i\partial_t-H_0-iQ\end{bmatrix}_{\mathbf{r},\mathbf{r'}}\begin{bmatrix}\psi^1\\\psi^2\end{bmatrix}_\mathbf{r'}.
\end{equation}
In the simplest situation, the single-particle matrices are local differential operators, so that $H_0(\mathbf{r},\mathbf{r'})=\delta(\mathbf{r}-\mathbf{r'})H_0(\mathbf{r},-i\partial_\mathbf{r})$ and similarly for the dissipative matrices.  When this is the case, the action is local in spacetime,
\begin{equation}\label{ActionGas}
S=\int\dif x\begin{bmatrix}\bar\psi^1(x)&\bar\psi^2(x)\end{bmatrix}\begin{bmatrix}i\partial_t-H_0+iQ&iD\\0&i\partial_t-H_0-iQ\end{bmatrix}\begin{bmatrix}\psi^1(x)\\\psi^2(x)\end{bmatrix}.
\end{equation}
where $x=(t,\mathbf{r})$ denotes a spacetime coordinate.

One obtains a semi-classical kinetic equation by Wigner transform and truncation of the gradient expansion of matrix products.  Upon Wigner transform, the parameter matrices become functions on the single-particle phase space,
\begin{equation}
H_0(\mathbf{r},\mathbf{k})=\int\dif\mathbf{r'}e^{-i\mathbf{k}\mathbf{r'}}H_0\bigg(\mathbf{r}+\frac{\mathbf{r'}}{2},\mathbf{r}-\frac{\mathbf{r'}}{2}\bigg),
\end{equation}
and similarly for $Q(\mathbf{r},\mathbf{k})$ and $D(\mathbf{r},\mathbf{k})$.  In the limit of slow variations, it is often appropriate to truncate the Wigner expansion to first order in gradients.
This amounts to a semiclassical treatment of the problem and is known as the Wigner approximation.  In this approximation, kinetic equation becomes a Boltzmann equation with a linear collision integral,
\begin{equation}\label{Boltzmann}
\partial_tF-\partial_\mathbf{r}H_0\partial_\mathbf{k}F+\partial_\mathbf{k}H_0\partial_\mathbf{r}F=-2QF+D.
\end{equation}
Note that in a bosonic system, this kinetic equation will be of exactly the same form.  Going beyond this approximation by incorporating higher-order terms, one finds the leading quantum corrections to the collision integral,
\begin{equation}
-\frac{1}{2}\partial_\mathbf{r_1}\partial_\mathbf{r_2}Q\partial_\mathbf{k_1}\partial_\mathbf{k_2}F-\frac{1}{2}\partial_\mathbf{k_1}\partial_\mathbf{k_2}Q\partial_\mathbf{r_1}\partial_\mathbf{r_2}F+\partial_\mathbf{r_1}\partial_\mathbf{k_2}Q\partial_\mathbf{k_1}\partial_\mathbf{r_2}F.
\end{equation}
The inclusion of these terms brings the Boltzmann equation to the form of a Fokker-Plank equation.  This contrasts to the purely coherent situation, in which there are no leading quantum corrections to the Boltzmann equation in the absence of multiple bands.

For concreteness, specialize to a local single-particle Hamiltonian that is a generalized Schr\"odinger operator, $H_0(\mathbf{r},-i\partial_\mathbf{r})=\varepsilon(-i\partial_\mathbf{r})+V(\mathbf{r})$.  Then the phase space function $H_0(\mathbf{r},\mathbf{k})=\varepsilon(\mathbf{k})+V(\mathbf{r})$ will be of the form of a dispersion relation plus a potential.  In addition, suppose that the coupling to bath is translationally invariant, so that the dissipation matrices are only functions of $\mathbf{k}$ in phase space.  Then the kinetic equation is identifiable as a Boltzmann equation in the relaxation time approximation,
\begin{equation}\label{BoltzmannRelaxationTime}
\big(\partial_t+\mathbf{v}_\mathbf{k}\partial_\mathbf{r}-(\partial_\mathbf{r}V)\partial_\mathbf{k}\big)F=\frac{F_0-F}{\tau},
\end{equation}
where $\mathbf{v}_\mathbf{k}=\partial_\mathbf{k}\varepsilon$ is the group velocity and $\tau(\mathbf{k})=1/2Q(\mathbf{k})$ is the relaxation time.  The distribution $F_0(\mathbf{k})=D(\mathbf{k})/2Q(\mathbf{k})$ takes the place of the equilibrium distribution; in the absence of an external potential $V=0$, one finds $F_\mathrm{st}(\mathbf{k})=F_0(\mathbf{k})$.

\subsection{Mean Field Theory}
This section illustrates how the above formalism can be extended to treat non-linear systems.  Using a mean-field approach, the semiclassical kinetics of the previous section can be extended to self-consistently accommodate interactions.  For specificity, bosonic systems will be focused on, though as with the previous section the details are similar for the fermionic analogue.

Like the previous section, no additional matrix structure is considered beyond the spatial degrees of freedom.  Only terms respecting the $U(1)$ particle number symmetry are considered, so that no Nambu space structure has to be dealt with.  Additionally, the Hamiltonian and jump operators are assumed to be invariant under spatial translations and rotations.  The non-interacting Hamiltonian and jump operators are given by the bosonic versions of eq.s~(\ref{HamGas}) and (\ref{JumpGas}).  Because of translational symmetry, in the Wigner representation $H_0,\ Q$ and $D$ are only functions of $\mathbf k$,
\begin{subequations}
\begin{equation}
Q(\mathbf k)=\frac{1}{2}\sum_v\Big(\mu_v^*(\mathbf k)\mu_v(\mathbf k)-\nu_v(\mathbf k)\nu^*_v(\mathbf k)\Big),
\end{equation}
\begin{equation}
D(\mathbf k)=\sum_v\Big(\mu_v^*(\mathbf k)\mu_v(\mathbf k)+\nu_v(\mathbf k)\nu^*_v(\mathbf k)\Big).
\end{equation}
\end{subequations}
Non-linearity can occur either on the level of the Hamiltonian or in the jump operators.  For simplicity, consider a contact interaction,
\begin{equation}
\hat\fH_\mathrm{int}=\frac{U}{2}\int\dif\mathbf{r}\,\hat a^\dagger(\mathbf{r})\hat a^\dagger(\mathbf{r})\hat a(\mathbf{r})\hat a(\mathbf{r}).
\end{equation}
For the jump operators, consider a local two-body loss and gain,
\begin{equation}
\hat\fL_3(\mathbf{r})=\frac{M}{\sqrt2}\, \hat a(\mathbf{r})\hat a(\mathbf{r}),\qquad\hat\fL_4(\mathbf{r})=\frac{N}{\sqrt2}\, \hat a^\dagger(\mathbf{r})\hat a^\dagger(\mathbf{r}).
\end{equation}
Note that the corresponding Lindbladian defined this way is similar to various models for driven-dissipative condensates \cite{DiehlKeldyshLindblad,BEC,BEC2,BEC3}.  Here the nonlinear interactions are consider only as a weak perturbation to a stable linear theory.  A condensate occurs when the theory is unstable on the linear level and is not treated here.

\begin{figure}
\begin{center}
\scalebox{.15}{\includegraphics{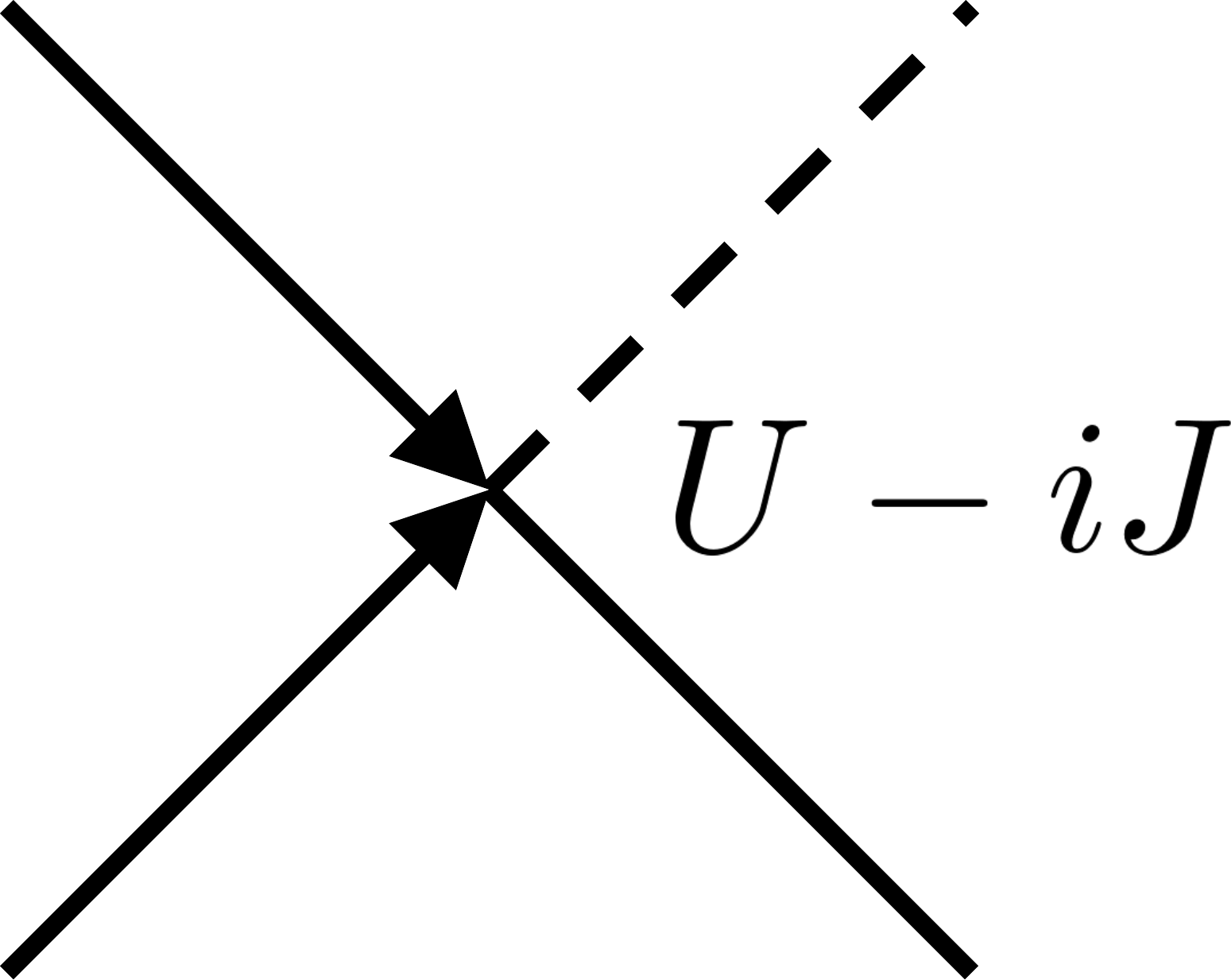}}\quad\qquad
\scalebox{.15}{\includegraphics{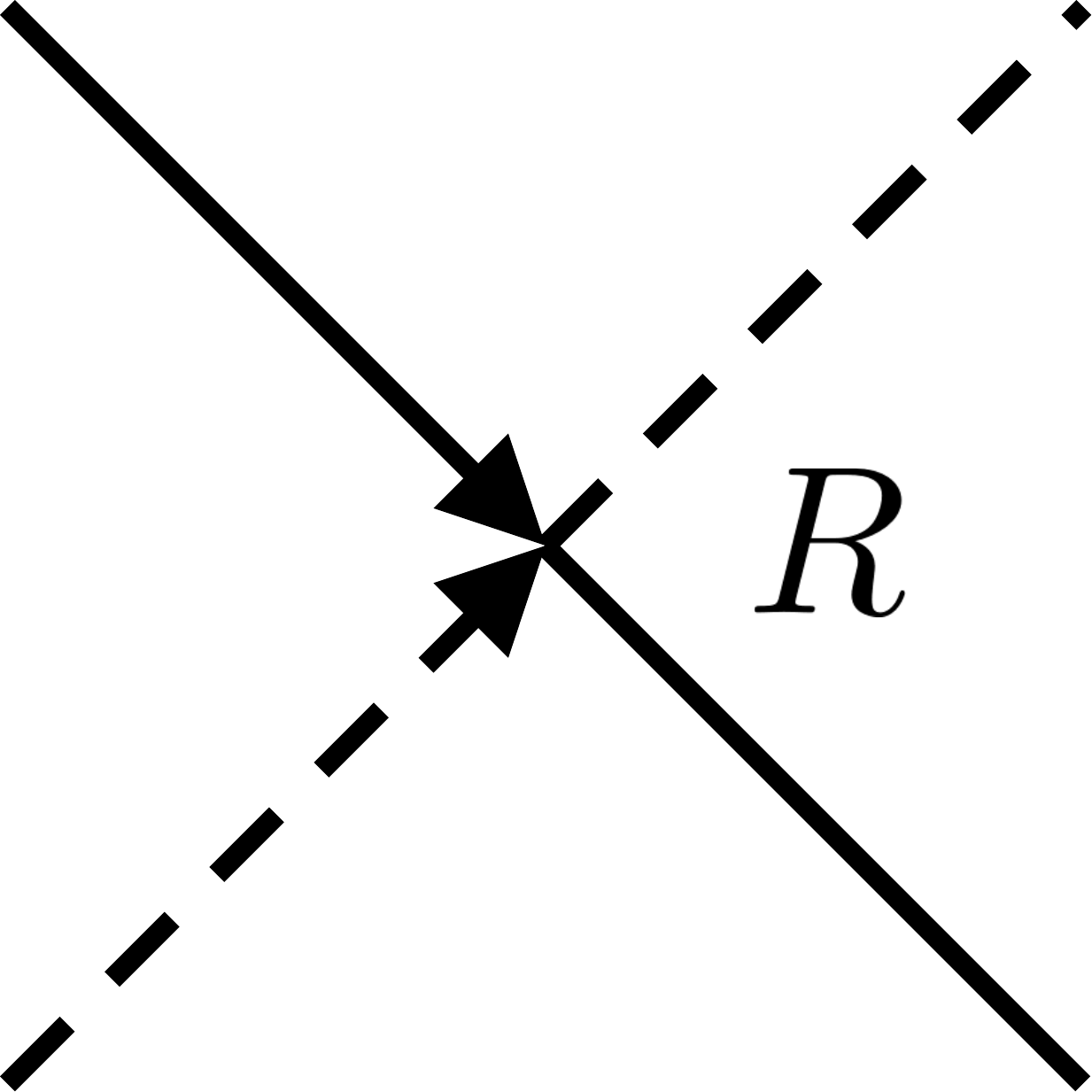}}\quad\qquad
\scalebox{.15}{\includegraphics{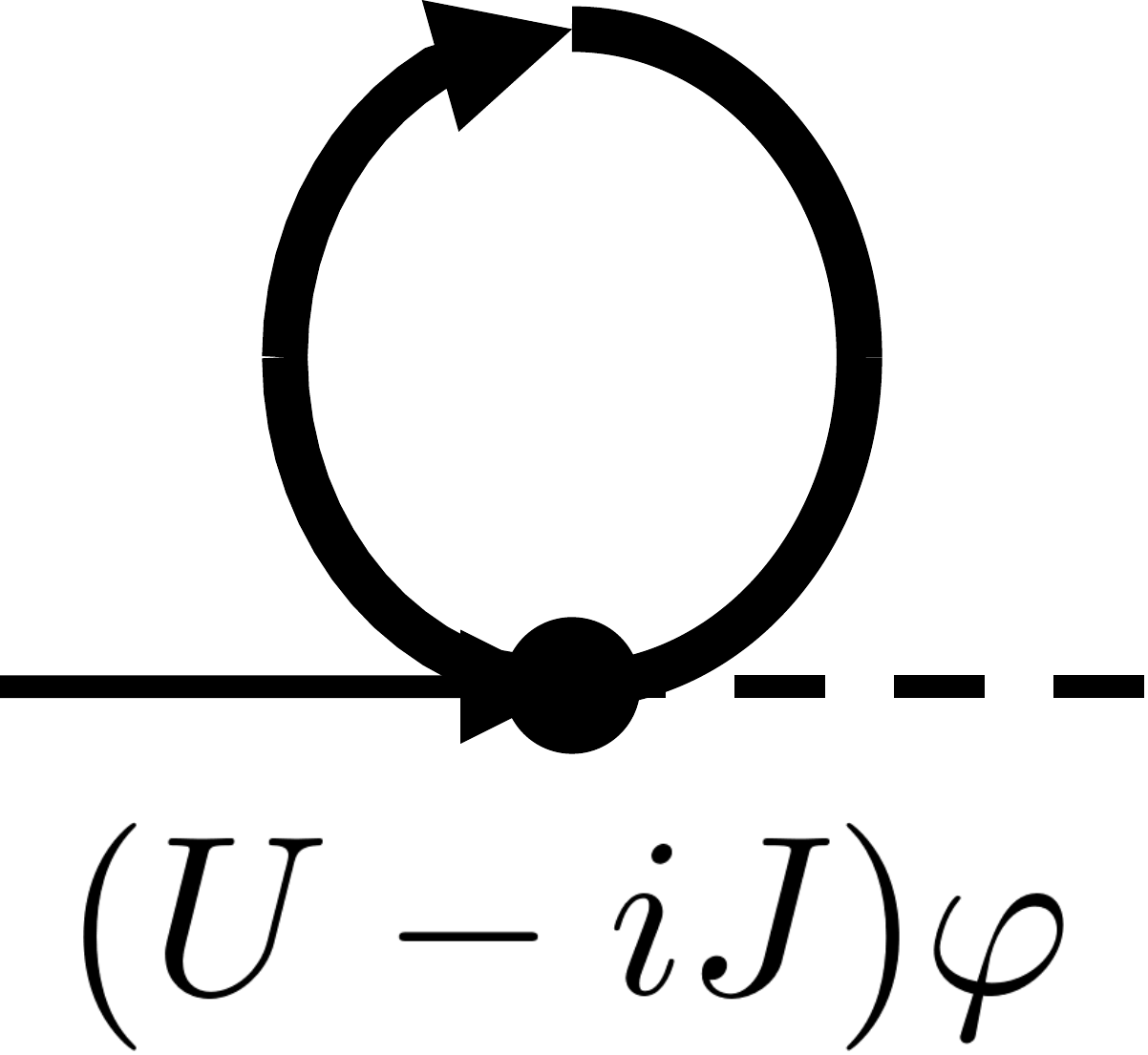}}\quad\qquad
\scalebox{.15}{\includegraphics{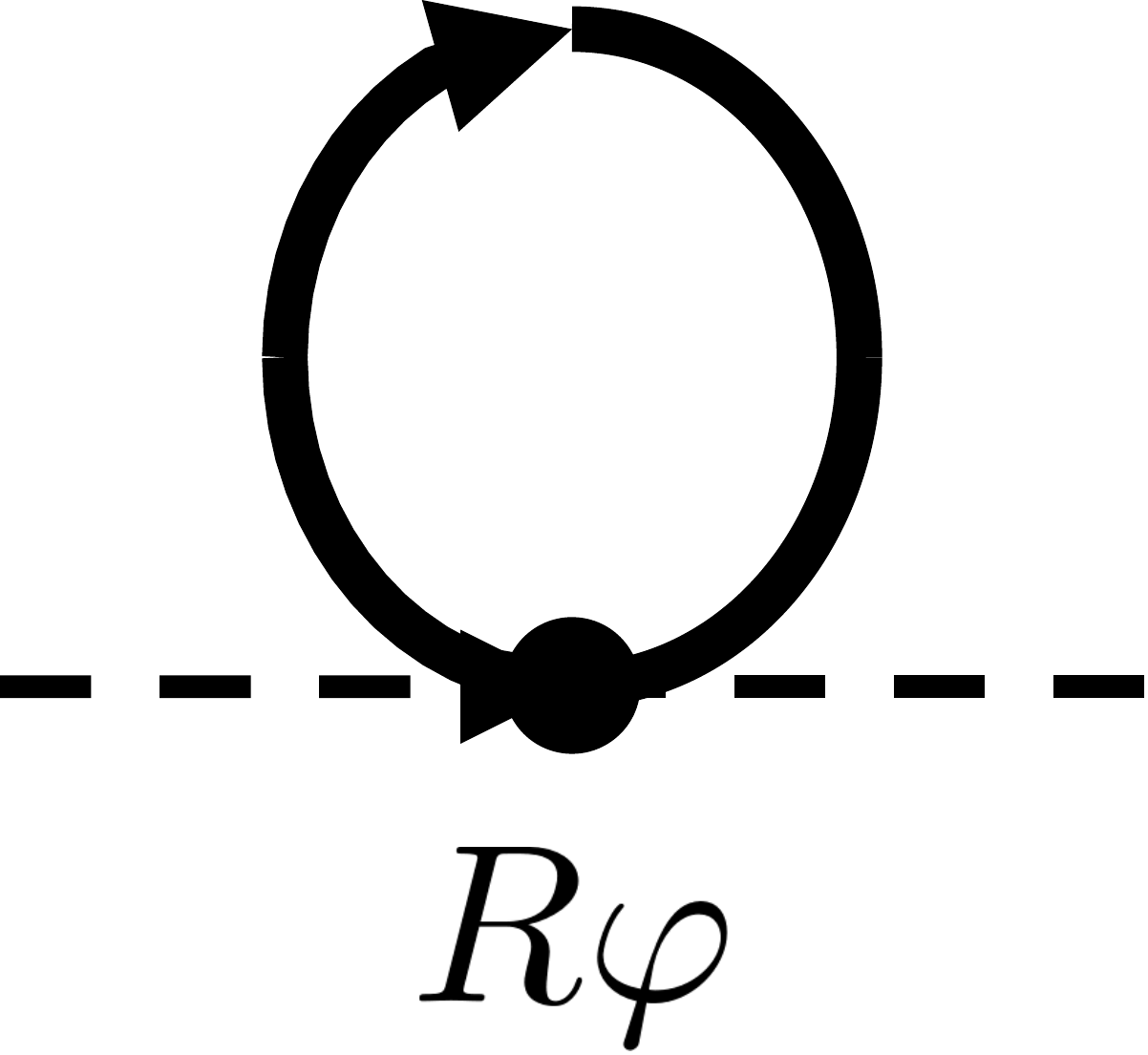}}\quad\qquad
\caption{The left two figures depict the vertices depicting the interactions in the nonlinear action eq.~(\ref{ActionNonlinear}).  The leftmost is comparable to one of the vertices that appears in the equilibrium theory, but come with additional imaginary factor.  The $R$ vertex has no equilibrium analogue and is purely dissipative.  The left two figures are diagrammatic representations of the mean-field contributions to the action from the interactions.  The collective field $\varphi$ enters as a fully renormalized bubble formed from two classical legs.  The left and right respectively renormalize the single-particle dispersion and distribution.  Note that the latter is uniquely a feature of the Lindbladian dynamics; in equilibrium theory there is no modification to the Keldysh component of eq.~(\ref{ActionNonlinear}) on the mean-field level.}
\label{Vertices}
\end{center}
\end{figure}

The Keldysh action has the form $S=S_0+S_\mathrm{int}$, where quadratic part of the action $S_0$ is given by the bosonic version of eq.~(\ref{ActionGas}),
\begin{equation}\label{ActionGasBose}
S_0=\int\dif x\begin{bmatrix}\bar\phi^\cl(x)&\bar\phi^\q(x)\end{bmatrix}\begin{bmatrix}0&i\partial_t-H_0-iQ\\i\partial_t-H_0+iQ&iD\end{bmatrix}\begin{bmatrix}\phi^\cl(x)\\\phi^\q(x)\end{bmatrix}.
\end{equation}
The non-linear part of the action is:
\begin{multline}\label{ActionNonlinear}
S_{\mathrm {int}}=-\frac{1}{2}\int\dif x\Big(U(\bar\phi^\cl\phi^\cl+\bar\phi^\q\phi^\q)(\bar\phi^\q\phi^\cl+\bar\phi^\cl\phi^\q)\\
-iJ(\bar\phi^\cl\phi^\cl-\bar\phi^\q\phi^\q)(\bar\phi^\q\phi^\cl-\bar\phi^\cl\phi^\q)-iR\bar\phi^\cl\phi^\cl\bar\phi^\q\phi^\q\Big),
\end{multline}
where $J$ and $R$ are defined similarly to $Q$ and $D$,
\begin{equation}
J=\frac{1}{2}\big(|N|^2-|M|^2\big),\qquad R=2\big(|N|^2+|M|^2\big)
\end{equation}
Diagrammatically, these interactions are four-point vertices as represented in Fig.~(\ref{Vertices}).

While in principle the full range of diagrammatic techniques can be applied to study this model, here a mean-field treatment is discussed as an extension of the quadratic formalism.  To achieve this, one should replace factors of $\bar\phi^\cl\phi^\cl$ in the action with their expectation value.  This reduces the nonlinearities to a quadratic coupling to the collective field $\varphi(x)=\frac{1}{2}\braket{\phi^\cl(x)\bar\phi^\cl(x)}$ whose value can be determined self-consistently from this definition.  To be specific, all three of the parameter matrices are modified,
\begin{equation}
\delta H_0(x)=U\varphi(x),\qquad\delta Q(x)=J\varphi(x),\qquad\delta D(x)=R\varphi(x).
\end{equation}
Following section \ref{2.4}, one can seek a solution to the now time-dependent kinetic equation eq.~(\ref{KinEqBoseRed}).  In the Wigner approximation, this will be of the same form as eq.~(\ref{BoltzmannRelaxationTime}),
\begin{equation}\label{BoltzmannMeanField}
\big(\partial_t+\mathbf{v}_\mathbf{k}\partial_\mathbf{r}-(\partial_\mathbf{r}V)\partial_\mathbf{k}\big)F=\frac{F_0-F}{\tau},
\end{equation}
with $\mathbf{v}_\mathbf{k}=\partial_\mathbf{k}H_0(\mathbf{k})$, $V(x)=U\varphi(x)$, $\tau^{-1}=2Q(\mathbf{k})+2J\varphi(x)$ and $F_0/\tau=D(\mathbf{k})+R\varphi(x)$.

Solutions to the kinetic equation determine $F$ as a function of $\varphi$.  This in turn can be fed into the definition of $\varphi$ to determine its value self-consistently.  To be precise, in the Wigner approximation one can express the spectral Green's function as:
\begin{equation}
G^\tR(x,p)\simeq\frac{1}{\epsilon-H_0(\mathbf k)+iQ(\mathbf k)-(U-iJ)\varphi(x)},
\end{equation}
where $p=(\epsilon,\mathbf k)$.  Comparably, the Keldysh Green's function at equal spacetime points is given approximately by:
\begin{equation}
iG^\tK(x,x)\simeq i\int\frac{\dif\epsilon}{2\pi}\frac{\dif\mathbf k}{(2\pi)^d}F(x,\mathbf k)\big(G^\tR(x,p)-G^\tA(x,p)\big).
\end{equation}
The integral over $\epsilon$ is just the frequency integral of the spectral function and is thus equal to $1$.  Note that unlike in the equilibrium theory, there is no assumption that the spectral function is sharply peaked.  A quasi-particle approximation is generally not valid, as for general Lindbladian systems due to the presence of dissipative terms which are not generically small.  Despite this, the frequency dependence of the distribution function may anyways be dropped in this mean-field treatment due to the Markovian nature of the dynamics.  This gives the self-consistency condition for the collective field $\varphi(x)=\frac{i}{2}G^\tK(x,x)$,
\begin{equation}
\varphi(x)=\frac{1}{2}\int\frac{\dif\mathbf k}{(2\pi)^d}F(x,\mathbf k).
\end{equation}
This together with the Boltzmann equation eq.~(\ref{BoltzmannMeanField}) constitute a closed system of two equations for $F$ and $\varphi$.

In the stationary limit, $F_\mathrm{st}$ and $\varphi_\mathrm{st}$ are fully isotropic.  The stationary solution can be read off from the Boltzmann equation as $F_0$.  Together with the self-consistency condition, this gives the set of equations:
\begin{equation}
F_\mathrm{st}(\mathbf k)=\frac{D(\mathbf k)+R\varphi_\mathrm{st}}{2Q(\mathbf k)+2J\varphi_\mathrm{st}},\qquad\varphi_\mathrm{st}=\frac{1}{2}\int\frac{\dif\mathbf k}{(2\pi)^d}F_\mathrm{st}(\mathbf k).
\end{equation}
Slow relaxation can be studied by linearizing the kinetic equation around this stationary solution.  Writing $F=F_\mathrm{st}+\delta F$ and $\varphi=\varphi_\mathrm{st}+\delta\varphi$, one has the closed system of equations:
\begin{subequations}
\begin{equation}
\big(\partial_t+\mathbf{v}_\mathbf{k}\partial_\mathbf{r}+2Q+2J\varphi_\mathrm{st}\big)\delta F=\partial_\mathbf{r}\delta\varphi\partial_\mathbf{k}F_\mathrm{st}+(R-2JF_\mathrm{st})\delta\varphi,
\end{equation}
\begin{equation}
\delta\varphi(x)=\frac{1}{2}\int\frac{\dif\mathbf k}{(2\pi)^d}\delta F(x,\mathbf k).
\end{equation}
\end{subequations}
By Fourier transforming in the spacetime coordinate $x=(t,\mathbf r)$ to $(\omega,\mathbf q)$, one may algebraically solve for $\delta F$.  In doing so, the self-consistency condition gives the condition for a non-trivial solution $\varphi$,
\begin{equation}\label{SelfConsistentDispersion}
1+\frac{1}{2}\int\frac{\dif\mathbf k}{(2\pi)^d}\frac{U\mathbf q\partial_\mathbf{k}F_\mathrm{st}-i(R-2JF_\mathrm{st})}{\omega-\mathbf q\mathbf{v}_\mathbf{k}+2i(Q+J\varphi_\mathrm{st})}=0.
\end{equation}
This relation fixes $\omega$ as a function of $\mathbf q$, which specifies the dispersion of collective modes which govern the relaxation the density field $\varphi$.

In the absence of dissipation $Q=0=D$ and $J=0=R$, this equation determines the dispersion of a coherent sound mode $\omega(\mathbf q)\simeq c|\mathbf q|$ at small momenta, where the speed of sound $c$ is determined from the relation
\begin{equation}\label{cSound}
1+\frac{U}{2}\int\frac{\dif\mathbf k}{(2\pi)^d}\frac{\mathbf q\partial_\mathbf{k}F_\mathrm{st}}{c|\mathbf q|-\mathbf q\mathbf{v}_\mathbf{k}}=0,
\end{equation}
where in the equilibrium theory $F_\mathrm{st}$ is the equilibrium quasi-particle distribution function.  This is the familiar equilibrium zero-sound mode.  To examine the how this is modified due to the presence of dissipation, consider first the simpler situation of a weak purely linear dissipation that is independent of $\mathbf k$, so that $R=0=J$ and $Q(\mathbf k)=1/2\tau$.  Then for small $\mathbf q$ one finds,
\begin{equation}
\omega(\mathbf q)\simeq-i/\tau+c|\mathbf q|,
\end{equation}
where $c$ is the speed of sound determined from eq.~(\ref{cSound}) with $F_\mathrm{st}(\mathbf k)=\tau D(\mathbf k)$.  Thus, for momenta small compared to the inverse relaxation time, $|\mathbf q|\ll1/c\tau$, the sound mode becomes over-damped.

In the more generic setting with non-linear dissipation, it is difficult to make general statements without a specific form of the single-particle dispersion.  However, one can see that the above behaviour is generic for small momenta.  The zero momentum limit of eq.~(\ref{SelfConsistentDispersion}) gives the relation:
\begin{equation}
1+\frac{1}{2}\int\frac{\dif\mathbf k}{(2\pi)^d}\frac{2JF_\mathrm{st}-R}{2(Q+J\varphi_\mathrm{st})-\Gamma}=0,
\end{equation}
where $\Gamma=i\omega(\mathbf q=0)$, demonstrated by this equation to be purely real.  Thus, while the specific form of the collective mode dispersion $\omega(\mathbf q)$ for finite $\mathbf q$ depends on the microscopic details of the single-particle dispersion, it is always over-damped for sufficiently small momenta.

\section{Conclusion}

We have presented a tutorial treatment of a many-body Lindbladian dynamics of driven-dissipative systems. 
We have employed the functional formalism, which naturally follows from the generic closed time contour formalism, under the assumption of  Markovian (i.e. time-local) bath correlators.  
As demonstrated, it allows one to evaluate local observables, various correlation functions, linear response characteristics, and collective modes spectra.

One of the major goals of this review is to emphasize the existence of two distinct quantities, characterizing dynamics of these non-equilibrium models: the complex effective Hamiltonian, $\check H$, and the stationary distribution function $\check F_\mathrm{st}$.  The complex effective Hamiltonian, determines the transient relaxation spectrum as well as the linear response to certain perturbations. On the other hand,  $\check F_\mathrm{st}$ dictates steady-state observables, shows up in spectra of collective modes, and participates in some linear responses. While in equilibrium the two are rigidly related through the fluctuation-dissipation theorem, they 
are essentially independent within the Lindbladian dynamics framework. Moreover, as is repeatedly  demonstrated above,  they exhibit qualitatively different properties. For example, 
complex spectra of the effective Hamiltonian generically exhibit exceptional points, where two or more  eigenvalues collide. The relaxation characteristics feature non-analytic behavior in the vicinity of such exceptional points in the parameter space. Yet, $\check F_\mathrm{st}$ and the long time stationary properties are completely smooth.  We provide other examples illustrating qualitative differences between the two quantities in the non-equilibrium setting. 

While spectra (and to some extent eigenfunctions) of the complex Hamiltonian received some attention, the stationary distribution went largely unexplored. We have shown here that it is determined by the effective kinetic equation. In the particular case of linear systems, such kinetic theory acquires the form of the so-called Lyapunov equation of the matrix algebra. Although there aren't many standard analytic tools to deal with it, it may be treated with stable and efficient numerical algorithms.     

The tools, outlined here, allow one to completely solve quadratic many-body 
Lindbladian problems by diagonalizing 
$N\times N$ complex Hamiltonian and 
solving $N\times N$ Lyapunov equation for the stationary distribution. Notice that the dimensionality of the corresponding fermionic Hilbert space is $2^N$. Therefore one achieves the exponential reduction in the problem's complexity. 
An immediate extension of the quadratic theory is 
 the mean-field approximation, which deals with the linearized treatment near a certain (self-consistent) state. 
 
Still a lot to be done yet for better understanding truly non-linear many-body Lindbladian dynamics. In our opinion, the techniques presented here are indispensable for this goal. One of the most exciting applications of the functional methods is in the study of non-perturbative (instanton) effects  \cite{SP4}, which provide, eg., an ultimate floor for the qubit decoherence rate. 
Other examples of essentially non-linear phenomena include  
studies of non-equilibrium phase transitions \cite{SP1,SP2,SP3}, and various applications of the functional renormalization group to driven-dissipative problems \cite{DiehlKeldyshLindblad,BEC3,RG1,RG2,RG3}.

\appendix

\section{Keldysh-Nambu Diagonalization}\label{AppA}
This appendix examines the classical mechanics of the Keldysh action.  One can diagonalize the quadratic form by means of a complex coordinate transformation.  This is a generalized form of Bogoliubov rotation, extended to the Keldysh phase space space.  In this new set of coordinates, the dynamics can be understood through semiclassical quantization.

For bosons, one must perform a complex canonical transformation on the Keldysh phase space.  To this end, one should write the Keldysh action in the form of a Hamiltonian-Lagrangian.  Ignoring the constant term, the action from eq.~(\ref{ActionBoson}) is:
\begin{equation}
S=\int\dif t\Big(\Phi^\q i\partial_t\Phi^\cl-\frac{1}{2}\Phi^\alpha\check K_{\alpha\beta}\Phi^\beta\Big),
\end{equation}
where here $\Phi^\q=[\bar\phi^\q\ -\phi^\q]$ is defined differently than in the main text.  The classical field $\Phi^\q$ plays the role of the canonical position and the quantum field $\Phi^\q$, its conjugate momentum.  The matrix $\check K_{\alpha\beta}$ is given by:
\begin{equation}
\check K=\begin{bmatrix}0&\check H^\tT\\\check H&-i\check\tau^3\check D\check\tau^3\end{bmatrix}.
\end{equation}
Note that this is a symmetric matrix on the full $4N\times4N$ Keldysh-Nambu space.

The classical mechanics of the Keldysh action is a generalized Hamiltonian mechanics on the $4N$-dimensional Keldysh phase space.  The equations of motion are given by Hamilton's equations 
\begin{equation}
\partial_t\Phi^\alpha=-[\fK,\Phi^\alpha]_\mathrm{PB},
\end{equation}
where the Keldysh Poisson bracket is defined by:
\begin{subequations}\label{KeldyshPoissonBose}
\begin{equation}
[\cdot,\cdot]_\mathrm{PB}=\check\sigma^2_{\alpha\beta}\cev\partial_{\Phi^\alpha}\vec\partial_{\Phi^\beta},
\end{equation}
\begin{equation}
[\Phi^\cl_s,\Phi^\q_{s'}]_\mathrm{PB}=-i\delta_{ss'}.
\end{equation}
\end{subequations}
The matrix $\check J=i\check\sigma^2$ defines the symplectic form of the Keldysh classical mechanics.

With the correct choice of coordinates, one can express $\fK$ in a diagonal form.  Because $\check K$ is a symmetric matrix, $\check J\check K$ is an element of the complex symplectic algebra $\mathfrak{sp}(4N)$, defined through the relation $\check J(\check  J\check K)+(\check J\check K)^\tT\check J=0$.  It can be brought to Jordan canonical form by a (generically complex) symplectic matrix $\check V\in\mathrm{Sp}(4N,\C)$,
\begin{equation}
\check V\check J\check K\check V^{-1}=\begin{bmatrix}\check U\check H\check U^{-1}&0\\0&-(\check U\check H\check U^{-1})^\tT\end{bmatrix}.
\end{equation}
As a symplectic matrix, $\check V$ obeys $\check V^\tT\check J\check V=\check J$.  The $2N\times2N$ blocks of this matrix $\check V_{\alpha\beta}$ are given by:
\begin{equation}
\check V_{\cl\cl}=\check U,\quad \check V_{\cl\q}=-\check F_\mathrm{st}\check\tau^1\check U^{\tT-1},\quad \check V_{\q\cl}=0,\quad\check V_{\q\q}=\check U^{\tT-1},
\end{equation}
where $\check V$ obeys the symplectic condition $\check V^\tT\check J\check V=\check J$.
This matrix defines a complex canonical transformation to a new set of coordinates,
\begin{equation}\label{VBose}
\begin{bmatrix}\zeta\\\bar\zeta\end{bmatrix}=\check V\begin{bmatrix}\Phi^\cl\\\Phi^\q\end{bmatrix},
\end{equation}
Note that each of the new $2N$ fields not related by complex conjugation, $\bar\zeta_s\neq\zeta_s^*$.  They are however symplectic conjugates, obeying the relation:
\begin{equation}\label{DiagonalPoissonBose}
[\zeta_s,\bar\zeta_{s'}]_\mathrm{PB}=\delta_{ss'}.
\end{equation}

In these coordinates, the action is brought to a canonical form in terms of decoupled fields.  In the diagonalizable case, this is:
\begin{equation}
S=\sum_s\int\dif t\bar\zeta_s(i\partial_t-\epsilon_s)\zeta_s.
\end{equation}
Written in this form, one can see that there are $2N$ integrals of the classical motion, $I_s=\bar\zeta_s\zeta_s$.  They are related to the Keldysh Hamiltonian through $\fK=\sum_s\epsilon_sI_s$.  Applying the Bohr-Sommerfeld quantization rule, one puts $I_s=n_s$ with $n_s$ a set of positive integers.  The Lindbladian spectrum is given by the quantized values taken by the Keldysh Hamiltonian, $-i\fK=-\sum_sn_s\epsilon_s$.  In the non-diagonalizable, there are less than $2N$ integrals $I_s$, corresponding to the degeneracy of eigenvalues.  In this situation, the action will have additional terms of the form $\bar\zeta_s\zeta_{s+1}$ depending on the Jordan block structure of $\check H$.

A similar argument can be made for fermions.  Writing the fermion action in a Hamiltonian form, one has:
\begin{equation}
S=\int\dif t\Big(\Psi^2 i\partial_t\Psi^1-\frac{1}{2}\Psi^a\check K_{ab}\Psi^b\Big),
\end{equation}
with $\Psi^2=[\bar\psi^1\ \psi^2]$ defined differently than in the main text.  The matrix $\check K$ is given by:
\begin{equation}
\check K=\begin{bmatrix}0&-\check H^\tT\\\check H&-i\check D\check\tau^1\end{bmatrix}.
\end{equation}
Note that this is an antisymmetric matrix on the Keldysh-Nambu space.

The pseudo-classical fermionic equations of motion can be defined in terms of a fermionic Poisson bracket,
\begin{equation}
\partial_t\Psi^a=-\{\fK,\Psi^a\}_\mathrm{PB}.
\end{equation}
The bracket is symmetric and defined through the Grassmann derivatives:
\begin{subequations}\label{KeldyshPoissonFerm}
\begin{equation}
\{\cdot,\cdot\}_\mathrm{PB}=-i\check\sigma^1_{ab}\cev\partial_{\Psi^a}\vec\partial_{\Psi^b},
\end{equation}
\begin{equation}
\{\Psi^1_s,\Psi^2_{s'}\}=-i\delta_{ss'}.
\end{equation}
\end{subequations}
The matrix $\check\sigma^1$ defines the inner product on the Keldysh Grassmann algebra.  As such, the product $\check\sigma^1\check K$ is an element of the complex orthogonal algebra $\mathfrak{so}(4N)$, obeying the relation $\check\sigma^1(\check\sigma^1\check K)+(\check\sigma^1\check K)^\tT\check\sigma^1=0$.  It can be brought to Jordan canonical form by a complex orthogonal transformation $\check V\in\mathrm{SO}(4N,\C)$, which preserves the inner product $\check V^\tT\check\sigma^1\check V=\check\sigma^1$.  The block components of $\check V$ are the same as in eq.~(\ref{VBose}), by replacing index names $\cl\to1$ and $\q\to2$.  The new set of Grassmann fields are given by:
\begin{equation}\label{VFerm}
\begin{bmatrix}\xi\\\bar\xi\end{bmatrix}=\check V\begin{bmatrix}\Psi^1\\\Psi^2\end{bmatrix}.
\end{equation}
The semiclassical quantization of the moments $I_s=\bar\xi_s\xi_s$ gives the same result as for bosons, with the occupation numbers $n_s$ restricted to $0$ or $1$.

The usefulness of this representation beyond quadratic theory is limited.  Because the diagonal basis incorporates the stationary distribution matrix $\check F_\mathrm{st}$ into its definition, it is not obvious how to use this representation in an interacting theory or in the presence of time-dependent perturbations.  Conventionally $\check F$ is defined self-consistently on the quadratic level to derive a renormalized quantum kinetic equation that is non-linear in $\check F$.  It is unclear how this procedure could be performed, if at all, with $\check F$ absorbed into the definition of the fields.

\section{Superoperator Quantization}\label{AppB}
In this appendix, the connection between the formalism presented here and the third quantization superoperator formalism of \cite{3rdQuantBose,3rdQuantFerm} is established.  The quantization of the Keldysh action reproduces the superoperator formulation of the Lindbladian dynamics.  The connection between the two approaches is analogous to the relation between the Hilbert space and path integral formulations of standard quantum theory, but has an added wrinkle: due to the complex nature of the Keldysh classical mechanics, the quantization must be ``non-canonical" in that the generators of the boson and fermion superoperator algebras are not related by adjoint.

For the bosonic theory, one can introduce the right/left boson superoperators defined by:
\begin{equation}\label{SuperoperatorBose}
\doublehat a_{+j}\rho=\hat a_j\rho;\quad\doublehat a_{-j}\rho=\rho\hat a_j.
\end{equation}
These superoperators are bosonic in that they obey the bosonic commutation algebra:
\begin{equation}
[\doublehat a_{+j},\doublehat a_{+i}^\dagger]=\delta_{ij}=[\doublehat a_{-j}^\dagger,\doublehat a_{-i}],\quad[\doublehat a_{+j},\doublehat a_{-i}]=0.
\end{equation}
With these, the Keldysh action is quantized by replacing the Keldysh fields with the Keldysh-rotated superoperators, $\doublehat a_{\cl,\q j}=(\doublehat a_{+j}\pm\doublehat a_{-j})/\sqrt2$, and the Keldysh Poisson structure from Appendix \ref{AppA} eq.~(\ref{KeldyshPoissonBose}) with the quantum commutator,
\begin{subequations}\label{KeldyshCCR}
\begin{equation}
[\cdot,\cdot]_\mathrm{PB}\to-i[\cdot,\cdot],
\end{equation}
\begin{equation}
\phi^\alpha\to\doublehat a_\alpha;\quad\bar\phi^\alpha\to\doublehat a_\alpha^\dagger.
\end{equation}
\end{subequations}
Using this procedure, the Keldysh Hamiltonian $-i\fK$ of the action eq.~(\ref{ActionBoson}) is translated back into the Lindbladian defined by eq.~(\ref{Ham&JumpBose}).

The quantum and classical superoperators define the bosonic algebra:
\begin{equation}
[\doublehat a_{\cl j},\doublehat a^\dagger_{\q i}]=\delta_{ij},\quad [\doublehat a_{\cl j},\doublehat a^\dagger_{\cl i}]=0=[\doublehat a_{\cl j},\doublehat a_{\q i}],
\end{equation}
The superoperator algebra in this basis reflects the complex nature of the Keldysh classical mechanics.  The canonical conjugate and complex conjugate fields are note equivalent; the canonical conjugate of the field $\phi^\cl$ is $\bar\phi^\q$, not $\bar\phi^\cl$.  As a consequence, the bosonic superoperator algebra is non-conjugate.  The classical superoperators $\doublehat a_{\cl j}$ and $\doublehat a^\dagger_{\cl j}$ act as creation operators, while the quantum superoperators $\doublehat a^\dagger_{\q j}$ and $-\doublehat a_{\q j}$ are their corresponding annihilation operators respectively, despite not being their adjoints.

The Lindbladian can be brought to Jordan canonical form by a generalized Bogoliubov transformation of the bosonic superoperator algebra.  This is achieved by a change of basis implemented by $\check V$ from eq.~(\ref{VBose}).  The resulting set of bosonic superoperators $(\doublehat b,\doublehat b')$ obey a non-conjugate version of the bosonic commutation algebra given by the quantization of eq.~(\ref{DiagonalPoissonBose}),
\begin{equation}
[\doublehat b_s,\doublehat b'_{s'}]=\delta_{ss'},\quad [\doublehat b_s,\doublehat b_{s'}]=0=[\doublehat b'_s,\doublehat b'_{s'}],
\end{equation}
where the prime denotes the creation superoperator that in general is not related to the annihilation superoperator through adjoint, $\doublehat b_s^\dagger\neq\doublehat b_s'$.  A diagonalizable Lindbladian written in this basis adopts the familiar form of a sum over number operators:
\begin{equation}\label{LindDiag}
\doublehat\fL=-i\sum_s\epsilon_s\doublehat b_s'\doublehat b_s.
\end{equation}
For a non-diagonalizable Lindbladian, there will be additional terms of the form $\doublehat b_s'\doublehat b_{s+1}$.  This results in a sequence of Jordan blocks in full Lindbladian of ascending size for each $n$-particle sector.

For the fermionic theory, the situation is messier still.  This owes to the fact that the naive definition of fermionic left/right superoperators akin to eq.~(\ref{SuperoperatorBose}) turns out to give the wrong particle statistics.  With such a construction, left and right superoperators will commute instead of anti-commuting, yielding a parafermion algebra.  This issue can be fixed with non-conjugate version of a Klein transformation, amounting to the addition of an additional factor in the definition of one of the superoperators:
\begin{equation}
\doublehat c_{+j}\rho=\hat c_j\rho,\quad\doublehat c_{+j}'\rho=\hat c_j^\dagger\rho,\quad\doublehat c_{-j}\rho=\hat P\rho\hat P\hat c_j,\quad\doublehat c_{-j}'\rho=\hat P\rho\hat P\hat c_j^\dagger,
\end{equation}
where $\hat P=\exp(i\pi\sum_j\hat c_j^\dagger\hat c_j)$ is the fermion parity operator.  This definition ensures the superoperators are fermionic, obeying the fermionic commutation relations:
\begin{equation}
\{\doublehat c_{\pm j},\doublehat c_{\pm i}'\}=\delta_{ji},\quad\{\doublehat c_{+j},\doublehat c_{-i}\}=0=\{\doublehat c_{+j}',\doublehat c_{-i}'\}.
\end{equation}

With this construction, the Keldysh rotated fermionic superoperators, $\doublehat c_{1,2j}=(\doublehat c_{+j}\pm\doublehat c_{-j})/\sqrt2$ and $\doublehat c_{1,2j}'=(\doublehat c_{+j}'\mp\doublehat c_{-j}')/\sqrt2$ replace the Keldysh Grassmann fields and the anti-commutator replaces the fermionic Poisson bracket from Appendix \ref{AppA} eq.~(\ref{KeldyshPoissonFerm}) in the quantization of the fermionic theory,
\begin{subequations}\label{KeldyshCCRFerm}
\begin{equation}
\{\cdot,\cdot\}_\mathrm{PB}\to-i\{\cdot,\cdot\},
\end{equation}
\begin{equation}
\psi^a\to\doublehat c_a;\quad\bar\psi^a\to\doublehat c_a'.
\end{equation}
\end{subequations}
The fermionic superoperators obey the fermionic canonical anti-commutation relations:
\begin{equation}
\{\doublehat c_{aj},\doublehat c_{bi}'\}=\delta_{ij}\delta_{ab},\quad \{\doublehat c_{aj},\doublehat c_{bi}\}=0=\{\doublehat c_{aj}',\doublehat c_{bi}'\}.
\end{equation}
This is a non-conjugate representation of the fermion anti-commutation algebra.  Diagonalizing the Lindbladian can be achieved through the quantization of the diagonal Grassmann fields of eq.~(\ref{VFerm}) to a new set of fermion superoperators $(\doublehat d,\doublehat d')$, obeying the fermionic algebra:
\begin{equation}
\{\doublehat d_s,\doublehat d_{s'}'\}=\delta_{ss'},\quad\{\doublehat d_s,\doublehat d_{s'}\}=0=\{\doublehat d_s',\doublehat d_{s'}'\}.
\end{equation}
Like eq.~(\ref{LindDiag}), the Lindbladian is diagonal expressed in this basis,
\begin{equation}
\doublehat\fL=-i\sum_s\epsilon_s\doublehat d_s'\doublehat d_s.
\end{equation}

In this formulation, the stationary state $\rho_\mathrm{st}$ is the superoperator Fock vacuum.  All other eigenvalues of the Lindbladian can be explicitly constructed by creating on the vacuum with the boson or fermion creation superoperators, $\doublehat b'_s$ or $\doublehat d'_s$ respectively.  These eigenvectors are not states in the sense of density matrices but rather are traceless operators.

\section{Disconnected Diagrams}\label{AppC}
\begin{figure}
\begin{center}
\scalebox{.15}{\includegraphics{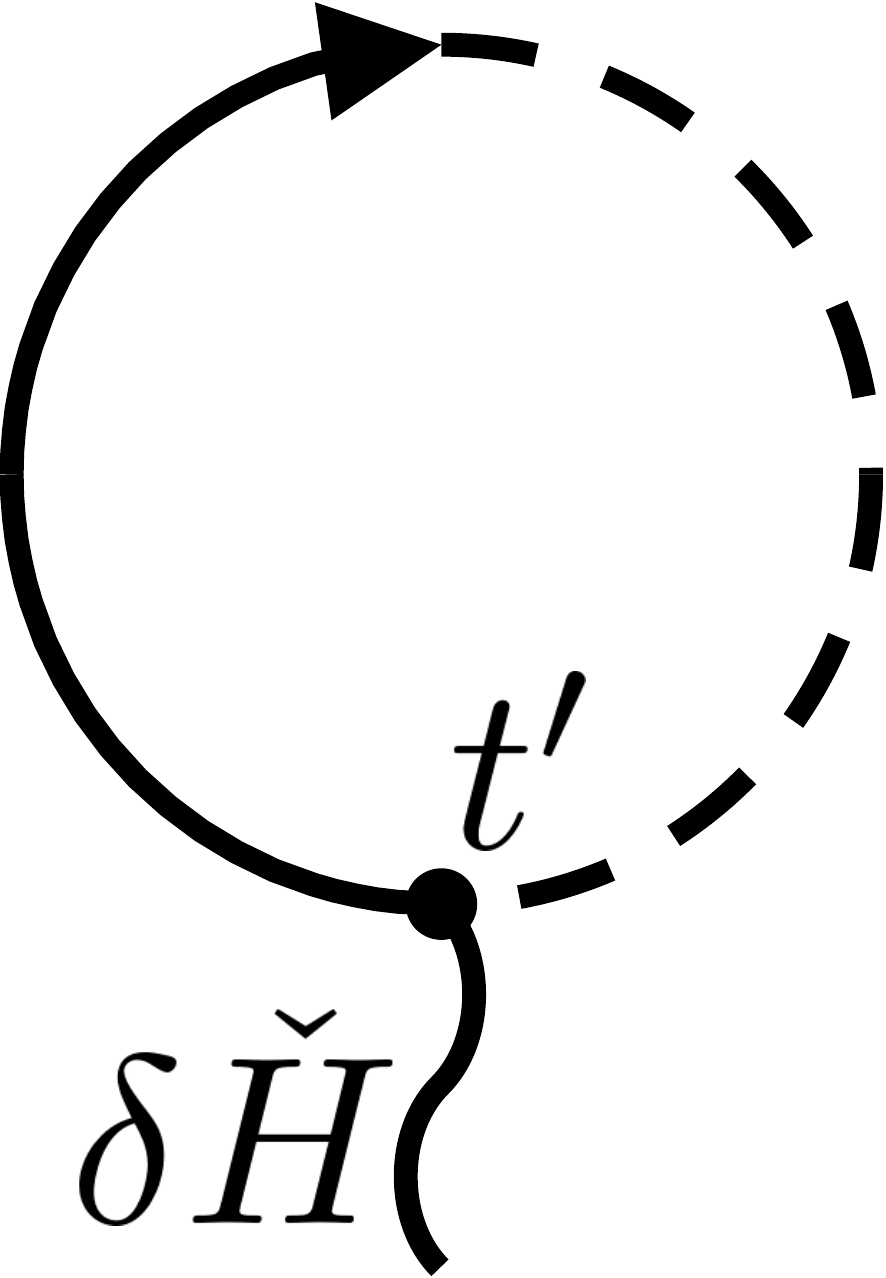}}
\qquad\qquad
\scalebox{.15}{\includegraphics{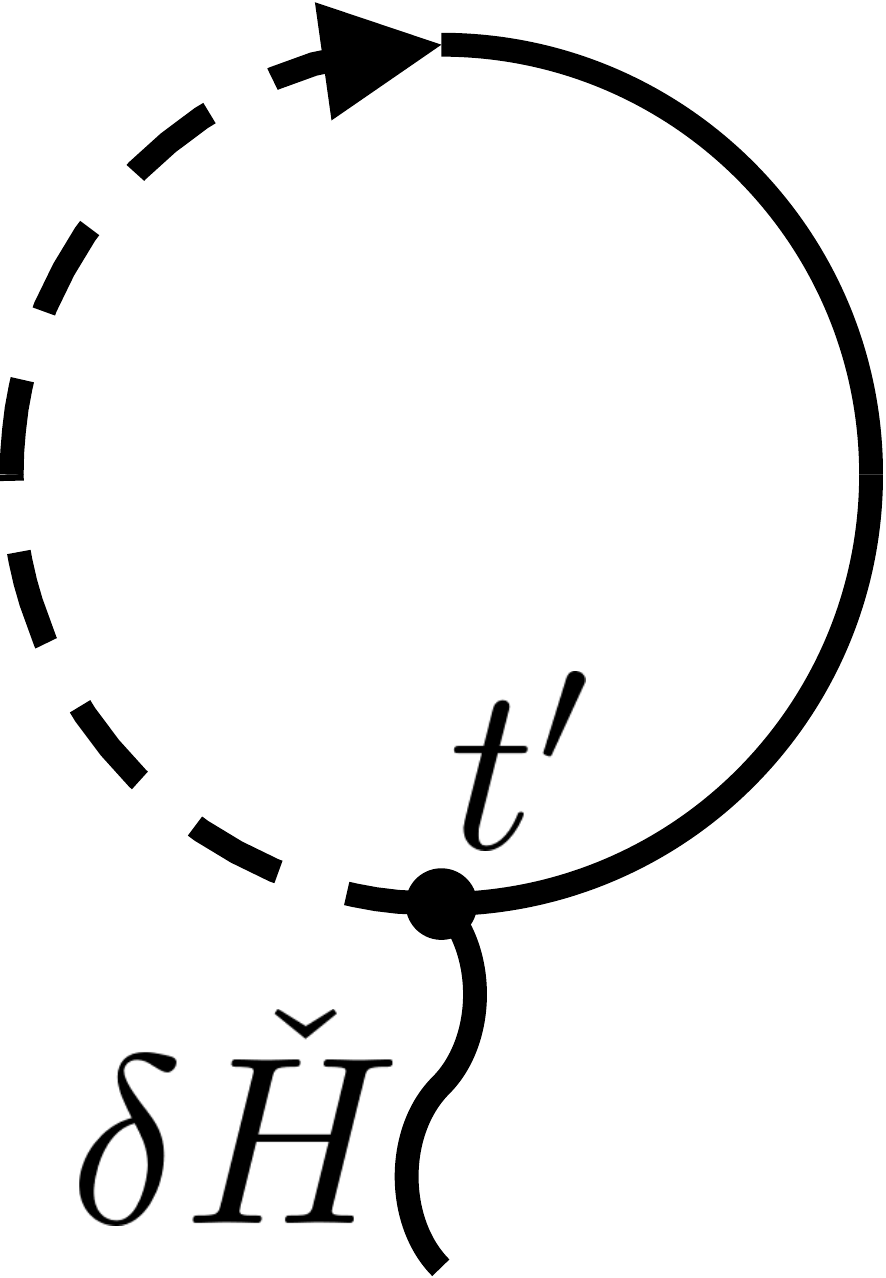}}
\qquad\qquad
\scalebox{.15}{\includegraphics{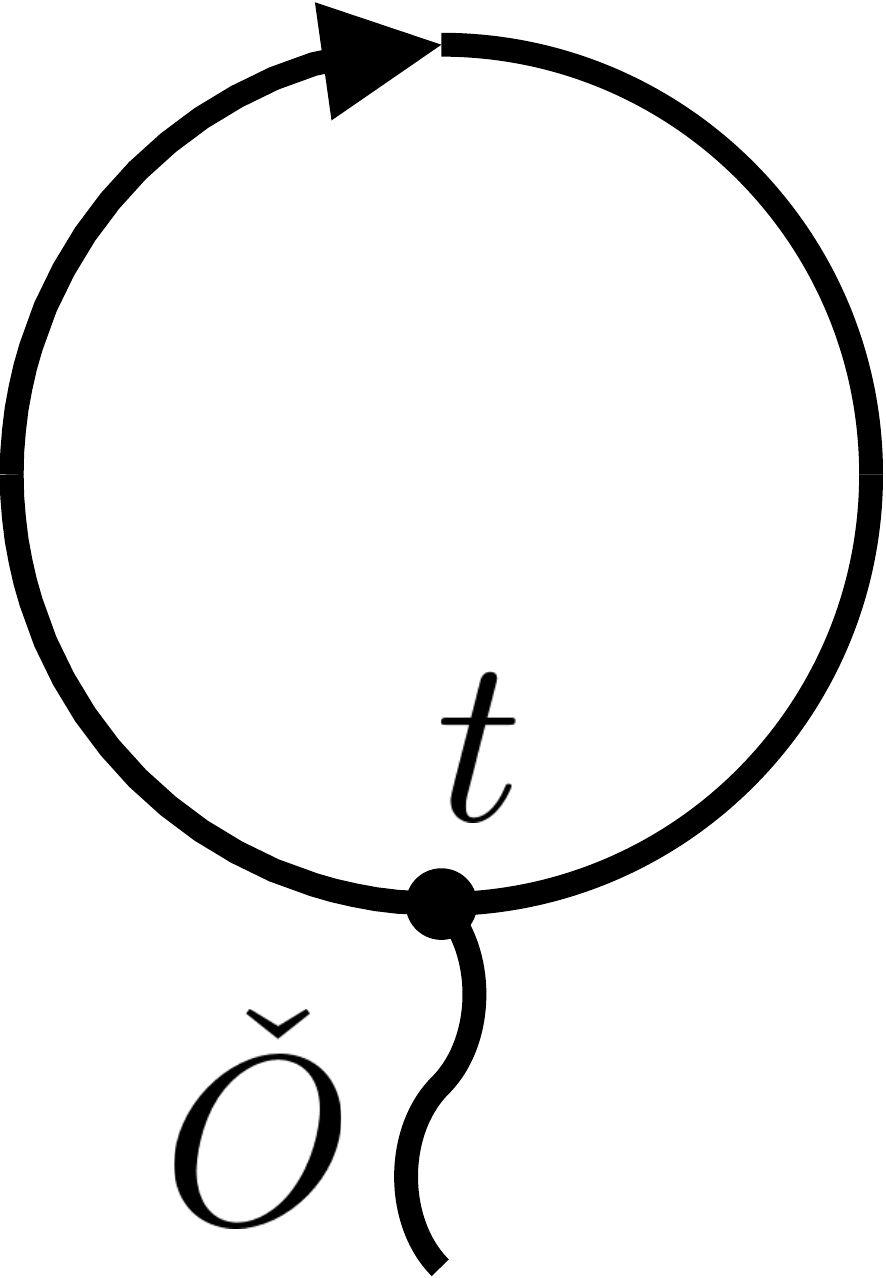}}
\caption{Disconnected loop diagrams featuring in the linear response formulas.  The left two diagrams are the retarded and advanced loops, the sums of which cancel.  The rightmost diagram is the Keldysh loop, which generically always appears as a prefactor in front of cancelling terms.}
\label{DisconnectedLoops}
\end{center}
\end{figure}

In this appendix, the cancellation of disconnected diagrams in the Lindblad Keldysh theory is addressed.  In the Keldysh diagrammatic theory, disconnected diagrams cancel identically as a consequence of the Keldysh causality structure.  This is in contrast, for example, to the Matsubara equilibrium theory in which cancellation happens due to competing contributions from the numerator and denominator of the partition function.  Disconnected diagrams in the Keldysh theory always come in pairs which translate to sums of retarded and advanced objects at equal times, which in turn vanish identically.  In the simplest case one may have factors of retarded and advanced Green's functions at equal times,
\begin{equation}\label{LoopSumZero}
\check G^\tR(t,t)+\check G^\tA(t,t)=0.
\end{equation}
This term appears for instance in the Wick contraction of the Hamiltonian linear response formula from the left-hand side of eq.~(\ref{LinRespH}).  Diagrammatically it is sum of disconnected loops; see Fig.~(\ref{DisconnectedLoops}).  

As a check on the validity of the Lindblad theory, one can verify that the normalization of the partition function in eq.~(\ref{KeldZBose}) remains $Z=1$ under the response to a perturbation.  The normalization of $\rho$, and therefore $Z$, should receive no perturbative corrections at any order.  Thus, expanding the exponentiated perturbation to the action in $Z$, $\exp(i\delta S)=1+i\,\delta S+...\,$, one expects all sub-leading terms to vanish identically.  To leading order, this mandates $\braket{\delta S}_\mathrm{st}=0$, or equivalently $\braket{\delta\fK(t)}_\mathrm{st}=0$, where $\braket{\cdot}_\mathrm{st}$ denotes the averaging with respect to the unperturbed system in its stationary state.

For simplicity, consider a quadratic perturbation as discussed in section \ref{2.4}.  The expectation of a Hamiltonian perturbation $\delta_{H_0}\fK$ possesses a sum of disconnected retarded and advanced loops as in eq.~(\ref{LoopSumZero}).  This is zero by virtue of being a sum of equal-time retarded and advanced objects as noted above.  The perturbation to $\check D$ gives only a factor of quantum-quantum correlation $\braket{\Phi^q\bar\Phi^q}$ and so is trivially also zero.  The perturbation to $\check Q$ however appears to include the difference of the retarded and advanced loops, $\check G^\tR(t,t)-\check G^\tA(t,t)$, which naively appears to be non-vanishing.  It is tempting to replace this difference with a factor of the constant matrix $\check\tau^3$, thus suggesting a non-vanishing contribution from disconnected terms of the form $\tr(\check\tau^3\delta\check Q)$.  This is of course erroneous.  The origin of the cancellation of these terms can be traced to the physical origin of the Lindbladian dynamics as arising from integrating out the environment of an open system.

Schematically, this procedure begins with a system coupled to a large number of bath degrees of freedom.  Upon integrating out the bath, bare system quantities are renormalization by the bath.  In the case of a quadratic theory, interactions with the bath leads to modification of the bare Green function by a self-energy, yielding an effective action for the system of the form:
\begin{equation}
S=\int\dif t\dif t'\bar\Phi(t) \Big(\check G_0^{-1}(t,t')-\check\Sigma(t,t')\Big)\Phi(t'),
\end{equation}
where the bare inverse Green's function is of the form of the quadratic form in eq.~(\ref{ActionBoson}) without the dissipative terms.  Under the Markovian approximation for the Lindbladian theory, the self-energy $\check\Sigma$ is local in time $\check\Sigma(t,t')\propto\delta(t-t')$.  In the operator language, the imaginary parts of self energy generate the dissipative part of the Lindbladian evolution ($\fD$ from eq.~(\ref{KeldHam&Diss})).  The retarded and advanced components specify $\check Q$,
\begin{equation}
\mathbb{I}\mathrm{m}\Big(\check\Sigma^{\tR,\tA}(t,t')\Big)\simeq\pm\delta(t-t')\check Q.
\end{equation}
Note however that the continuum notation is deceptive here.  The regularization of the retarded and advanced components of $\check\Sigma$ are different even in the Markovian approximation: the arguments of the delta functions should be understood as differing by an infinitesimal time step in opposite directions $\delta^{\tR,\tA}(t-t')\sim\delta_{t,t\pm\delta t}$.  As such, the perturbative corrections to the dissipative part of the Lindbladian action should be understood as a modifications to the spectral components of the self-energy.  Leading-order contributions are thus {\em not} of the form of just the difference of the retarded and advanced loops from Fig.~(\ref{DisconnectedLoops}), but rather should be understood implicitly as:
\begin{equation}
\tr\Big(\check G^\tR\circ\check\Sigma^\tR+\check G^\tA\circ\check\Sigma^\tA\Big),
\end{equation}
where the trace is taken over both the time and matrix spaces.  This is appropriately the sum of retarded and advanced objects, and as such should be understood as zero.  In practice, one can safely use the continuum notation for calculations using the formalism presented in the main body with the understanding that disconnected diagrams should always cancel due to the underlying regularization.

 \bibliographystyle{elsarticle-num} 
 \bibliography{biblio}

\end{document}